\documentclass[twocolumn]{aastex631}
\usepackage{amsmath}
\usepackage{booktabs}
\usepackage{multirow}
\usepackage{tikz}
\usepackage{bm}         

\newcommand{\ra}[1]{\renewcommand{\arraystretch}{#1}}  

\newcommand*\circled[1]{\tikz[baseline=(char.base)]{
    \node[shape=circle,draw,inner sep=2pt] (char) {#1};}}

\makeatletter
\newcommand*{\linktocite}[2]{%
  \hyper@natlinkstart{#1}#2\hyper@natlinkend}
\makeatother

\submitjournal{AAS Journals}

\shorttitle{3D Transmission Spectra with TRIDENT}
\shortauthors{MacDonald \& Lewis}

\graphicspath{{./}{figures/}}

\begin{document}

\title{TRIDENT: A Rapid 3D Radiative Transfer Model for Exoplanet Transmission Spectra}

\correspondingauthor{Ryan MacDonald}
\email{rmacdonald@astro.cornell.edu}

\author[0000-0003-4816-3469]{Ryan J. MacDonald}
\author[0000-0002-8507-1304]{Nikole K. Lewis}
\affiliation{Department of Astronomy and Carl Sagan Institute, Cornell University, 122 Sciences Drive, Ithaca, NY 14853, USA}

\begin{abstract}

\noindent Transmission spectroscopy is one of the premier methods used to probe the temperature, composition, and cloud properties of exoplanet atmospheres. Recent studies have demonstrated that the multidimensional nature of exoplanet atmospheres --- due to non-uniformities across the day-night transition and between the morning and evening terminators --- can strongly influence transmission spectra. However, the computational demands of 3D radiative transfer techniques have precluded their usage within atmospheric retrievals. Here we introduce TRIDENT, a new 3D radiative transfer model which rapidly computes transmission spectra of exoplanet atmospheres with day-night, morning-evening, and vertical variations in temperature, chemical abundances, and cloud properties. We also derive a general equation for transmission spectra, accounting for 3D atmospheres, refraction, multiple scattering, ingress/egress, grazing transits, stellar heterogeneities, and nightside thermal emission. After introducing TRIDENT's linear algebra-based approach to 3D radiative transfer, we propose new parametric prescriptions for 3D temperature and abundance profiles and 3D clouds. We show that multidimensional transmission spectra exhibit two significant observational signatures: (i) day-night composition gradients alter the relative amplitudes of absorption features; and (ii) morning-evening composition gradients distort the peak-to-wing contrast of absorption features. Finally, we demonstrate that these signatures of multidimensional atmospheres incur residuals $> 100$\,ppm compared to 1D models, rendering them potentially detectable with JWST. TRIDENT's rapid radiative transfer, coupled with parametric multidimensional atmospheres, unlocks the final barrier to 3D atmospheric retrievals.

\end{abstract}


\section{Introduction}
\label{sec:intro}

Transiting exoplanets are a fortuitous gift of nature, providing distant observers a regular glimpse into the atmospheres of these distant worlds. Transmission spectroscopy \citep{Seager2000,Brown2001,Hubbard2001} compares the effective occulting area of a planet, relative to its host star, before and during primary transit. This differential measurement, constituting a spectrum when observed at multiple wavelengths, encodes a wide variety of atmospheric properties. For giant planets, infrared transmission spectra contain a profusion of absorption features from light molecules \citep[e.g.][]{Snellen2010,Deming2013,Spake2021} and condensate vibrational modes \citep[e.g.][]{Wakeford2015,Pinhas2017}. Visible wavelength transmission spectra contain signatures of aerosol scattering \citep[e.g.][]{Pont2013,Sing2016,Ohno2020}, alkali metals \citep[e.g.][]{Charbonneau2002,Chen2018,Alam2021}, and heavy metal oxides / hydrides \citep[e.g.][]{Sharp2007,Sedaghati2017}. Near-UV transmission spectra offer a promising window into heavy atoms \citep[e.g.][]{Hoeijmakers2018,Lothringer2020} and ionic species \citep[e.g.][]{Vidal-Madjar2004,Lewis2020}. These signatures, in turn, can be affected by atmospheric winds \citep[e.g.][]{Seidel2020}, rotation \citep[e.g.][]{Brogi2016}, escape processes \citep[e.g.][]{Allart2018}, and magnetic fields \citep[e.g.][]{Oklopcic2020}.

Over 50 exoplanets have observed transmission spectra to date \citep{Zhang2020}. Atmospheric detections now cover planets ranging in mass from ultra-hot Jupiters \citep[e.g.][]{Evans2018,Casasayas-Barris2019} through exo-Neptunes \citep[e.g.][]{Fraine2014,Wakeford2017} and recently into the sub-Neptune regime \citep{Benneke2019,Tsiaras2019}. The \emph{James Webb Space Telescope} (JWST) will open the realm of terrestrial planet transmission spectroscopy \citep[e.g.][]{Greene2016,Lustig-Yaeger2019,Lin2021}.

The theory of exoplanet transmission spectroscopy has seen significant development over the last 20 years. Early studies focused on 1D model atmospheres, concluding that transmission spectra of giant planets can be calculated assuming stellar rays traverse the day-night terminator region in straight lines \citep{Seager2000,Brown2001,Hubbard2001}. Later, General Circulation Models (GCMs) predicted that hot Jupiters exhibit day-night temperature gradients and eastward-flowing equatorial jets \citep[e.g.][]{Showman2002,Cho2003,Menou2009}. Consequently, the terminator --- the region probed by transmission spectra --- can feature strong gradients in both temperature \citep[e.g.][]{Dobbs-Dixon2008} and chemical composition \citep[e.g.][]{Showman2009}. This prompted the development of 3D radiative transfer codes for transmission spectra \citep{Fortney2010,Burrows2010,Dobbs-Dixon2012}. 3D transmission spectra can differ substantially from 1D models, since rays travelling through the dayside accrue absorption features from a region with a different composition than the terminator plane \citep{Fortney2010}. Alongside mid-transit spectroscopy, spectra taken during transit ingress/egress can probe temperature, composition, and aerosol differences between the morning and evening terminators \citep{Fortney2010,Burrows2010,Dobbs-Dixon2012,Kempton2017}. 

More recently, many studies have shown that previously-neglected processes can shape transmission spectra. For terrestrial exoplanets, refraction can be important \citep[e.g.][]{GarciaMunoz2012,Betremieux2014}. Other studies have revisited the impact of multiple scattering \citep[e.g.][]{Hubbard2001,Robinson2017a,Lee2019}. A consensus is now emerging that transmission spectra can be contaminated by sources distinct from transmitted starlight. Hot giant planet transmission spectra contain a contribution from nightside thermal emission \citep{Kipping2010}, which can be significant in the mid-infrared \citep{Chakrabarty2020,Morello2021}. Stellar heterogeneities (spots or faculae) outside the transit chord can imprint non-planetary signatures into transmission spectra \citep[e.g.][]{Rackham2018,Iyer2020}. However, many studies investigate only a subset of these effects, potentially limiting our ability to understand their competing contributions to observed spectra. Our first goal in this paper is thus to present a unified theoretical framework for transmission spectra.

The interpretation of observed exoplanet spectra follows two paradigms, the first of which is termed forward modelling \citep[e.g.][]{Seager1998,Sudarsky2003,Morley2013,Goyal2019}. A forward model generates an atmospheric spectrum from first-principles, self-consistently calculating the temperature structure and composition (traditionally assuming radiative-convective and thermochemical equilibrium). One then compares theoretical spectra generated from forward models with observed spectra to determine which model provides the best match to the data. Recently, forward modelling studies have folded processes such as disequilibrium chemistry and cloud microphysics into 3D GCM frameworks \citep[e.g.][]{Parmentier2016,Lines2018,Venot2020}. Forward models provide vital insights into the interplay between chemistry, winds, and radiation in planetary atmospheres, but they come with a steep computational cost.

The second modelling paradigm is the inverse approach known as atmospheric retrieval \citep[e.g.][]{Madhusudhan2009,Benneke2012,Line2013,Waldmann2015,MacDonald2017a,Molliere2019,Cubillos2021}. Retrievals wrap an atmospheric model, described by a set of parameters (mixing ratios, temperature, cloud pressure etc.), inside a Bayesian sampling algorithm \citep[see][]{Trotta2017}. Retrieval codes output statistical constraints on the parameters underpinning the model, alongside detection significances for model components \citep[see][for a review]{Madhusudhan2018}. While retrievals can explore a wider range of solutions than forward models, they rely on rapid evaluation (10$^5$ models in $<$ 1 day) to fully map the parameter space. This speed requirement has precluded many important model considerations from transmission spectra retrievals, with most retrieval codes assuming uniform 1D atmospheres.

Recent years have seen a renewed focus on the limitations of 1D models in interpreting exoplanet transmission spectra. \citet{Line2016} demonstrated that inhomogenous terminator clouds can mimic a 1D cloud-free spectrum with a high mean molecular weight. \citet{MacDonald2017a} subsequently found evidence of patchy clouds at HD 209458b's terminator, confirming that high-quality \emph{Hubble} transmission spectra already contain signatures of multidimensional clouds \citep[see also][]{Pinhas2019,Barstow2020}. Later, \citet{Caldas2019}, \citet{Pluriel2020}, and \citet{Pluriel2021} showed that transmission spectra with day-night temperature or abundance gradients result in biased inferences when retrieved by a 1D model. We showed in \citet{MacDonald2020} that morning-evening abundance gradients can produce erroneously low retrieved temperatures when using 1D models.

Initial efforts are now underway to develop 2D retrieval techniques to mitigate some of these biases. \citet{Lacy2020a} demonstrated that distinct dayside and nightside temperatures can be retrieved from simulated JWST transmission spectra. \citet{Espinoza2021} found the same for distinct morning and evening terminator temperatures. With the launch of JWST, the need to transition to multidimensional retrievals is now clear. The horizon goal of these research endeavours is a generalised 3D retrieval technique.

The development of 3D atmospheric retrieval has been hindered by two major obstacles. First, existing 3D radiative transfer models are too computationally demanding to be wrapped in retrieval frameworks. Second, it is unclear how to parametrise 3D models in a general way (i.e. avoiding assumptions such as chemical equilibrium, clear atmospheres), while maintaining a reasonable number of free parameters. Our second goal in this paper is to offer solutions to these problems.

Here we present TRIDENT\footnote{\textbf{T}hree-dimensional \textbf{R}adiative transfer via \textbf{I}ntegration of \textbf{D}iscretised \textbf{E}xoplanets with \textbf{N}on-uniform \textbf{T}erminators.}, a new 3D radiative transfer model designed from the outset to enable 3D retrievals of exoplanet transmission spectra. We develop a new radiative transfer algorithm for 3D atmospheres, generalising the path distribution approach introduced by \citet{Robinson2017a}, which uses efficient linear algebra operations to rapidly generate 3D transmission spectra. We also introduce parametric descriptions for 3D temperature and abundance variations and a 3D cloud model. These new temperature, abundance, and cloud prescriptions render the parameter space of multidimensional atmospheres tractable for 3D retrievals.

Our study is structured as follows. In Section~\ref{sec:unifed_model}, we present a unified model of transmission spectra. We introduce TRIDENT in Section~\ref{sec:TRIDENT_forward_model}, wherein we describe the computation of 3D transmission spectra and our parametrisations for multidimensional atmospheres. We validate TRIDENT against existing radiative transfer codes in Section~\ref{sec:validation}. We explore the influence of 3D atmospheric properties on transmission spectra in Section~\ref{sec:3D_signatures}. Finally, we summarise our results, discuss their implications, and note future directions in Section~\ref{sec:discussion}.

\section{A Unified Model of Exoplanet Transmission Spectra} \label{sec:unifed_model}

\begin{figure*}[ht!]
    \centering
    \includegraphics[width=\textwidth]{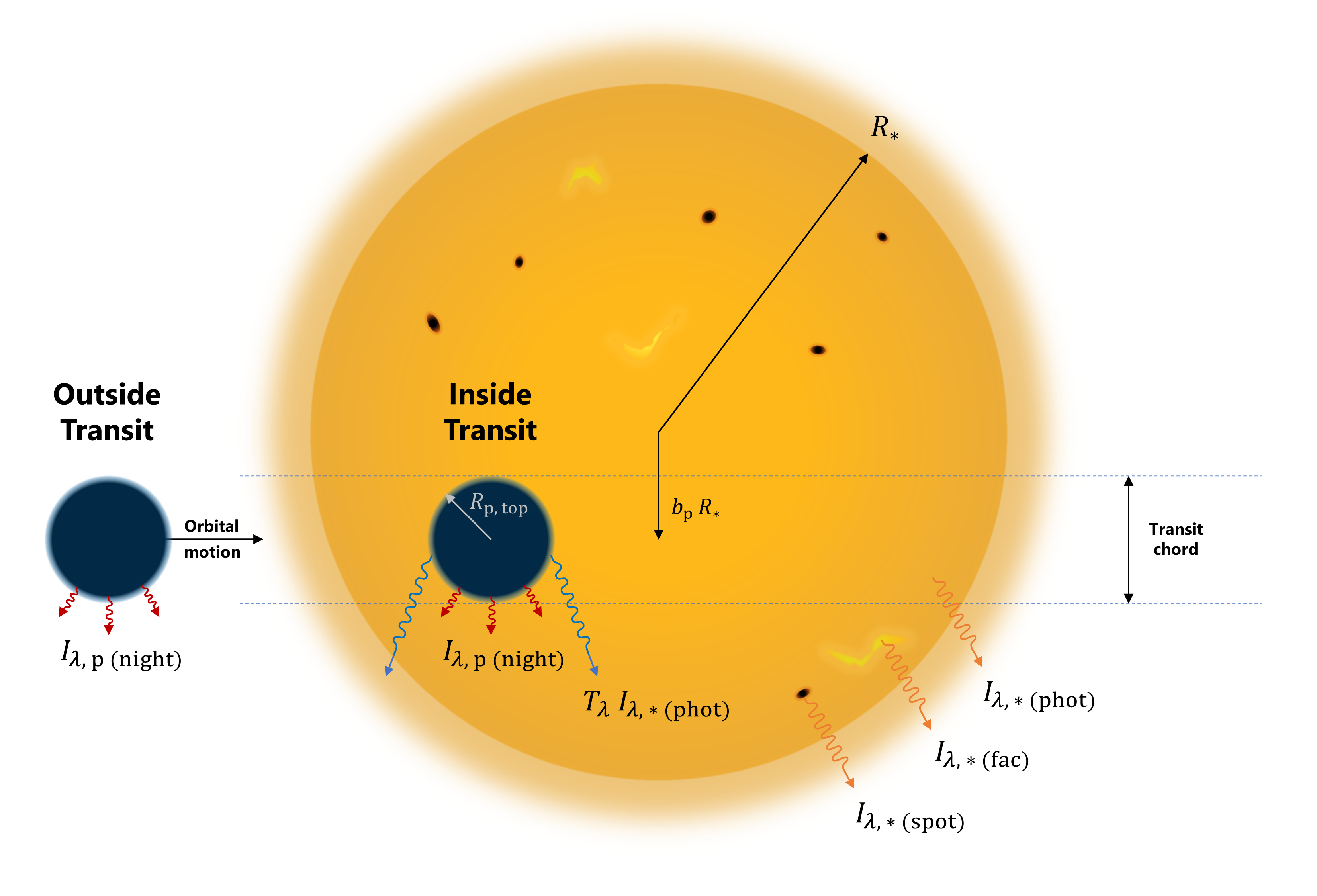}
    \caption{Schematic diagram of the processes encoded in an exoplanet transmission spectrum. The transit geometry is face-on from the observer perspective. Outside transit, the observed flux has two components: the integrated intensity of the stellar disc and thermal emission from the planetary nightside. The stellar flux can include contributions from spots and faculae outside the transit chord (centred on $b_{\rm{p}}$). Inside transit, the observed flux has three components: the integrated intensity of the unobscured portion of the stellar disc, thermal emission from the nightside, and stellar light transmitted through the terminator.}
\label{fig:schematic_diagram}
\end{figure*}

A transmission spectrum encodes how the spectral flux observed from an extrasolar system changes when an exoplanet transits its host star. In general, transmission spectra are shaped by three processes: (i) stellar flux from the host star, which can include contributions from areas of the stellar photosphere with cool spots or hot faculae; (ii) thermal flux emitted by the planetary nightside; and (iii) stellar flux transmitted through the day-night terminator of the transiting exoplanet. We illustrate these processes schematically in Figure~\ref{fig:schematic_diagram}.

We show in Appendix~\ref{appendix_A} that a transmission spectrum can be generally decomposed into three factors:

\begin{equation}
    \Delta_{\lambda} \equiv \frac{F_{\lambda, \, \rm{out}} - F_{\lambda, \, \rm{in}}}{F_{\lambda, \, \rm{out}}} = \delta_{\lambda, \, \rm{atm}} \, \epsilon_{\lambda, \, \rm{het}} \, \psi_{\lambda, \, \rm{night}}
\label{eq:transmission_spectrum_general}
\end{equation}
where
\begin{equation}
    \delta_{\lambda, \, \rm{atm}} = \frac{A_{\rm{p \, (overlap)}} -  \displaystyle\int_{A_{\rm{p}}} \overline{\delta_{\rm{ray}*} \, \mathcal{T}_{\lambda}} \, dA}{\pi R_{*}^2}
\label{eq:atmosphere_factor}
\end{equation}
\begin{equation}
    \epsilon_{\lambda, \, \rm{het}} = \frac{1}{1 - \displaystyle\sum_{i=1}^{N_{\rm{het}}} f_{\rm{het}, \, i} \left(1 - \frac{I_{\lambda, \, \rm{het}, \, i}}{I_{\lambda, \, *}} \right)}
\label{eq:stellar_contam_factor}   
\end{equation}
\begin{equation}
    \psi_{\lambda, \, \rm{night}} = \frac{1}{1 + \displaystyle\frac{F_{\lambda, \, \rm{p \, (night)}}}{F_{\lambda, \, *}}}
\label{eq:nightside_contam_factor}   
\end{equation}
The first factor, $\delta_{\lambda, \, \rm{atm}}$, encodes the transit depth of the planet and its atmosphere. The second factor, $\epsilon_{\lambda, \, \rm{het}}$, accounts for unocculted stellar heterogeneities (spots and faculae) outside the transit chord \citep[e.g.][]{Rackham2018}. The final factor arises from the thermal emission of the planetary nightside \citep[e.g.][]{Kipping2010}. We now elaborate on each of these expressions.

Equation~\ref{eq:atmosphere_factor} specifies the effective area ratio between a transiting planet and its host star. The first term gives the transit depth if the entire atmosphere were opaque, while the second term subtracts area elements weighted by the fraction of light transmitted. In this equation, $A_{\mathrm{p \, (overlap)}}$ is the projected area of a circular disc, encompassing the planet and its atmosphere, overlapping a star of radius $R_{*}$. In the case of a spherical planet completely overlapping its star, $A_{\mathrm{p \, (overlap)}} = \pi R_{\mathrm{p, \, top}}^2$ (where $R_{\mathrm{p, \, top}}$ is the planetary radius at the top of the modelled atmosphere). For more general transiting geometries --- such as grazing transits or during ingress/egress --- $A_{\mathrm{p \, (overlap)}}$ is determined analytically by the area of overlap between two circles of radii $R_{\mathrm{p, \, top}}$ and $R_{*}$ at the relevant projected distance between their centres. The \emph{transmission} of an area element, $\mathcal{T}_{\lambda}$, is the ratio between the stellar intensity transmitted through an atmospheric area element and the intensity incident on the atmosphere. It is given by
\begin{equation}
    \mathcal{T}_{\lambda} \equiv e^{-\tau_{\lambda, \, \rm{path}}}
\label{eq:transmission}
\end{equation}
where $\tau_{\lambda, \, \rm{path}}$ is the \emph{path optical depth} experienced by a ray or photon. The last term, $\delta_{\rm{ray}*}$, is defined in analogy with the Kronecker delta
\begin{equation}
    \delta_{\rm{ray}*} = 
    \begin{cases}
        1 \ , &\text{if ray intersects star}\\
        0 \ , &\text{else}
    \end{cases}
\label{eq:delta_ray}  
\end{equation}
i.e. $\delta_{\rm{ray}*}$ is unity only if a ray backtracked from an observer intersects the stellar surface. In the geometric limit (where refraction and scattering are neglected, such that rays travel on straight lines), $\delta_{\rm{ray}*} = 1$ for any area element of the planetary atmosphere occulting the star. Therefore, in the geometric limit, a full transit (after ingress but before egress) has $\delta_{\rm{ray}*} = 1 \, \forall \, dA$ and this term can be dropped from Equation~\ref{eq:atmosphere_factor}. In full generality, $\delta_{\rm{ray}*}$ can be numerically computed --- via ray tracing or backwards Monte Carlo prescriptions \citep[e.g.][]{Robinson2017a} --- to account for rays that are deflected onto ($\delta_{\rm{ray}*} = 1$) or off ($\delta_{\rm{ray}*} = 0$) the star. The overline above $\delta_{\rm{ray}*}$ and $\mathcal{T}_{\lambda}$ denotes a photon average; models excluding multiple scattering can neglect this average and use a single ray for each area element. Equation~\ref{eq:atmosphere_factor} holds for 1D, 2D, and 3D atmospheres, with the multidimensional nature of a planetary atmosphere encoded in the optical depth, $\tau_{\lambda, \, \rm{path}}$ (see Section~\ref{subsec:radiative_transfer}).

Equation~\ref{eq:stellar_contam_factor} accounts for any differences between the stellar intensity of the transit chord (the light incident on the atmosphere) and the average intensity of the stellar disc. This factor is also referred to as the `transit light source effect' \citep[e.g.][]{Rackham2018}. For a uniform stellar disc, $\epsilon_{\lambda, \, \rm{het}} = 1$. When unocculted \emph{stellar heterogeneities} (i.e. spots or faculae) are present, Equation~\ref{eq:stellar_contam_factor} models their influence according to the fractional coverage area of the $i^\text{th}$ heterogeneity, $f_{\rm{het}, \, i}$, and the intensities of the heterogeneities and background photosphere, $I_{\lambda, \, \rm{het}, \, i}$ and $I_{\lambda, \, *}$, respectively. Different active regions can be included via the summation in Equation~\ref{eq:stellar_contam_factor}. We note that this functional form assumes that the transit chord is representative of the stellar photosphere (i.e. no spots or faculae are occulted).

Finally, Equation~\ref{eq:nightside_contam_factor} models `nightside contamination' from the thermal emission of the planetary nightside \citep{Kipping2010}. Under the approximation of a `dark' nightside, $\psi_{\lambda, \, \rm{night}} = 1$.  Since the emergent flux from the nightside is wavelength-dependent (due to absorption and/or emission from chemical species on the nightside), $\psi_{\lambda, \, \rm{night}}$ can `contaminate' transmission spectra (if once considers a transmission spectrum only as Equation~\ref{eq:atmosphere_factor}). The influence of nightside contamination becomes more pronounced at longer wavelengths, especially the mid-infrared, where planet-star flux ratios are higher \citep{Kipping2010}.

The focus of this paper is to demonstrate how multidimensional atmospheric properties impact exoplanet transmission spectra. Our starting point for radiative transfer will thus be the general formula, Equation~\ref{eq:transmission_spectrum_general}, which holds for 1D, 2D, and 3D atmospheres. We now turn to the computation of transmission spectra for generalised multidimensional atmospheres.

\section{Transmission Spectra of 3D Exoplanet Atmospheres} \label{sec:TRIDENT_forward_model}

TRIDENT is a new 3D radiative transfer model designed to be sufficiently fast to serve as a forward model for multidimensional atmospheric retrievals. In this section, we first describe our method for rapid computation of 3D transmission spectra. We then introduce how we parametrise multidimensional atmospheres, summarise our opacity sources, and, finally, introduce a new prescription for 3D cloud coverage in transmission spectra.

\subsection{Radiative Transfer} \label{subsec:radiative_transfer}

\begin{figure*}[ht!]
    \centering
    \includegraphics[width=\textwidth]{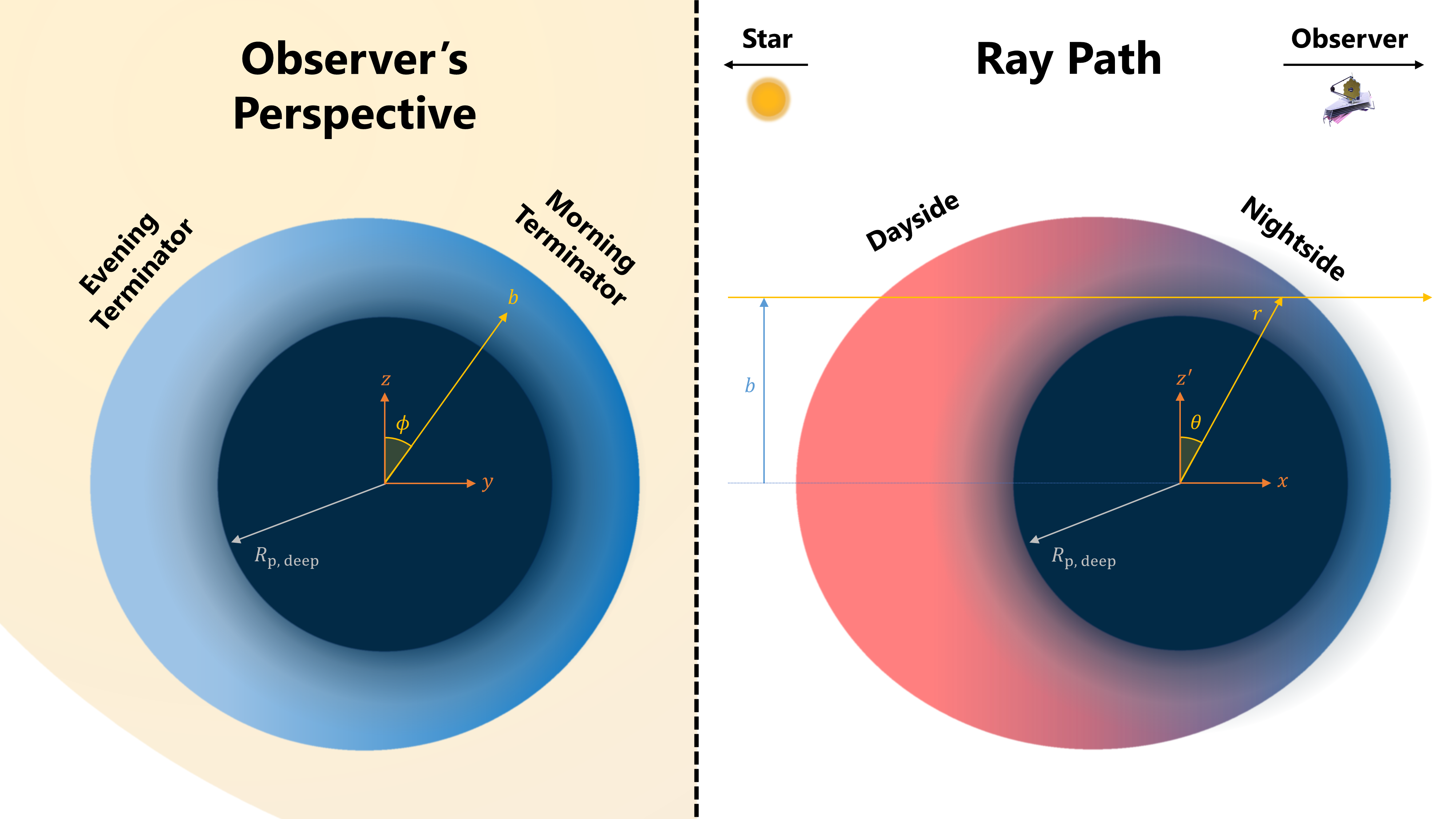}
    \caption{Geometry and coordinate systems used to compute 3D transmission spectra. Left: an observer sees a transiting planet as an opaque disc, of radius $R_{\mathrm{p, \, deep}}$, surrounded by a partially transmittable atmosphere. The star is shown as background shading. The morning terminator (first to cross the stellar disc) is not generally the same temperature as the evening terminator, distorting the projected shape of the atmosphere from an annulus. A Cartesian coordinate system $(z, y)$ has $z$ aligned with the north pole and $y$ with the path of orbital motion. A polar coordinate system $(b, \phi)$ specifies the atmospheric sector probed by a ray exiting the atmosphere with impact parameter $b$ and azimuthal angle $\phi$. Right: a ray with impact parameter $b$ enters the dayside, crosses the terminator plane, and emerges from the nightside. The dayside has a significantly greater radial extent than the nightside. A Cartesian coordinate system $(z', y)$ has $z'$ aligned with the impact parameter ($z' = z$ only when $\phi = 0$) and $x$ specifies the distance from the terminator plane. Local properties of the atmosphere are specified by $(r, \theta, \phi)$, where $\theta$ gives the angular separation from the terminator plane. Radiative transfer is calculated in the cylindrical coordinate system $(b, \phi, x)$. }
    \label{fig:coordinate_system}
\end{figure*}

Consider a tidally-locked transiting exoplanet, as illustrated in Figure~\ref{fig:coordinate_system}. The stellar flux incident on the planetary dayside causes a day-night temperature difference, resulting in an extended dayside atmosphere compared to the nightside. Atmospheric circulation similarly redistributes heat asymmetrically across the terminator region, typically resulting in a warmer evening terminator compared to the morning terminator \citep[e.g.][]{Showman2002}. This inhomogenous temperature distribution causes different regions to have distinct scale heights, hence making the atmosphere non-spherical. These temperature gradients can, in turn, drive differences in the chemical composition and cloud properties throughout the atmosphere. Since stellar rays sample atmospheric regions with substantially different properties as they traverse the day-night transition (Figure~\ref{fig:coordinate_system}, right panel), with different rays sampling different terminator sectors (Figure~\ref{fig:coordinate_system}, left panel), radiative transfer must be solved using a 3D coordinate system.

\subsubsection{Coordinate Systems} \label{subsubsec:coordinate_systems}

A cylindrical coordinate system is a natural choice for radiative transfer, given the symmetry of the problem \citep{Fortney2010}. Consider a Cartesian coordinate system $(x, y, z)$, with $x$ encoding the distance from the day-night boundary (increasing towards the observer), $y$ aligned with the direction of orbital motion, and $z$ pointing to the planetary north pole. From the perspective of a distant observer, $y-z$ defines the \emph{terminator plane}, located at $x = 0$. We define a polar coordinate system $(b, \phi)$ in the terminator plane (Figure~\ref{fig:coordinate_system}, left panel), with $b$ the radial coordinate and $\phi$ the azimuthal angle with respect to the north pole. The advantage of this coordinate system is that a ray has constant values of $b$ and $\phi$ in the geometric limit (i.e. rays travel in straight lines when refraction and scattering is negligible\footnote{The radial coordinate can deviate from $b$ due to refraction, but $\phi$ remains constant even with refraction. Scattering can be treated as a loss term from rays with constant $(b, \phi)$ or, in general, Monte Carlo approaches can track how these coordinates change following photons through the atmosphere \citep[e.g.][]{Robinson2017a}.}). The only changing coordinate for a ray is then $x$. The trajectory of a ray is thus readily specified by a cylindrical coordinate system $(b, \phi, x)$.

Atmospheric properties, meanwhile, are more naturally described by a spherical coordinate system. While spherical coordinate systems are often defined about the north pole (i.e. longitude and latitude), the day-night irradiation along the $x$ axis leads us to a different choice. We consider a coordinate system defined by three quantities: the radial distance from the planet's centre, $r$; the azimuthal angle in the terminator plane, $\phi$ (identical to the polar angle used in radiative transfer); and the local zenith angle with respect to the terminator plane, $\theta$. The coordinate system $(r, \theta, \phi)$ is closely related to a spherical coordinate system about the $x$ axis, differing only from our choice of $\theta$ being with respect to the terminator plane instead of the $x$ axis. We stress that $\theta$ is defined with respect to the terminator plane vertical for a \emph{specific} ray, $z'$, not the north pole axis, $z$ (see Figure~\ref{fig:coordinate_system}). The axis $z'$ is aligned with the impact parameter vector. Our $(r, \theta, \phi)$ coordinate system has the advantage of reducing the geometrical complexity of 3D radiative transfer to a series of 2D problems. For a given ray at $(b, \phi, x)$, the corresponding atmospheric coordinates $(r, \theta, \phi)$ are simply
\begin{equation}
    r = \sqrt{b^2 + x^2}
\end{equation}
and
\begin{equation}
    \theta = \tan^{-1} (x / b)
\end{equation}
with $\phi$ the same in both systems. For a given $\phi$, one then traces a series of rays, with different impact parameters, through a 2D day-night slice. The final 3D solution is then just a sum of 2D radiative transfer solutions.

We can now write the general transmission spectrum expression (Equation~\ref{eq:transmission_spectrum_general}) in our cylindrical coordinate system. In what follows, we assume a uniform stellar photosphere ($\epsilon_{\lambda, \, \rm{het}} = 1$) and negligible nightside emission ($\psi_{\lambda, \, \rm{night}} = 1$) to focus on 3D effects arising from atmospheric transmission. We consider only full transits (see Section~\ref{subsec:discussion_TRIDENT} for a discussion of ingress/egress and grazing transits), such that $A_{\mathrm{p \, (overlap)}} = \pi R_{\mathrm{p, \, top}}^2$. Given these assumptions, a transmission spectrum can be written as
\begin{equation}
    \Delta_{\lambda} = \frac{R_{\mathrm{p, \, top}}^2 - \frac{1}{\pi} \displaystyle\int_{-\pi}^{\pi} \displaystyle\int_{0}^{R_{\mathrm{p, \, top}}} \delta_{\rm{ray}*} (b, \phi) \, \mathcal{T}_{\lambda} (b, \phi) \, b \, db \, d\phi}{R_{*}^2}
\label{eq:transmission_spectrum_cyclindrical}
\end{equation}
where 
\begin{equation}
    \mathcal{T}_{\lambda} (b, \phi) = e^{-\tau_{\lambda, \, \rm{path}} (b, \phi)}
\label{eq:transmission_cylindrical}
\end{equation}
and 
\begin{equation}
    \tau_{\lambda, \, \rm{path}} (b, \phi) = \displaystyle\int_{0}^{\infty} \kappa_{\lambda} (s) \, ds
\label{eq:path_optical_depth}
\end{equation}
In the final expression, the path optical depth is expressed as an integral of the \emph{extinction coefficient}, $\kappa_{\lambda}$ (see Section~\ref{subsec:opacity}), over the path, $s$, taken by a ray through the atmosphere. This functional form assumes no scattering into each beam and negligible forward scattering, with all scattering opacity treated as a loss equivalent to absorption\footnote{In full generality, multiple scattering can be treated by replacing $\kappa_{\lambda}$ with the absorption coefficient and tracing packets of photons through an atmosphere \citep{Robinson2017a}. The scattering coefficient then only determines the path of each photon and does not need to be included in Equation~\ref{eq:path_optical_depth}.}. If one further assumes refraction to be negligible, the path optical depth is identical to the \emph{slant optical depth}
\begin{equation}
    \tau_{\lambda, \, \rm{slant}} (b, \phi) = \displaystyle\int_{-\infty}^{\infty} \kappa_{\lambda} (b, \phi, x) \, dx
\label{eq:slant_optical_depth}
\end{equation}
The evaluation of Equation~\ref{eq:slant_optical_depth} determines how an atmosphere shapes a transmission spectrum. We describe the computation of this integral in the following section.

Finally, we demonstrate that Equation~\ref{eq:transmission_spectrum_cyclindrical} is equivalent to a more familiar expression under two further simplifying assumptions. First, if one neglects ray deflection due to refraction or scattering, $\delta_{\rm{ray}*} = 1$ for all impact parameters and azimuthal angles. Second, if one assumes the planetary atmosphere has uniform properties (temperature, composition, etc.) at all longitudes and latitudes, varying only in the radial direction (i.e. the 1D atmosphere approximation), then the azimuthal integral in Equation~\ref{eq:transmission_spectrum_cyclindrical} evaluates to $2 \pi$. With these approximations, a 1D transmission spectrum is given by
\begin{equation}
    \Delta_{\lambda} \approx \frac{R_{\mathrm{p, \, top}}^2 - 2 \displaystyle\int_{0}^{R_{\mathrm{p, \, top}}} b \, e^{-\tau_{\lambda, \, \rm{slant}} (b)} \, db}{R_{\mathrm{*}}^{2}}
\label{eq:1D_transmission_spectrum}  
\end{equation} 
Expressions similar to Equation~\ref{eq:1D_transmission_spectrum} are widely seen in the transmission spectra literature \citep[e.g.][]{Brown2001,Tinetti2012}, forming the basis for many commonly used radiative transfer models. In this paper, we instead employ the more general expression given by Equation~\ref{eq:transmission_spectrum_cyclindrical}. We now show how to evaluate the optical depth for 3D, non-uniform, exoplanet atmospheres. 

\subsubsection{The Path Optical Depth} \label{subsubsec:slant_optical_depth}

The computation of path optical depths becomes non-trivial in 3D due to the need to interrelate two coordinate systems. In other words, the atmospheric properties (and hence the extinction coefficient) are defined in the spherical coordinate system $(r, \theta, \phi)$, whilst the path optical depth is evaluated in the cylindrical coordinate system $(b, \phi, x)$ (see Figure~\ref{fig:coordinate_system}). Here, we first illustrate how to solve this problem for 1D atmospheres (with properties varying only with $r$), following an approach introduced by \citet{Robinson2017a}, before generalising the method for 3D atmospheres.

\newpage

In 1D, the path optical depth is given by
\begin{equation}
    \tau_{\lambda, \, \rm{path}} (b) = \displaystyle\int_{0}^{\infty} \kappa_{\lambda} (s) \, ds = \displaystyle\int_{r_{\rm{min}}(b)}^{\infty} \kappa_{\lambda} (r) \, \mathcal{P}_{\rm{1D}} (b, r) \, dr
\label{eq:1D_path_optical_depth_1}
\end{equation}
where $r_{\rm{min}}$ is the minimum radius\footnote{For a 1D atmosphere, $r_{\rm{min}} = b/(1 + \eta(r_{\rm{min}}))$, where $\eta$ is the refractivity \citep{Betremieux2013}. When refraction is negligible, such as for H$_2$-He dominated planets, $r_{\rm{min}} = b$.} probed by a ray with impact parameter $b$, while $\mathcal{P}_{\rm{1D}} (b, r)$ is the one-dimensional \emph{path distribution} \citep{Robinson2017a}. The 1D path distribution is defined such that the distance travelled by a ray with impact parameter $b$ passing through an atmospheric layer extending from $r$ to $r + dr$ is
\begin{equation}
    ds = \mathcal{P}_{\rm{1D}} (b, r) \, dr
\label{eq:1D_path_distribution_1}
\end{equation}
Recognising that the product $\kappa_{\lambda} (r) \, dr$ in Equation~\ref{eq:1D_path_optical_depth_1} is just the differential vertical optical depth across a layer between $r$ and $r + dr$, we can thus write
\begin{equation}
    \tau_{\lambda, \, \rm{path}} (b) = \displaystyle\int \mathcal{P}_{\rm{1D}} (b, r) \, d\tau_{\lambda, \, \rm{vert}}
\label{eq:1D_path_optical_depth_2}
\end{equation}
where the integral is evaluated between the lowest and highest layers in the atmosphere, and we leave the $r$ dependence explicit to denote the layer corresponding to each differential vertical optical depth. We thus see that the path distribution acts to map the vertical optical depth in an atmosphere to the path optical depth. For a continuous atmosphere, the 1D path distribution can be analytically calculated in the geometric limit. From the geometry in Figure~\ref{fig:coordinate_system}, we have
\begin{equation}
    \mathcal{P}_{\rm{1D}} (b, r) = \frac{ds}{dr} = 2 \frac{dx}{dr} = \frac{2 \, r}{\sqrt{r^2 - b^2}}
\label{eq:1D_path_distribution_2}
\end{equation}
where the factor of 2 arises from the equal distance travelled on either side of the terminator plane in 1D models. 

To numerically compute transmission spectra, atmospheres must be discretised. Consider an atmosphere comprised of distinct layers, $r_{l}$, impinged by rays spanning a grid of impact parameters, $b_{i}$. The path optical depth for the $i^\text{th}$ impact parameter is then given by 
\begin{equation}
    \tau_{\lambda, \, \rm{path}, \, i}  = \sum_{l=1}^{N_{\rm{layer}}} \mathcal{P}_{\rm{1D}, \, il} \, \Delta\tau_{\lambda, \, \rm{vert}, \, l}
\label{eq:1D_path_optical_depth_3}
\end{equation}
where $\Delta\tau_{\lambda, \, \rm{vert}, \, l} = \kappa_{\lambda, \, l} \, \Delta r_{l}$ and the 1D path distribution becomes a matrix, $\boldsymbol{\mathcal{P}}_{\rm{1D}}$. In the geometric limit, the elements of the 1D path distribution matrix are given by (see Appendix~\ref{appendix_B})
\begin{equation}
    \mathcal{P}_{\rm{1D}, \, il} =
    \begin{cases}
        0, & \circled{1} \\
        \frac{2}{\Delta r_{l}} \left( \sqrt{ r_{\rm{up}, \, l}^2 - b_{i}^2 } \right), & \circled{2} \\
        \frac{2}{\Delta r_{l}} \left( \sqrt{ r_{\rm{up}, \, l}^2 - b_{i}^2 } - \sqrt{ r_{\rm{low}, \, l}^2 - b_{i}^2 } \right), & \circled{3} \\ 
    \end{cases}
\label{eq:1D_path_distribution_elements}
\end{equation}
where $r_{\rm{up}, \, l}$ and $r_{\rm{low}, \, l}$ are the upper and lower boundaries of the layer centred on $r_{l}$. The conditions of applicability in Equation~\ref{eq:1D_path_distribution_elements} are
\begin{align}  
    & \circled{1}: \ r_{\rm{up}, \, l} \leq b_{i} \nonumber \\
    & \circled{2}: \ r_{\rm{low}, \, l} < b_{i} < r_{\rm{up}, \, l} \nonumber \\ 
    & \circled{3}: \ r_{\rm{low}, \, l} \geq b_{i}
\label{eq:1D_path_distribution_conditions}
\end{align}
Equation~\ref{eq:1D_path_optical_depth_3} can then be conveniently recast in vector notation to compute a vector of path optical depths
\begin{equation}
    \boldsymbol{\tau}_{\lambda, \, \rm{path}} = \boldsymbol{\mathcal{P}}_{\rm{1D}} \cdot \boldsymbol{\Delta\tau}_{\lambda, \, \rm{vert}}
\label{eq:1D_path_optical_depth_vector_notation}
\end{equation}
As noted in \citet{Robinson2017a}, the main advantage of computing path optical depths in this manner is that the path distribution matrix is \emph{wavelength independent}\footnote{This is exact in the geometric limit. Models including refraction or multiple scattering can induce small wavelength dependencies in path distribution matrices.}. The path distribution matrix then need only be computed once for a given atmosphere, significantly reducing the computational burden of transmission spectra calculations. Given the need for quick spectral computations for atmospheric retrievals (which typically require $\gtrsim 10^5$ model spectra), the path distribution framework is an excellent foundation for retrieval forward models.

For 3D atmospheres, the path distribution technique must be generalised. The changing temperature across the terminator (see Figure~\ref{fig:coordinate_system}) causes the ray path length between two altitudes to become a function of $\theta$, breaking the symmetry between the dayside and nightside. We thus express the path optical depth as
\begin{equation}
    \tau_{\lambda, \, \rm{path}} (b, \phi) = \displaystyle\int_{-\frac{\pi}{2}}^{\frac{\pi}{2}} \displaystyle\int_{r_{\rm{min}}(b)}^{r (b, \theta)} \kappa_{\lambda} (r, \theta, \phi) \, \mathcal{P} (b, \phi, \theta, r) \, dr \, d\theta
\label{eq:3D_path_optical_depth_1}
\end{equation}
where $r (b, \theta)$ is the radial distance of a ray\footnote{Neglecting refraction, $r (b, \theta) = b / \cos \theta$ from the geometry in Figure~\ref{fig:coordinate_system}. More generally, $r (b, \theta)$ can be computed by ray tracing.} with impact parameter $b$ at zenith angle $\theta$. Here, we introduce the three-dimensional generalisation of the path distribution, $\mathcal{P} (b, \phi, \theta, r)$. We define the 3D path distribution such that the distance travelled by a ray (with impact parameter $b$ and azimuthal angle $\phi$) passing through an atmospheric region extending from $\theta$ to $\theta + d\theta$ and a layer extending from $r$ to $r + dr$ is
\begin{equation}
    ds = \mathcal{P} (b, \phi, \theta, r) \, dr \, d\theta
\label{eq:3D_path_distribution_1}
\end{equation}
With this definition, the 3D path optical depth is
\begin{equation}
    \tau_{\lambda, \, \rm{path}} (b, \phi) = \displaystyle\int_{-\frac{\pi}{2}}^{\frac{\pi}{2}} \displaystyle\int \mathcal{P} (b, \phi, \theta, r) \, d\tau_{\lambda, \, \rm{vert}} \, d\theta
\label{eq:3D_path_optical_depth_2}
\end{equation}

Consider now a 3D discretised atmosphere. We divide the atmosphere into columns specified by their azimuthal angle, $\phi_{j}$, and zenith angle, $\theta_{k}$. Since the temperature profile can be different in each column (see Section~\ref{subsec:atmosphere_profiles}), every column has a distinct radial array, $r_{jkl}$, with `$l$' denoting the layer in each column. This atmosphere is illuminated by a grid of impact parameters, $b_i$. For such a discretised atmosphere, Equation~\ref{eq:3D_path_optical_depth_2} becomes
\begin{equation}
    \tau_{\lambda, \, \rm{path}, \, ij}  = \sum_{k=1}^{N_{\rm{zenith}}} \sum_{l=1}^{N_{\rm{layer}}} \mathcal{P}_{ijkl} \, \Delta\tau_{\lambda, \, \rm{vert}, \, jkl} \, \Delta\theta_{k}
\label{eq:3D_path_optical_depth_3}
\end{equation}
where $\Delta\tau_{\lambda, \, \rm{vert}, \, jkl} = \kappa_{\lambda, \, jkl} \, \Delta r_{jkl}$. The path distribution is now a four-dimensional tensor, $\boldsymbol{\mathcal{P}}$, with elements given, in the geometric limit, by (see Appendix~\ref{appendix_B})
\begin{equation}
    \mathcal{P}_{ijkl} =
    \begin{cases}
        0, & \circled{1} \\
        \frac{1}{\Delta r_{jkl} \Delta\theta_k} \left( \sqrt{ r_{\rm{up}, \, jkl}^2 - b_{i}^2 } - \sqrt{ r_{\rm{min}, \, ijk}^2 - b_{i}^2 } \right), &  \circled{2} \\
        \frac{1}{\Delta r_{jkl} \Delta\theta_k} \left( \sqrt{ r_{\rm{max}, \, ijk}^2 - b_{i}^2 } - \sqrt{ r_{\rm{min}, \, ijk}^2 - b_{i}^2 } \right), &  \circled{3} \\
        \frac{1}{\Delta r_{jkl} \Delta\theta_k} \left( \sqrt{ r_{\rm{up}, \, jkl}^2 - b_{i}^2 } - \sqrt{ r_{\rm{low}, \, jkl}^2 - b_{i}^2 } \right), &  \circled{4} \\
        \frac{1}{\Delta r_{jkl} \Delta\theta_k} \left( \sqrt{ r_{\rm{max}, \, ijk}^2 - b_{i}^2 } - \sqrt{ r_{\rm{low}, \, jkl}^2 - b_{i}^2 } \right), &  \circled{5} \\
        0, & \circled{6} \\
    \end{cases}
\label{eq:3D_path_distribution_elements}
\end{equation}
where the conditions of applicability are
\begin{align}  
    \circled{1}: \ r_{\rm{low}, \, jkl} & \leq r_{\rm{min}, \, ijk} \nonumber \\ r_{\rm{up}, \, jkl} & \leq r_{\rm{min}, \, ijk} \nonumber \\ 
    \circled{2}: \ r_{\rm{low}, \, jkl} & \leq r_{\rm{min}, \, ijk} \nonumber \\ r_{\rm{min}, \, ijk} & < r_{\rm{up}, \, jkl} < r_{\rm{max}, \, ijk} \nonumber \\ 
    \circled{3}: \ r_{\rm{low}, \, jkl} & \leq r_{\rm{min}, \, ijk} \nonumber \\ r_{\rm{up}, \, jkl} & \geq r_{\rm{max}, \, ijk} \nonumber \\ 
    \circled{4}: \ r_{\rm{min}, \, ijk} & < r_{\rm{low}, \, jkl} < r_{\rm{max}, \, ijk}  \nonumber \\ r_{\rm{min}, \, ijk} & < r_{\rm{up}, \, jkl} < r_{\rm{max}, \, ijk} \nonumber \\ 
    \circled{5}: \ r_{\rm{min}, \, ijk} & < r_{\rm{low}, \, jkl} < r_{\rm{max}, \, ijk} \nonumber \\ r_{\rm{up}, \, jkl} & \geq r_{\rm{max}, \, ijk} \nonumber \\ 
    \circled{6}: \ r_{\rm{low}, \, jkl} & \geq r_{\rm{max}, \, ijk} \nonumber \\ r_{\rm{up}, \, jkl} & \geq r_{\rm{max}, \, ijk}
\label{eq:3D_path_distribution_conditions}
\end{align}
In the expressions above, $r_{\rm{up}, \, jkl}$ and $r_{\rm{low}, \, jkl}$ are the upper and lower boundaries of the $l^\text{th}$ layer in column $jk$, while $r_{\rm{min}, \, ijk}$ and $r_{\rm{max}, \, ijk}$ are the minimum and maximum radial coordinates encountered by a ray with impact parameter $b_{i}$ in column $jk$. They are given by  
\begin{equation}
    r_{\rm{min}, \, ijk} = \min \left( R_{\mathrm{p, \, top}, \, jk} \ , \ \frac{b_{i}}{\cos{\theta_{\rm{min}, \, k}}} \right)
\label{eq:r_min}
\end{equation}
\begin{equation}
    r_{\rm{max}, \, ijk} = \min \left( R_{\mathrm{p, \, top}, \, jk} \ , \ \frac{b_{i}}{\cos{\theta_{\rm{max}, \, k}}} \right)
\label{eq:r_max}
\end{equation}
where $R_{\mathrm{p, \, top}, \, jk}$ is the radial coordinate at the top of column $jk$, $\theta_{\rm{min}, \, k}$ is the minimum zenith angle of the column, and $\theta_{\rm{max}, \, k}$ is the maximum zenith angle of the column. The conditions in Equation~\ref{eq:3D_path_distribution_conditions} account for all possible ways in which a ray can traverse zenith slices before leaving the atmosphere (see Appendix~\ref{appendix_B}). 

We efficiently numerically compute 3D path optical depths via matrix operations. By defining an ancillary tensor $\boldsymbol{\tilde{\mathcal{P}}}$, with elements given by $\mathcal{P}_{ijkl} \Delta\theta_k$, we can then succinctly write Equation~\ref{eq:3D_path_optical_depth_3} as 
\begin{equation}
    \boldsymbol{\tau}_{\lambda, \, \rm{path}, \, j} = \boldsymbol{\tilde{\mathcal{P}}_{j}} : \boldsymbol{\Delta\tau}_{\lambda, \, \rm{vert}, \, j}
\label{eq:3D_path_optical_depth_vector_notation}
\end{equation}
where `:' denotes the double dot product. In other words, the path distribution tensor for sector `$j$' is contracted with the vertical optical depth matrix for sector `$j$' over their common indices `$lk$'. The result is a vector of size $N_b$ containing the slant optical depths for all the impact parameters in sector `$j$'. Computing this double dot product for all azimuthal sectors yields the path optical depth matrix, $\boldsymbol{\tau}_{\lambda, \, \rm{path}}$ (dimensions $N_b \times N_{\rm{sectors}}$). This equation is the 3D generalisation of Equation~\ref{eq:1D_path_optical_depth_vector_notation}. We stress that the functional form in Equation~\ref{eq:3D_path_optical_depth_vector_notation} holds for models both with and without refraction. However, the analytical expressions we derive for the path distribution tensor (Equation~\ref{eq:3D_path_distribution_elements}) are specific to the geometric limit\footnote{To include refraction, the path distribution tensor must instead be numerically computed \citep[see][]{Robinson2017a}.}. With the path optical depth determined, one can readily compute 3D transmission spectra.

\subsubsection{3D Radiative Transfer with TRIDENT} \label{subsubsec:TRIDENT_radiative_transfer}

\begin{figure*}[ht!]
    \centering
    \includegraphics[width=\textwidth]{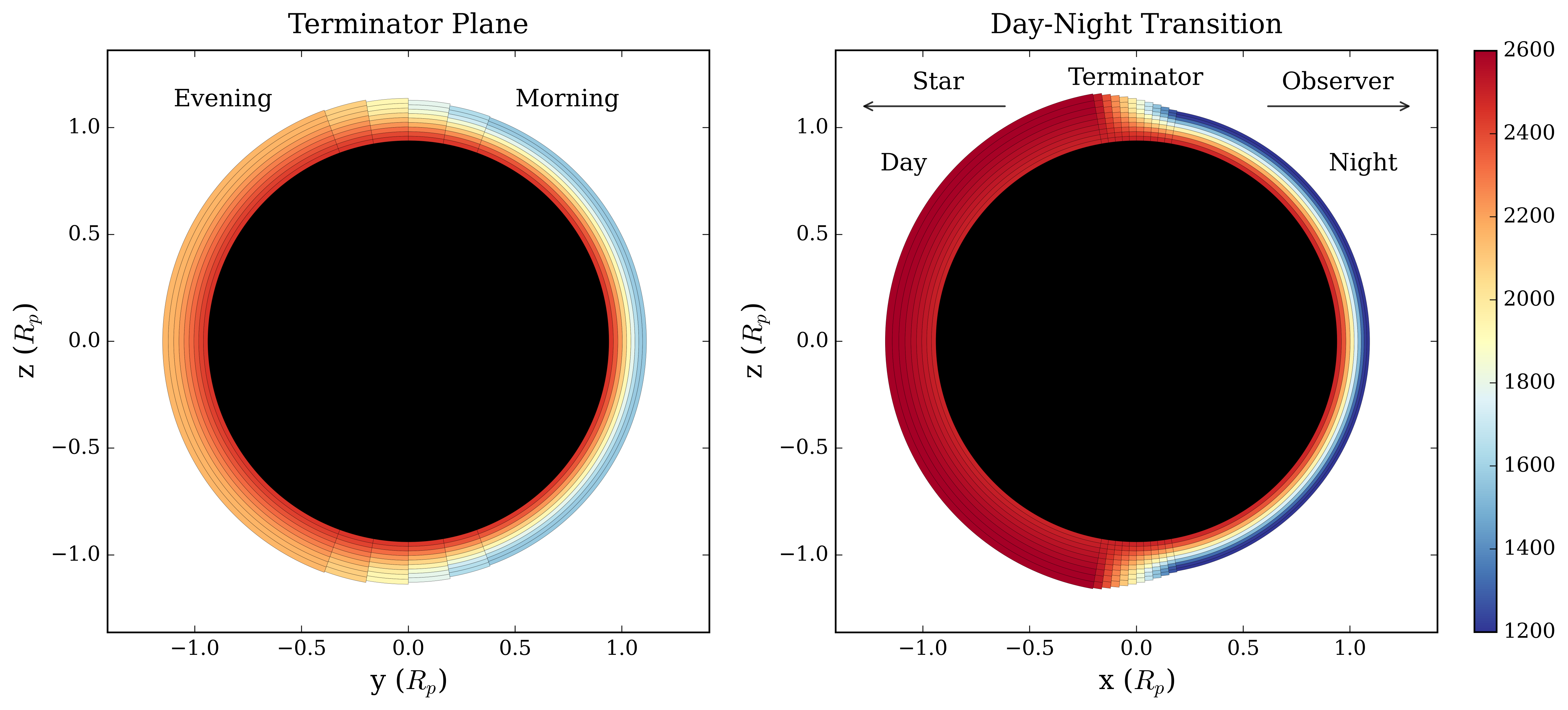}
    \caption{2D representation of the temperature field for a parametrised ultra-hot Jupiter. Left: slice through the terminator plane ($\theta = 0$), showing a morning-evening terminator temperature gradient. Right: slice through the north-south polar plane ($\phi = 0$), highlighting an extreme day-night temperature gradient. The radial extent of the atmosphere ranges from $10^{2}$--$10^{-7}$\,bar (shown to scale), demonstrating the highly non-spherical atmospheric structure. The southern hemisphere is assumed to symmetrically mirror the northern hemisphere. The temperature field for this example planet (spanning the colourbar range) is defined by: $\overline{T}_{\rm term} = 1900$\,K, $\Delta T_{\rm term} = 600$\,K, $\Delta T_{\rm DN} = 1400$\,K, and $T_{\rm deep} = 2500$\,K (see Figure~\ref{fig:1D_atm_structure}).}
\label{fig:2D_atm_structure}
\end{figure*}

TRIDENT computes transmission spectra in a discretised representation of the geometry in Figure~\ref{fig:coordinate_system}. We divide the atmosphere into $N_{\rm{sector}}$ azimuthal sectors in the terminator plane, $N_{\rm{zenith}}$ zenith slices from the dayside to nightside, and $N_{\rm{layer}}$ layers in each column. We assume the day-night transition region has a zenith angular width of $\beta$, with atmospheric properties linearly varying over the transition region (see Section~\ref{subsec:atmosphere_profiles} and Figure~\ref{fig:2D_atm_structure}). Similarly, we assume a morning-evening transition region with an azimuthal angular width of $\alpha$. Outside the transition regions (i.e. $|\theta| > \beta/2$ and $|\phi| > \alpha/2$) the atmosphere varies only with altitude. By convention, the dayside is the first zenith slice ($k = 1$) and the nightside is the last ($k = N_{\rm{zenith}}$); the evening terminator is the first azimuthal sector ($j = 1$) and the morning terminator is the last ($j = N_{\rm{sector}}$). The day-night and morning-evening transition regions thus span $N_{\rm{zenith}} - 2$ zenith slices and $N_{\rm{sector}} - 2$ azimuthal sectors, respectively, with uniform angular spacing over the transitions. We further assume north-south symmetry for the background atmosphere (a common assumption, especially in cases with zero obliquity), such that sectors need only be defined for the northern hemisphere.

TRIDENT computes transmission spectra by solving the discretised representation of Equation~\ref{eq:transmission_spectrum_cyclindrical}. For a collection of rays ($b$, $\phi$), with dimensions $N_{\rm{b}} \times N_{\rm{sector}}$, the transmission spectrum is given by
\begin{equation}
    \Delta_{\lambda} = \frac{R_{\mathrm{p, \, top}}^2 - \frac{1}{\pi} \sum\limits_{j=1}^{N_{\rm{sector}}} \sum\limits_{i=1}^{N_{b}} \delta_{\rm{ray}*, \, ij} \, \mathcal{T}_{\lambda, \, ij} \, b_i \, \Delta b_i \, \Delta \phi_j}{R_{*}^2}
\label{eq:transmission_spectrum_discrete}
\end{equation}
where we take $R_{\mathrm{p, \, top}}$ to be the maximum radial extent of the dayside\footnote{This choice draws a projected disc with radius equal to the highest point in the modelled atmosphere. This ensures that no part of the atmosphere is missed during radiative transfer. Any rays with impact parameters that never intersect the atmosphere do not change Equation~\ref{eq:transmission_spectrum_discrete}, since they equally add to the two terms on the numerator ($\mathcal{T}_{\lambda} = 1$ and $\delta_{\rm{ray}*} = 1$). Choosing a disc radius higher than $R_{\mathrm{p, \, top}}$ thus does not alter the spectrum.}. We take the impact parameter array to coincide with the lower layer boundaries of the column with greatest radial extent (i.e. $b_i$ = $\max_{jk} r_{\rm{low}, \, jkl}$). Defining the projected area matrix\footnote{We absorb $\delta_{\rm{ray}*}$ into the atmosphere area matrix, such that only rays tracing back to the star contribute. This can thus be considered the `effective' area matrix for the atmosphere.} for the atmosphere, $\boldsymbol{\Delta A}_{\rm{atm}}$, via $\Delta A_{\rm{atm}, \, ij} = b_i \Delta b_i \Delta \phi_j \delta_{\rm{ray}*, \, ij}$ we finally have 
\begin{equation}
    \Delta_{\lambda} = \frac{R_{\mathrm{p, \, top}}^2 - \frac{1}{\pi} \left( \boldsymbol{\mathcal{T}}_{\lambda} : \boldsymbol{\Delta A}_{\rm{atm}} \right) }{R_{*}^2}
\label{eq:transmission_spectrum_discrete_final}
\end{equation}
where the transmission matrix is
\begin{equation}
    \boldsymbol{\mathcal{T}}_{\lambda} = e^{- \boldsymbol{\tau}_{\lambda, \, \rm{path}}}
\label{eq:3D_transmission}
\end{equation}
Although deceptively simple, Equation~\ref{eq:transmission_spectrum_discrete_final} provides the practical basis for rapid 3D radiative transfer through non-uniform exoplanet atmospheres. In the geometric limit, which we adopt in the following sections, the optical depth matrix is given by Equation~\ref{eq:3D_path_optical_depth_vector_notation}. With our radiative transfer framework established, we proceed to describe the construction of 3D model atmospheres.

\subsection{Atmospheric Structure} \label{subsec:atmosphere_profiles}

TRIDENT models 3D atmospheres via distinct columns, ($\phi, \theta$), each locally obeying hydrostatic equilibrium and the ideal gas law (both valid in the atmospheric regions probed by transmission spectra). The columns are divided into layers, uniformly spaced in log-pressure, with each column sharing the same pressure grid. However, the temperature and composition can vary both vertically (within a column) and between each column. Consequently, each column possess a different radial grid and the planetary atmosphere is non-spherical. Here we describe each of these aspects in turn.

\subsubsection{A 3D Temperature Parametrisation} \label{subsubsec:P-T_profiles}

\begin{figure*}[!htb]
    \centering
    \includegraphics[width=0.485\textwidth]{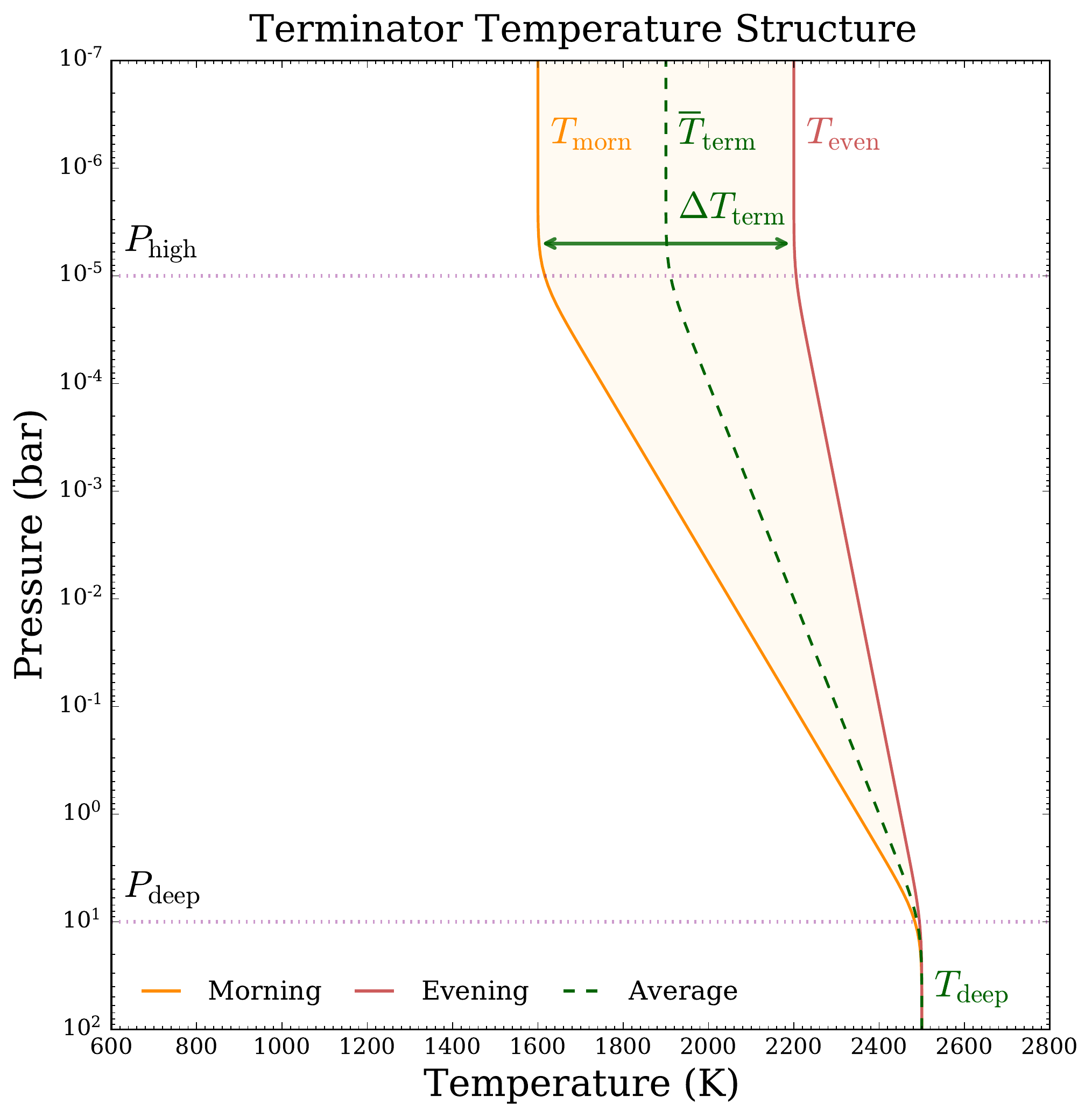}
    \includegraphics[width=0.485\textwidth]{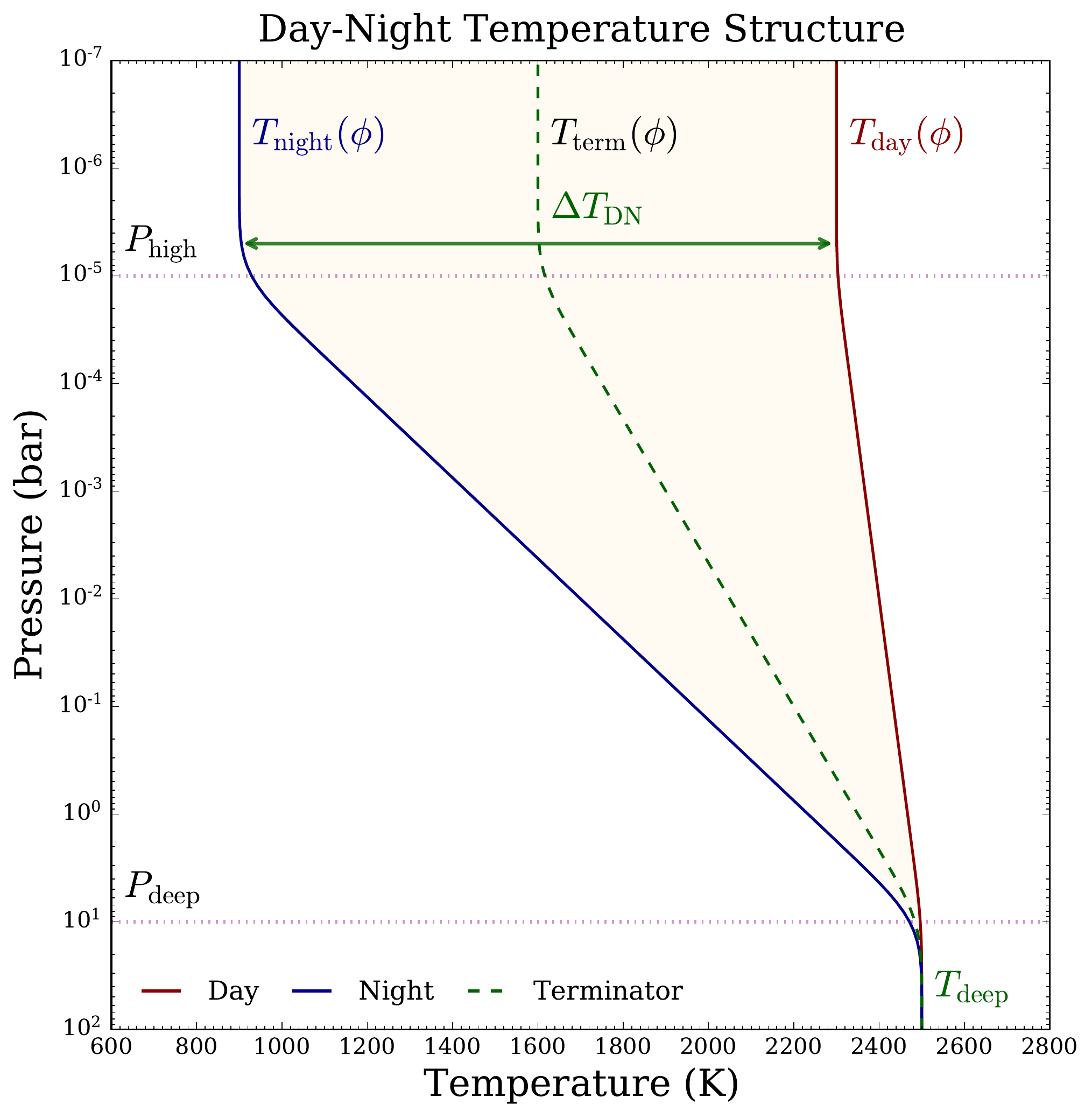}
    \caption{A temperature structure parametrisation for 3D exoplanet atmospheres. Left: in the terminator plane the top-of-atmosphere temperature ranges from $T_{\rm morn}$ to $T_{\rm even}$. Each temperature profile varies linearly in log-pressure from $P_{\rm high}$ to $P_{\rm deep}$, converging in the deep atmosphere to a common temperature, $T_{\rm deep}$. Within the morning-evening opening angle ($|\phi| < \alpha/2$) the temperature profile varies linearly between the morning and evening terminators (orange shading). Right: a ray (constant $\phi$) samples temperatures ranging from $T_{\rm day} (\phi)$ to $T_{\rm night} (\phi)$ (orange shading) within the day-night opening angle ($|\theta| < \beta/2$). The terminator temperature profile (black dashes) corresponds to the profile for the angle $\phi$ in the left panel (here, $T_{\rm day} (\phi) = T_{\rm morn}$). The 3D temperature structure is fully specified by 4 quantities (green labels): $\overline{T}_{\rm term}$, $\Delta T_{\rm term}$, $\Delta T_{\rm DN}$, and $T_{\rm deep}$.}
\label{fig:1D_atm_structure}
\end{figure*}

Model atmosphere temperature structures are generally constructed by one of two approaches. Forward modelling studies typically calculate pressure-temperature (P-T) profiles self-consistently with various physical principles (e.g. radiative-convective equilibrium) via an iterative procedure \citep[e.g.][]{Sudarsky2003,Molliere2015}. Retrieval studies, however, routinely assume an isotherm or a parametric P-T profile to rapidly explore a wide parameter space of potential temperature structures \citep[e.g.][]{Madhusudhan2009,Line2013}. Here, we introduce a temperature structure parametrisation appropriate for 3D retrievals of exoplanet transmission spectra.  

Multidimensional retrievals require a \emph{temperature field} prescription, $T (P, \phi, \theta)$, which must satisfy several requirements. First, it must be sufficiently flexible to capture realistic non-uniform temperature structures --- potentially varying on the order of $\sim 1000$\,K between different regions of hot Jupiter atmospheres. Second, it must account for the vertical dependence of temperature with pressure, since transmission spectra are sensitive to temperature gradients \citep{Rocchetto2016,Heng2017}. Third, the temperature profiles in each column should converge in the deep atmosphere where external irradiation has no influence. Finally, the temperature field should employ the minimal set of free parameters required to capture the information content of 3D transmission spectra. This latter requirement renders common 1D P-T profiles \citep[e.g.][]{Madhusudhan2009,Line2013} suboptimal\footnote{Transmission spectra are only weakly sensitive to curvature in P-T profiles, leading to several largely unconstrained parameters \citep[e.g.][their Figure 6]{MacDonald2017a}.} for multidimensional retrievals. Instead, we propose a new parametrisation that represents a simple construction appropriate for 3D transmission spectra retrievals.

\begin{figure*}[!ht]
    \centering
    \includegraphics[width=0.84\textwidth, trim={0.0cm 0.0cm 0.0cm -0.2cm}]{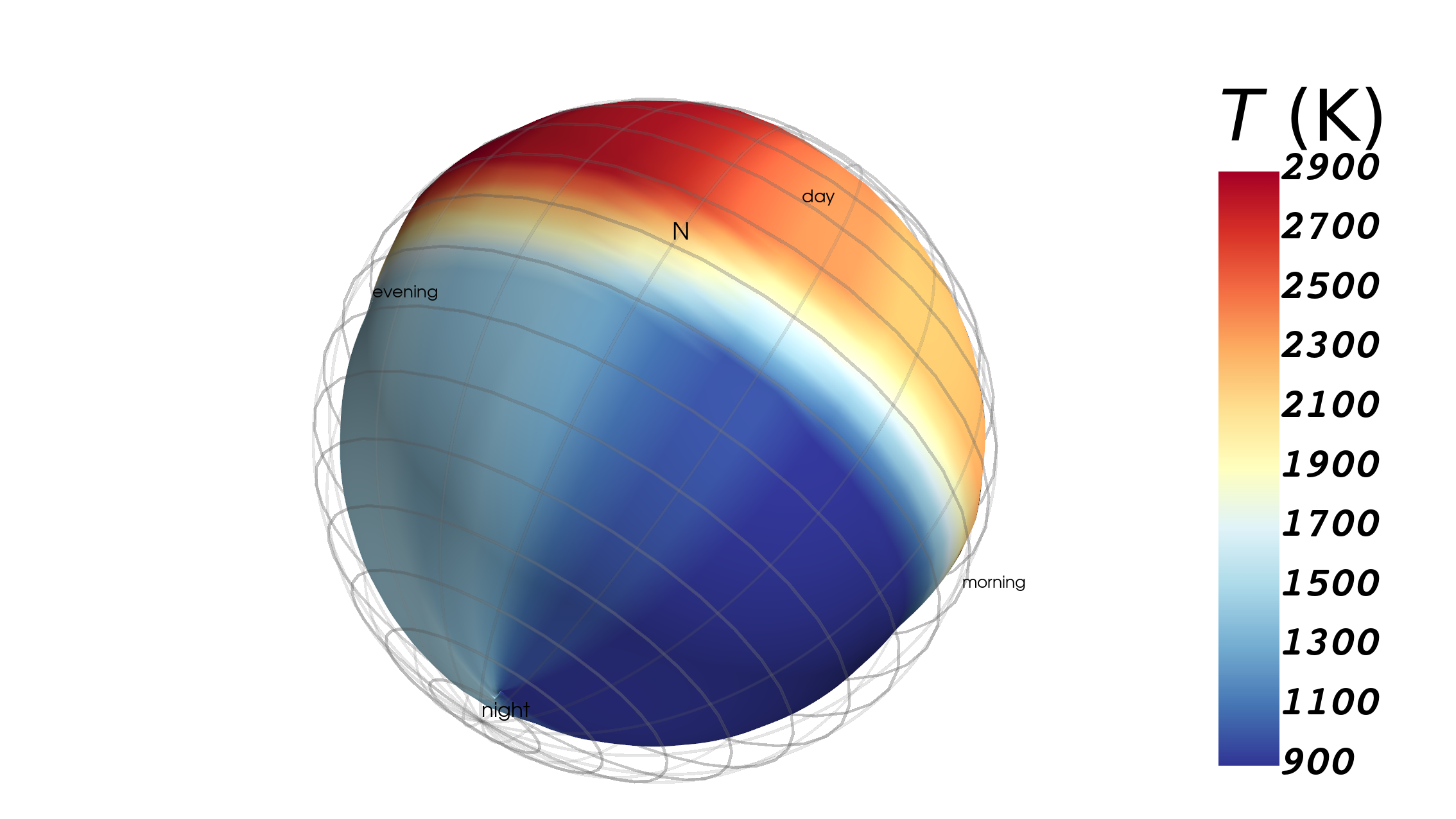}
    \caption{3D representation of the top-of-atmosphere temperature field for a parametrised ultra-hot Jupiter. Temperature inhomogeneities can be prescribed across the day-night transition (simulating the incident irradiation difference) and between the morning and evening terminators (accounting for hotspot advection due to winds). The temperature field shown here corresponds to the same model as the 2D slice diagram in Figure~\ref{fig:2D_atm_structure} and 1D temperature profiles in Figure~\ref{fig:1D_atm_structure}. The discretised temperature field (see Figure~\ref{fig:2D_atm_structure}) has been interpolated to a finer $(\phi, \theta)$ grid in this representation. The spherical outline (grey frame) shows the maximum radius of the dayside, highlighting the smaller column height of the nightside. An animated version of this figure, showing a $360\degr$ rotation from the nightside through the dayside, is available in the HTML version of this article.}
\label{fig:3D_atm_structure}
\end{figure*}

We propose a 4-parameter temperature field. Consider a slice through the terminator plane, with the top-of-atmosphere temperature varying linearly between a morning terminator temperature, $T_{\rm morn}$, and an evening terminator temperature, $T_{\rm even}$, over a transition region with an angular width $\alpha$. This can be expressed via
\begin{equation}
    T_{\rm term} (\phi) =
    \begin{cases}
        T_{\rm even}, & \circled{1} \\
        \overline{T}_{\rm term} - \left(\frac{\phi}{\alpha/2}\right) \frac{\Delta T_{\rm term}}{2}, & \circled{2} \\
        T_{\rm morn}, & \circled{3} \\
    \end{cases}
\label{eq:T_field_1}
\end{equation}
\begin{align*}  
    & \circled{1}: \ \phi \leq -\alpha/2 \\ 
    & \circled{2}: \ -\alpha/2 < \phi < \alpha/2 \\ 
    & \circled{3}: \ \phi \geq \alpha/2 
\end{align*}
where $\overline{T}_{\rm term} \equiv \frac{1}{2}(T_{\rm even} + T_{\rm morn})$ and $\Delta T_{\rm term} \equiv T_{\rm even} - T_{\rm morn}$. Once the terminator temperature, $T_{\rm term} (\phi)$, is determined for a given azimuthal angle, we can account for the day-night variation. Assuming a linear transition between the dayside and nightside temperatures over a region of angular width $\beta$, we have
\begin{equation}
    T_{\rm high} (\phi, \theta) =
    \begin{cases}
        T_{\rm day} (\phi), & \circled{1} \\
        \overline{T}_{\rm term} (\phi) - \left(\frac{\theta}{\beta/2}\right) \frac{\Delta T_{\rm DN}}{2}, & \circled{2} \\
        T_{\rm night} (\phi), & \circled{3}
    \end{cases}
\label{eq:T_field_2}
\end{equation}
\begin{align*}  
    & \circled{1}: \ \theta \leq -\beta/2 \\ 
    & \circled{2}: \ -\beta/2 < \theta < \beta/2 \\ 
    & \circled{3}: \ \theta \geq \beta/2 
\end{align*}
where we assume the day-night temperature contrast, $\Delta T_{\rm DN} \equiv T_{\rm day} - T_{\rm night}$, is constant for all $\phi$. Therefore, the dayside and nightside temperatures are automatically specified by $T_{\rm day} (\phi) = T_{\rm term} (\phi) + \frac{1}{2}\Delta T_{\rm DN}$ and $T_{\rm night} (\phi) = T_{\rm term} (\phi) - \frac{1}{2}\Delta T_{\rm DN}$. We note that if morning-evening gradients are neglected ($\Delta T_{\rm term} = 0$), Equation~\ref{eq:T_field_2} is equivalent to the day-night gradient prescription studied by \citet{Caldas2019}. Finally, each column has a vertical temperature gradient, assumed linear in log-pressure, running from $T_{\rm high} (\phi, \theta)$ to $T_{\rm deep}$ 
\begin{equation}
    T (P, \phi, \theta) =
    \begin{cases}
        T_{\rm high} (\phi, \theta), & \hspace{-0.8em} \circled{1} \\
        T_{\rm high} (\phi, \theta) + \left[\frac{T_{\rm deep} - T_{\rm high} (\phi, \theta)}{\log(P_{\rm deep}/P_{\rm high})}\right] \log\left(\frac{P}{P_{\rm high}}\right), & \hspace{-0.8em} \circled{2} \\
        T_{\rm deep}, & \hspace{-0.8em} \circled{3} \\
    \end{cases}
\label{eq:T_field_3}
\end{equation}
\begin{align*}  
    & \circled{1}: \ P \leq P_{\rm high} \nonumber \\ 
    & \circled{2}: \ P_{\rm high} < P < P_{\rm deep} \nonumber \\ 
    & \circled{3}: \ P \geq P_{\rm deep} \nonumber
\end{align*}
where $P_{\rm high}$ denotes where higher altitudes are fixed to $T_{\rm high} (\phi, \theta)$, whilst $P_{\rm deep}$ specifies where lower altitudes are at $T_{\rm deep}$ (common for all columns). We set $P_{\rm high} = 10^{-5}$\,bar and $P_{\rm deep} = 10$\,bar, serving as anchors outside the pressure range probed by low-resolution transmission spectra. Finally, TRIDENT slightly Gaussian smooths each profile (with a standard deviation of 3 layers), avoiding a discontinuous temperature gradient at $P_{\rm high}$ and $P_{\rm deep}$ (not applying this smoothing negligibly alters transmission spectra). Our temperature field prescription has the strength of being fully specified by just four quantities: $\overline{T}_{\rm term}$, $\Delta T_{\rm term}$, $\Delta T_{\rm DN}$, and $T_{\rm deep}$. This simplicity avoids over-parametrisation of the temperature field in multidimensional retrievals.   

We visualise our temperature field prescription for a typical ultra-hot Jupiter in Figure~\ref{fig:1D_atm_structure}. This demonstration shows P-T profiles for a model with $\overline{T}_{\rm term} = 1900$\,K, $\Delta T_{\rm term} = 600$\,K, $\Delta T_{\rm DN} = 1400$\,K, and $T_{\rm deep} = 2500$\,K. The left panel shows the range of profiles in the terminator plane, while the right panel shows the range of profiles sampled by a ray crossing the day-night transition. We also show a 2D representation for the same planet in Figure~\ref{fig:2D_atm_structure}, illustrating how this temperature field translates into local altitude grid differences for each column (see section~\ref{subsubsec:radial_grids}). Finally, we provide a 3D representation of the same temperature field in Figure~\ref{fig:3D_atm_structure}.

\subsubsection{Atmospheric Composition} \label{subsubsec:X_profiles}

We encode the composition of an atmosphere via the \emph{volume mixing ratios} ($X_{q} \equiv n_{q} / n_{\rm tot}$) for a set of chemical species $(q = 1, 2, ..., N_{\rm species})$. The total number density is determined, assuming the ideal gas law, solely from the temperature field
\begin{equation}
    n_{\rm tot} (P, \phi, \theta) = \frac{P}{k_B T (P, \phi, \theta)}
\label{eq:ideal_gas}
\end{equation}
where $k_B$ is the Boltzmann constant. In general, each chemical species can have its own \emph{mixing ratio field}, using the same prescription as Section~\ref{subsubsec:P-T_profiles}, by replacing the temperature parameters with log-mixing ratio parameters in Equations~\ref{eq:T_field_1}, \ref{eq:T_field_2}, and \ref{eq:T_field_3} (i.e. $\overline{\log X}_{\rm q, \, term}$, $\Delta \log X_{\rm q, \, term}$, $\Delta \log X_{\rm q, \, DN}$, and $\log X_{\rm q, \, deep}$). If the abundance of a given species does not vary in the terminator plane or across the day-night transition, $\Delta \log X_{\rm q, \, term} = 0$ or $\Delta \log X_{\rm q, \, DN} = 0$, respectively. Similarly, isocompositional vertical profiles can be obtained by setting $\log X_{\rm q, \, deep} = \log X_{\rm{H_{2}O, \, high}} (\theta, \phi)$. Eliminating redundant parameters readily allows a 3D, 2D, or 1D mixing ratio prescription, allowing the adaptation of model complexity in light of data quality. In our convention, $\Delta \log X_{\rm q, \, term} > 0$ implies a higher abundance on the evening terminator, while $\Delta \log X_{\rm q, \, DN} > 0$ implies a higher dayside abundance.

TRIDENT currently supports over 50 chemical species, covering a panoply of molecules, atoms, and ions expected in exoplanet atmospheres (see Appendix~\ref{appendix_C}). The main atmospheric constituents for planets ranging from ultra-hot Jupiters to temperate terrestrial planets are included (see Section~\ref{subsec:opacity}). When initialising a model atmosphere, we assign one gas as the bulk constituent, with its mixing ratio in each layer and column determined by the requirement that $\sum_q X_q = 1$. For giant planets, we treat H$_2$ and He as a single component with a fixed abundance ratio of $X_{\rm He} / X_{\rm H_2} = 0.17$. 

\subsubsection{Radial Grids} \label{subsubsec:radial_grids}

The radial grid of an atmosphere is readily computed from the temperature and mixing ratio fields. Under the assumptions of hydrostatic equilibrium and inverse square gravity, the distance from the planetary centre to a layer at pressure $P$ in each column is given by
\begin{equation}
    r (P, \phi, \theta) = \left( \frac{1}{R_{\mathrm{p, \, deep}}} + \int_{P_{\rm deep}}^{P} \frac{k_B \, T(P, \phi, \theta)}{G \, M_{\rm p} \, \mu(P, \phi, \theta)} \, \frac{dP}{P} \right)^{-1}
\label{eq:hydrostatic_equilibrium_solution}
\end{equation}
where $G$ is the gravitational constant, $M_{\rm p}$ is the planetary mass, and $\mu (P, \phi, \theta)= \sum_q m_q \, X_q (P, \phi, \theta)$ is the mean molecular mass in a given layer and column. For multidimensional models, one no longer has the freedom to chose between a reference pressure and a reference radius \citep[e.g.][]{Heng2017,Welbanks2019a}. Choosing a reference pressure in the upper atmosphere (e.g. 1\,mbar) would necessitate specifying a \emph{different} reference radius for every column, due to the non-spherical atmosphere (see Figures~\ref{fig:2D_atm_structure} and \ref{fig:3D_atm_structure}). The converse is not true if one specifies a radius at an altitude where the pressure is the same in each column. We therefore assume in Equation~\ref{eq:hydrostatic_equilibrium_solution} that the deep interior of the atmosphere (here, 10\,bar) is homogeneous, such that each column shares a common spherical surface defined by $r (P_{\rm deep}, \phi, \theta) = R_{\mathrm{p, \, deep}}$. Once the radial grid for a model is determined, one can readily compute the elements of the path distribution tensor (Equation~\ref{eq:3D_path_distribution_elements}) required for 3D radiative transfer.

\subsection{Opacity Sources} \label{subsec:opacity}

The morphology of a transmission spectrum is ultimately shaped by the wavelength dependence of absorption and scattering processes. These phenomena are encoded via the extinction coefficient
\begin{equation}
    \kappa_{\lambda} = \kappa_{\lambda, \, \rm{chem}} + \kappa_{\lambda, \, \rm{Rayleigh}} + \kappa_{\lambda, \, \rm{pair}} + \kappa_{\lambda, \, \rm{aerosol}}
\label{eq:extinction_coefficient}
\end{equation}
where $\kappa_{\lambda, \, \rm{chem}}$ is the extinction (in m$^{-1}$) due to absorption by chemical species (atoms, ions, and molecules), $\kappa_{\lambda, \, \rm{Rayleigh}}$ is the extinction from Rayleigh scattering, $\kappa_{\lambda, \, \rm{pair}}$ is the extinction from pair processes (collision-induced absorption and free-free opacity), and $\kappa_{\lambda, \, \rm{aerosol}}$ is the extinction from aerosols (clouds and hazes). We omit the $(P, \phi, \theta)$ dependence of each term in Equation~\ref{eq:extinction_coefficient} for brevity. We now briefly describe each of these sources of atmospheric extinction.

\subsubsection{Absorption Cross Sections} \label{subsubsec:cross_sections}

The extinction coefficient due to absorption by the chemical species in an atmosphere is given by
\begin{equation}
    \kappa_{\lambda, \, \rm{chem}} (P, \phi, \theta) = \sum_{q=1}^{N_{\rm species}} n_q (P, \phi, \theta) \, \sigma_{\lambda, \, \rm{abs}, \, q} (P, T)
\label{eq:extinction_chem}
\end{equation}
where $n_q = X_q \, n_{\rm tot}$ is the number density of species `$q$' and $\sigma_{\lambda, \, \rm{abs}, \, q}$ is the corresponding absorption cross section (in m$^2$). TRIDENT uses a database of pre-computed cross sections shared with the POSEIDON retrieval code \citep{MacDonald2017a}. We calculated pressure and temperature broadened cross sections from various line list databases --- ExoMol \citep{Tennyson2020}, HITRAN \citep{Gordon2021}, and VALD3 \citep{Ryabchikova2015} --- ranging from 100--3500\,K and 10$^{-6}$--100\,bar. We treat each line as a Voigt profile\footnote{Excepting the Na and K resonance lines, for which a sub-Voigt treatment is applied \citep[see][]{MacDonald_Thesis_2019}.} from the line core up to a wing cutoff (at the minimum of 500 Voigt half-width half maxima or 30\,cm$^{-1}$, see \citealt{Hedges2016}), accounting for H$_2$ and He broadening where data is available. Our cross sections are calculated at a spectral resolution of $\Delta \nu = 0.01$\,cm$^{-1}$, making them suitable for high-resolution models, and generally range from 0.4--50\,$\micron$ (we extend down to 0.2\,$\micron$ for some atoms and ions with strong near-UV opacities). Further details on our cross section computations can be found in \citet{MacDonald_Thesis_2019}, where we further find excellent agreement with EXOCROSS \citep{Yurchenko2018a}. We also include bound-free absorption (presently for H$^{-}$) in $\kappa_{\lambda, \, \rm{chem}}$. We show representative cross sections from our database in Figure~\ref{fig:cross_sections} (top panels); see Appendix~\ref{appendix_C} for a full summary of our line list sources.

\begin{figure*}[ht!]
    \centering
    \includegraphics[width=0.49\textwidth]{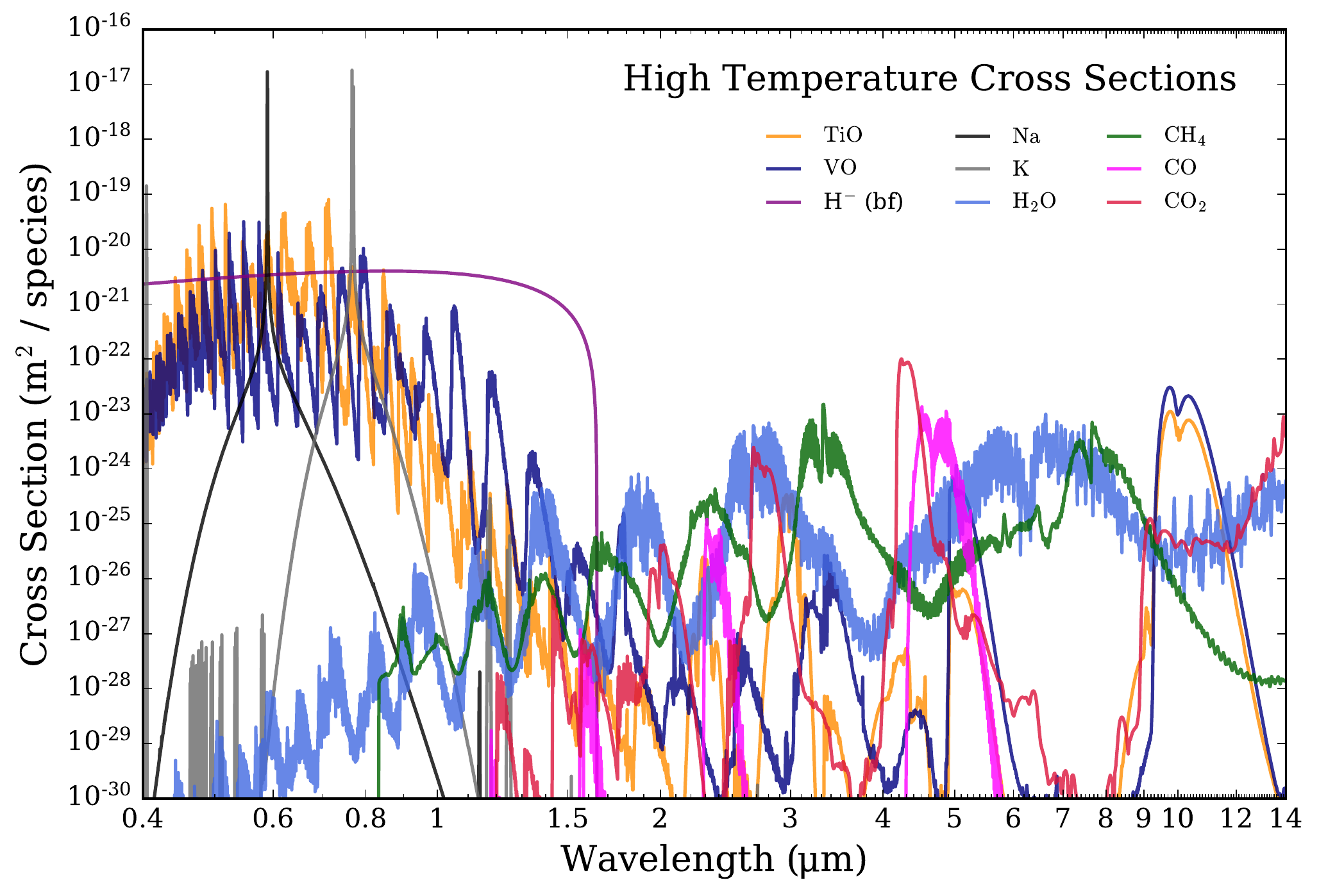}
    \includegraphics[width=0.49\textwidth]{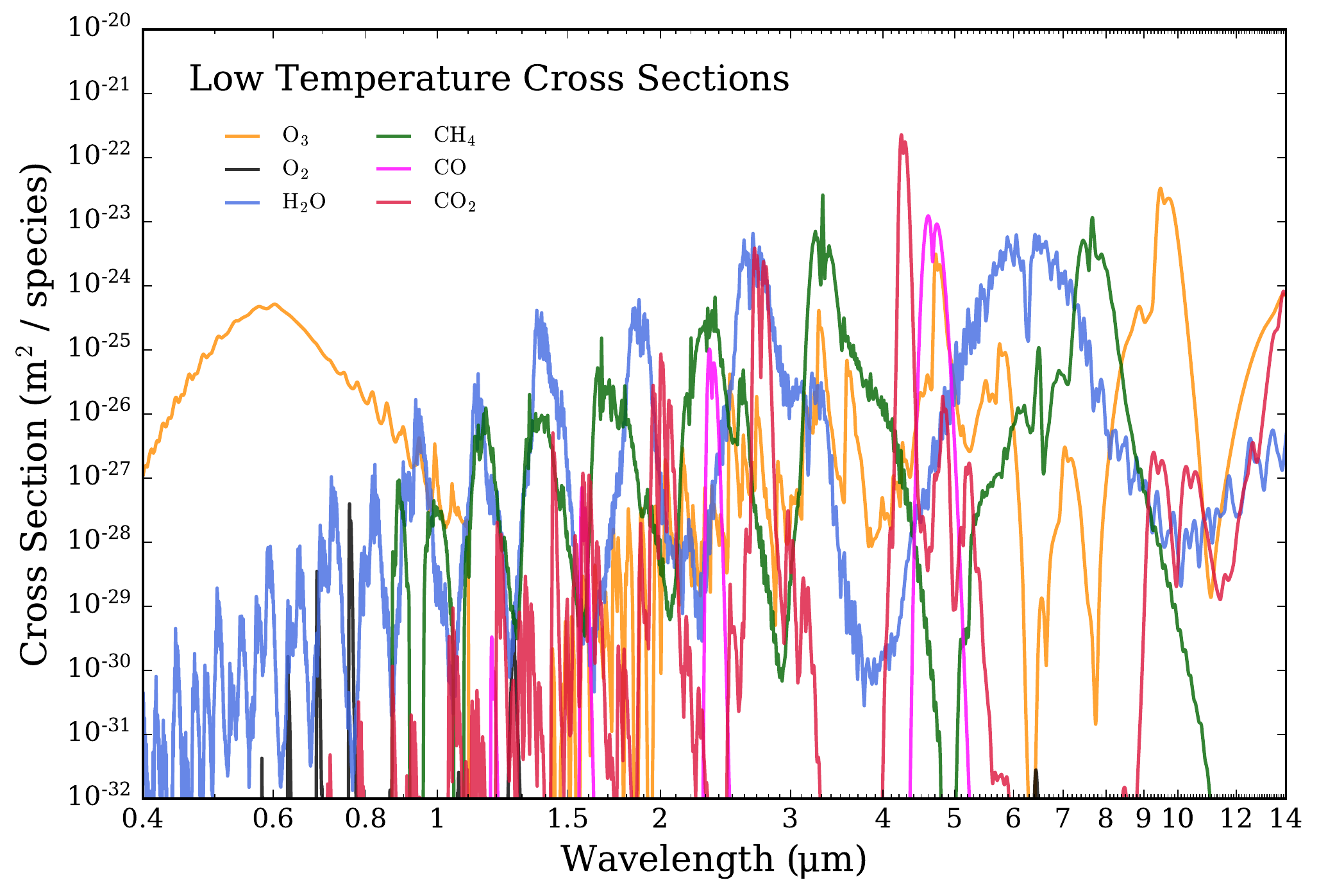}
    \includegraphics[width=0.49\textwidth]{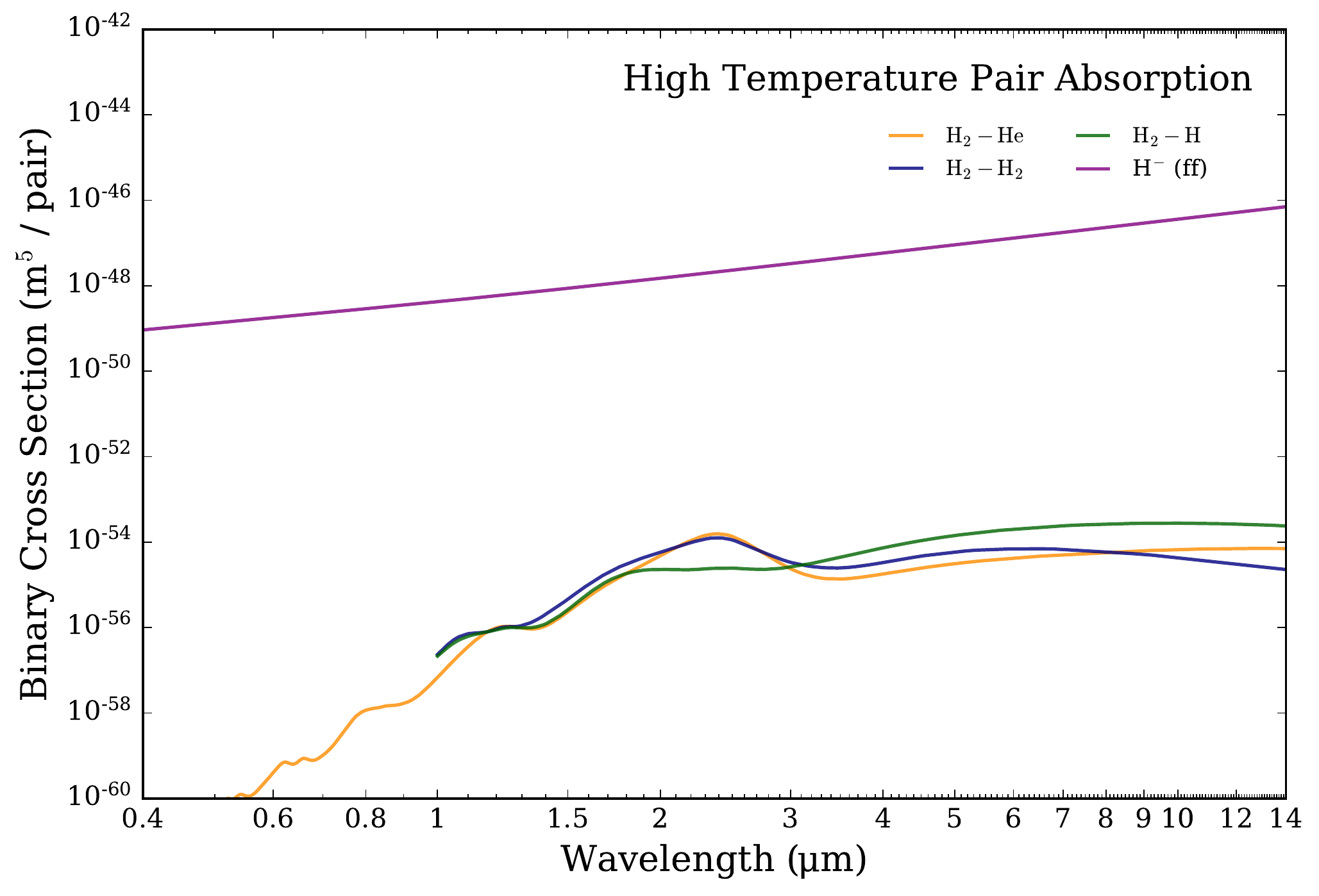}
    \includegraphics[width=0.49\textwidth]{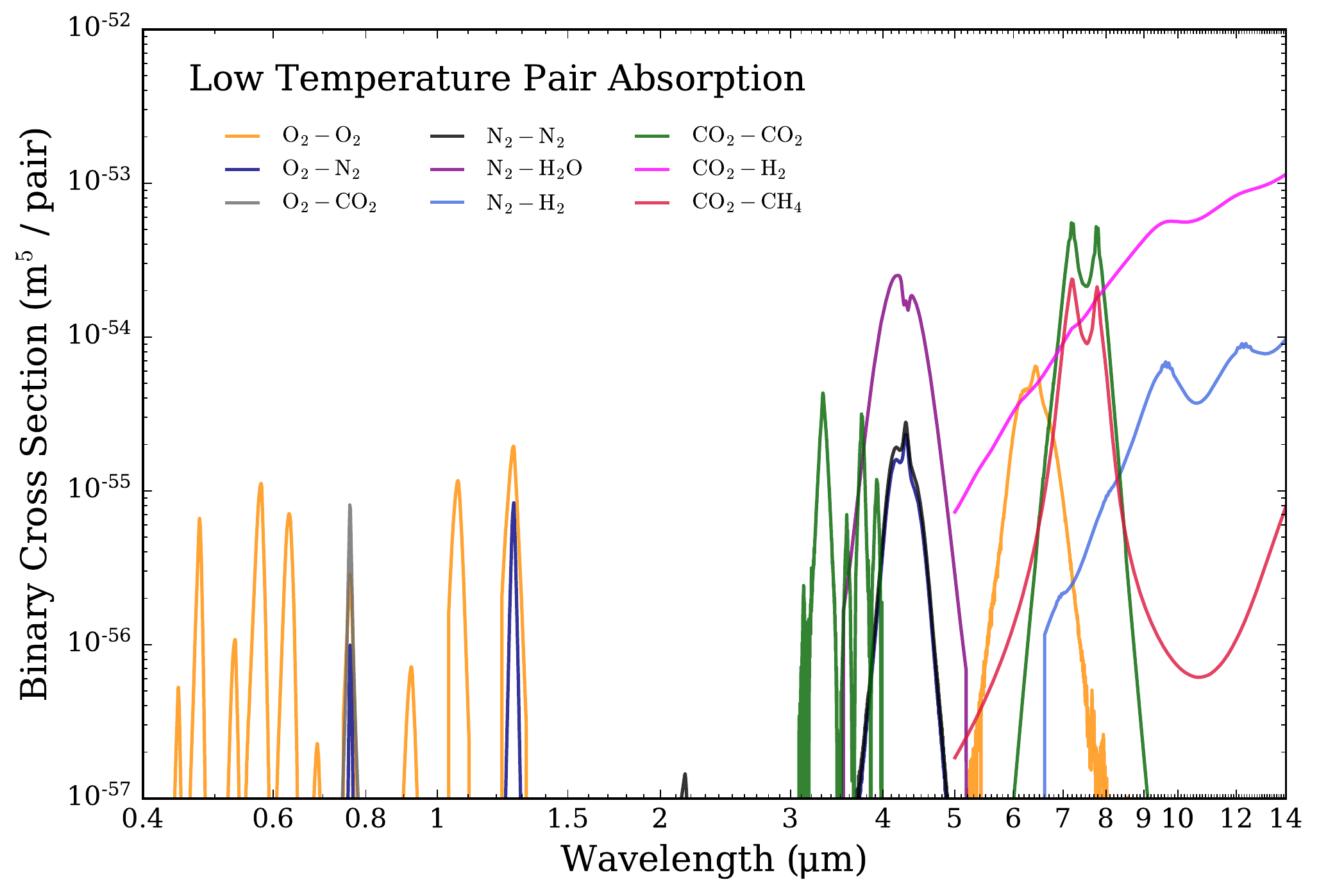}
    \caption{Representative opacity sources for exoplanet atmospheres. Top left: absorption cross sections for various chemical species important for high-temperature giant planets. The cross sections are shown at $T = 2,000$\,K and $P = 1$\,bar. Each cross section is Gaussian-smoothed from its native resolution ($\Delta \nu = 0.01$\,cm$^{-1}$) to $R \sim 1,000$ for clarity. Top right: absorption cross sections for several species important for low-temperature terrestrial planets ($T = 300$\,K and $P = 1$\,bar). The line lists used to compute these cross sections are summarised in Appendix~\ref{appendix_C}. Bottom left: binary cross sections arising from pair processes (collision-induced absorption and free-free opacity) important for high-temperature giant planets ($T = 2,000$\,K). Bottom right: binary cross sections from collision-induced absorption pairs important for low-temperature terrestrial planets ($T = 300$\,K).}
\label{fig:cross_sections}
\end{figure*}

\subsubsection{Rayleigh Scattering} \label{subsubsec:Rayleigh_scattering}

We model Rayleigh scattering as a pure loss term\footnote{Forward scattering is negligible for Rayleigh phase functions, so Rayleigh scattering can be treated as additional absorption.} with extinction expressed as
\begin{equation}
    \kappa_{\lambda, \, \rm{Rayleigh}} (P, \phi, \theta) = \sum_{q=1}^{N_{\rm species}} n_q (P, \phi, \theta) \, \sigma_{\lambda, \, \rm{Rayleigh}, \, q}
\label{eq:extinction_Rayleigh}
\end{equation}
where the Rayleigh cross section can be written as either
\begin{equation}
    \sigma_{\lambda, \, \rm{Rayleigh}, \, q} = \frac{24 \pi^3}{n_{\rm{ref}, \, q}^2 \, \lambda^4} \, \left(\frac{\eta^2_{q}\,(\lambda) - 1}{\eta^2_{q}\,(\lambda) + 2}\right)^2 \, F_{\rm{King}, \, q}\, (\lambda)
\label{eq:Rayleigh_1}
\end{equation}
or, equivalently
\begin{equation}
    \sigma_{\lambda, \, \rm{Rayleigh}, \, q} = \frac{128}{3} \pi^5 \, \bar{\alpha}_{q}^2\,(\lambda) \, \lambda^{-4} \, F_{\rm{King}, \, q}\, (\lambda)
\label{eq:Rayleigh_2}
\end{equation}
In the expressions above $\eta_q$ is the refractive index of species `$q$' (measured at reference number density $n_{\rm{ref}, \, q}$), $F_{\rm{King}, \, q}$ is the King correction factor (accounting for non-spherical charge distributions), and $\bar{\alpha}_{q}$ is the polarisability. We employ various fitting functions for the refractive indices and King correction factors \citep[e.g.][]{Sneep2005,Hohm1994}. Where such data is not available, we use static polarisabilities from the CRC handbook \citep{Haynes2014} and assume $F_{\rm{King}, \, q} = 1$.

\subsubsection{Pair Absorption} \label{subsubsec:pair_absorption}

Continuum absorption can also arise from interactions between pairs of chemical species. The most common pair process opacity in exoplanet atmospheres is collision-induced absorption (CIA), though free-free absorption (e.g. from H$^{-}$) is also a pair process \citep[e.g. see][]{Sharp2007}. Unlike single-species absorption or scattering, pair processes are proportional to $n_{\rm tot}^2$ with the following functional form
\begin{equation}
    \kappa_{\lambda, \, \rm{pair}} (P, \phi, \theta) = \sum_{\rm pairs} n_{q_1} (P, \phi, \theta) \, n_{q_2} (P, \phi, \theta) \, \sigma_{\lambda, \, \rm{pair}, \, q} \, (T)
\label{eq:extinction_pair}
\end{equation}
where $n_{q_1}$ and $n_{q_2}$ are the number densities of the two species constituting the pair (e.g. for H$_2$-He CIA they are $n_{\rm{H}_2}$ and $n_{\rm{He}}$) and $\sigma_{\lambda, \, \rm{pair}, \, q}$ is the binary absorption cross section (in m$^5$) for the interaction. We use temperature-dependent CIA data from the HITRAN CIA database \citep{Karman2019} and free-free absorption data for H$^{-}$ from \citet{John1988}. Where data is not available for a given temperature, we assume the cross section is equal to the nearest temperature with tabulated data. We further assume the cross section is zero at wavelengths outside the data range. We show the range of pair process cross sections in our opacity database, relevant for both gas giant and terrestrial planets, in Figure~\ref{fig:cross_sections} (bottom panels).

\subsubsection{Aerosol Extinction} \label{subsubsec:aerosol_extinction}

Aerosols contribute to atmospheric extinction via both absorption and scattering. In the limit where forward scattering is neglected\footnote{As in Section~\ref{subsubsec:coordinate_systems}, we note that the scattering and absorption coefficients should be treated separately in a full Monte Carlo multiple scattering calculation.} we can generally write the aerosol extinction as  
\begin{equation}
    \kappa_{\lambda, \, \rm{aerosol}} (P, \phi, \theta) = \sum_{q=1}^{N_{\rm aerosol}} n_q (P, \phi, \theta) \, \overline{\sigma}_{\lambda, \, \rm{ext}, \, q}
\label{eq:extinction_aerosol}
\end{equation}
where the mean extinction cross section is a weighted average over a (normalised) particle size distribution
\begin{equation}
    \overline{\sigma}_{\lambda, \, \rm{ext}, \, q} = \int_{0}^{\infty}  \sigma_{\lambda, \, \rm{ext}, \, q} (a) \, f(a) \, da
\label{eq:mean_aerosol_cross_section}
\end{equation}
The aerosol cross section for a specific particle size, $\sigma_{\lambda, \, \rm{ext}, \, q} (a)$, can be analytically determined via treatments such as Mie theory. Many studies have already considered the spectral impact of Mie scattering \citep[e.g.][]{Wakeford2015,Pinhas2017,Benneke2019}, so we focus here on how spatial variations in cloud opacity impact transmission spectra. 

\subsection{A 3D Cloud Prescription} \label{subsec:clouds}

Aerosols with inhomogenous spatial distributions are a common prediction of GCM models, driven by the non-uniform temperature field of tidally locked exoplanets \citep[e.g.][]{Parmentier2016}. Many previous studies have investigated the influence of such patchy clouds on transmission spectra \citep[e.g.][]{Line2016,MacDonald2017a,Welbanks2021}. These parametric models assume one or more aerosol layers occupy a given terminator fraction, such that a transmission spectrum with inhomogenous clouds is a linear combination of the 1D transmission spectra of each sector. However, these models make two key assumptions that render them unsuitable for multidimensional atmospheres. First, they assume the cloud is uniform from the dayside to the nightside. In reality, the region where aerosols manifest (i.e. cloud condensation or haze formation) can be a strong function of temperature. Second, previous models assume the background atmosphere has a uniform temperature and composition around the terminator. Consequently, the azimuthal location of a cloud need not be specified, since the spectrum for a 1D atmosphere is invariant under rotations about the line of sight. To illustrate the problem with this assumption, imagine a 2D atmosphere with a temperature gradient between the morning and evening terminators (e.g. Figure~\ref{fig:2D_atm_structure}, left panel). Due to the larger scale height of the warmer evening terminator, a cloud covering 40\% of the evening terminator will have a more pronounced effect than an identical cloud covering 40\% of the morning terminator. Our goal in this section is to propose a parametric prescription for 3D clouds that relaxes these limiting assumptions.

\begin{figure*}[!htb]
    \centering
    \includegraphics[width=\textwidth]{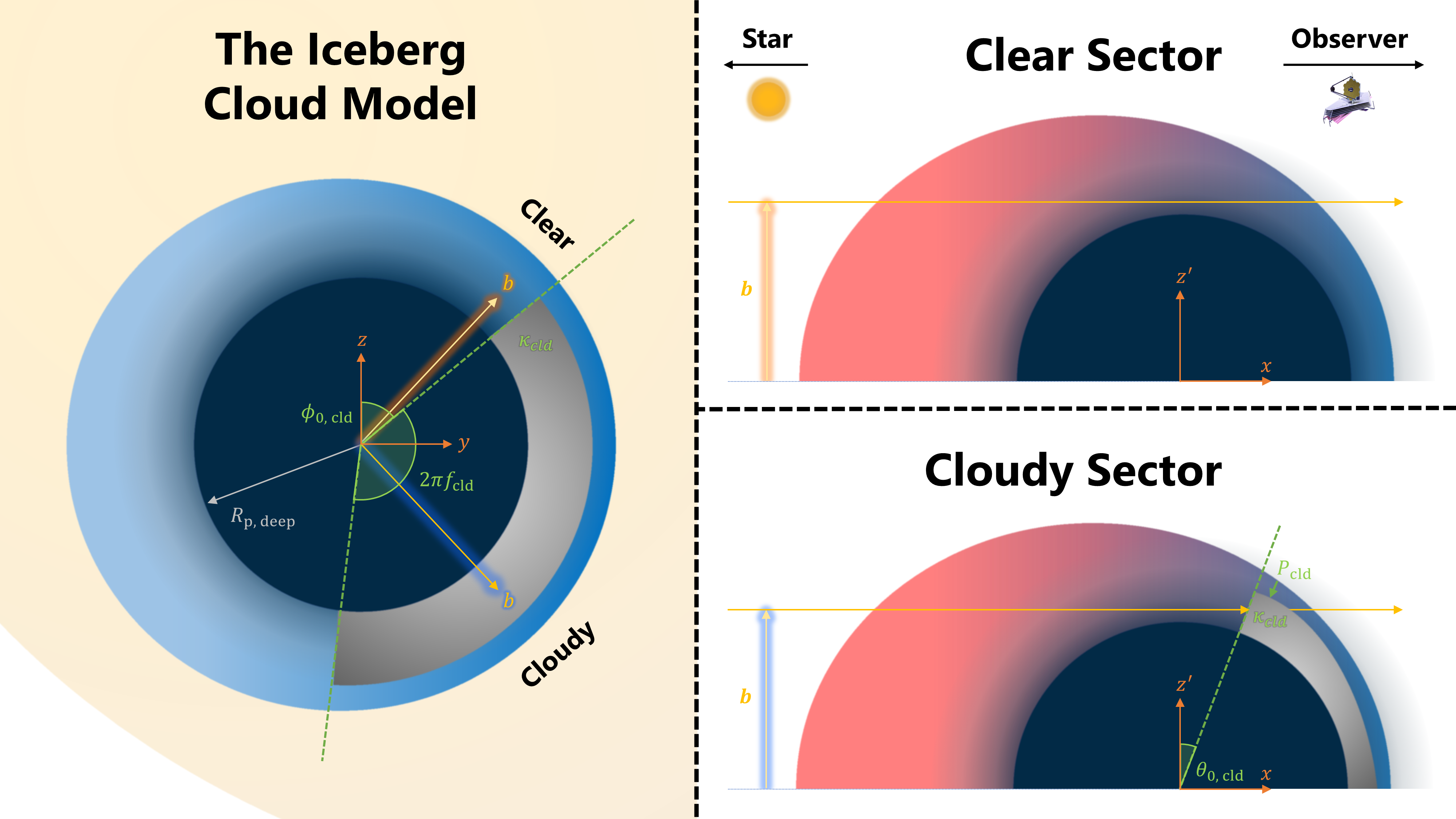}
    \caption{A parametric cloud model for 3D exoplanet atmospheres. Left (observer's perspective): an inhomogenous cloud, with coverage fraction $f_{\rm cld}$, ranges between azimuthal angles of $\phi_{0, \, \rm{cld}}$ and $\phi_{0, \, \rm{cld}} + 2 \pi f_{\rm cld}$. Two rays are highlighted with identical impact parameters, $b$, but different azimuthal angles, $\phi$, such that one ray passes through a clear sector (orange arrow) whilst the other encounters the cloud (blue arrow). Top right: the ray in the clear sector experiences no cloud opacity. Bottom right: the ray in the cloudy sector encounters a cloud ranging from zenith angle $\theta_{0, \, \rm{cld}}$ to the anti-stellar point. In cloudy atmospheric regions the cloud has an extinction of $\kappa_{\rm cld}$ for $P \geq P_{\rm cld}$ and no extinction in layers above the cloud deck. Our cloud model is thus defined by 5 parameters. We term this 3D generalisation of the classical cloud deck the `Iceberg cloud model'.}
\label{fig:iceberg_cloud_model}
\end{figure*}

Here we introduce a simple inhomogenous cloud prescription that is applicable for multidimensional atmospheres. Consider an inhomogenous cloud deck with an extinction specified by 
\begin{equation}
    \kappa_{\lambda, \, \rm{aerosol}} (P, \phi, \theta) =
    \begin{cases}
        \kappa_{\rm cld}, & \circled{1} \\
        0, & \circled{2} \\
    \end{cases}
\label{eq:iceberg_cloud_model}
\end{equation}
\begin{align*}  
    \circled{1}: \ & P \geq P_{\rm cld} \\
    & \phi_{0, \, \rm{cld}} \leq \phi \leq \phi_{0, \, \rm{cld}} + 2 \pi f_{\rm cld} \\ 
    & \theta \geq \theta_{0, \, \rm{cld}} \nonumber \\ 
    \circled{2}: \ & \rm{else} \nonumber \\ 
\end{align*}
This model has 4 parameters defining the geometry of the cloud: $\phi_{0, \, \rm{cld}}$ (the azimuthal angle where the cloud begins), $f_{\rm cld}$ (the terminator cloud coverage fraction), $\theta_{0, \, \rm{cld}}$ (the zenith angle where the cloud begins), and $P_{\rm cld}$ (the cloud-top pressure). For regions containing the cloud, we ascribe a 1-parameter grey extinction, $\kappa_{\rm cld}$. We term this cloud model --- the 3D generalisation of the classical cloud deck --- the \emph{Iceberg cloud model}. We illustrate the Iceberg cloud model in Figure~\ref{fig:iceberg_cloud_model}.

The Iceberg cloud model provides a natural foundation for more sophisticated 3D cloud models. In general, atmospheric regions containing clouds can have a wavelength-dependent cross section and an altitude-dependent number density profile \citep[e.g.][]{Mai2019,Lacy2020b}. These can be incorporated into the Iceberg cloud model by replacing $\kappa_{\rm cld}$ with Equation~\ref{eq:extinction_aerosol}. Similarly, diffuse hazes can be considered by setting $P_{\rm cld}$ to the top-of-atmosphere pressure and including a wavelength dependent extinction. Another advantage of the Iceberg model is that one can add additional aerosol layers by defining another four spatial parameters alongside the aerosol spectral properties\footnote{This provides a generalisation of the cloud model proposed by \citet{Welbanks2021}, but without the limitation of a 1D background atmosphere.}.

We numerically implement the Iceberg cloud model in TRIDENT within the radiative transfer framework described in Section~\ref{subsubsec:TRIDENT_radiative_transfer}. Once the background atmosphere is initialised, we add additional azimuthal sectors and zenith slices with boundaries where the cloud intersects the background atmosphere grid (i.e. at $\phi_{0, \, \rm{cld}}$, $\phi_{0, \, \rm{cld}} + 2 \pi f_{\rm cld}$, and $\theta_{0, \, \rm{cld}}$). This causes $N_{\rm sector} \rightarrow N_{\rm sector} + 2$ and $N_{\rm zenith} \rightarrow N_{\rm zenith} + 1$ (unless cloud edges fall exactly on any sector/zenith boundaries in the background atmosphere grid). Since Iceberg clouds can break the north-south hemisphere symmetry of the background atmosphere, we add additional sectors for cloudy models to treat each hemisphere separately. Cloudy atmospheric regions have cloud extinction added to Equation~\ref{eq:extinction_coefficient}, whilst clear regions have zero cloud extinction. Since these additional sub-sectors and sub-slices can re-use the path distribution from any regions with the same background atmosphere (since the elements in Equation~\ref{eq:3D_path_distribution_elements} only depend on temperature and composition), TRIDENT can efficiently handle 3D clouds without a linear increase in computation time --- an important advantage for multidimensional retrievals.

\section{Validation of TRIDENT} \label{sec:validation}

Here we validate TRIDENT's new radiative transfer prescription against existing models in the literature. We first validate our implementation of the path distribution method by comparing to a 1D model from \citet{Robinson2017a}. We then validate our 3D radiative transfer technique against a model from the open source code Pytmosph3R \citep{Caldas2019,Falco2021}.

\subsection{Validation for 1D Transmission Spectra} \label{subsec:1D_validation}

\citet{Robinson2017a} introduced the path distribution technique for 1D transmission spectra computation, which we have generalised for 3D atmospheres. We therefore first validate TRIDENT by verifying that our technique reproduces their results for a 1D atmosphere.

\citet{Robinson2017a} presented a comparison between their transmission spectrum code and a forward model from the CHIMERA retrieval code \citep{Line2013} (see \citealt{Robinson2017a}, their Figure~7). Their test case was a simple hot Jupiter with the following physical properties\footnote{Our intercomparison identified a difference in the values of $R_{\rm{J}}$ used in \citet{Robinson2017a} and in TRIDENT. To avoid confusion, we quote our stellar and planetary radii in terms of the solar and equatorial Jovian radii adopted by the IAU \citep{Prsa2016}: $\nom{R} = 6.957 \times 10^8$\,m and $\nomJovianEqRadius = 7.1492 \times 10^7$\,m.}: $R_{*} = 0.78078\,\nom{R}$, $R_{\rm{p}} = 1.1363\,\nomJovianEqRadius$ (at 10\,bar), and $M_{\rm{p}} = 1.14 M_{\rm{J}}$ (hence $g_{\rm{p}} = 21.883$\,ms$^{-2}$ at 10\,bar). The atmosphere has a H$_2$O mixing ratio of $4 \times 10^{-4}$ alongside H$_2$ and He with a relative number fraction of He/H$_2$ = 0.17647. The temperature profile is isothermal at 1500\,K, with the atmosphere divided into 126 layers spaced uniformly in log-pressure from 10$^{-9}$\,bar to 10\,bar. The atmosphere is assumed cloud-free.

\begin{figure}[!ht]
    \centering
    \includegraphics[width=\columnwidth]{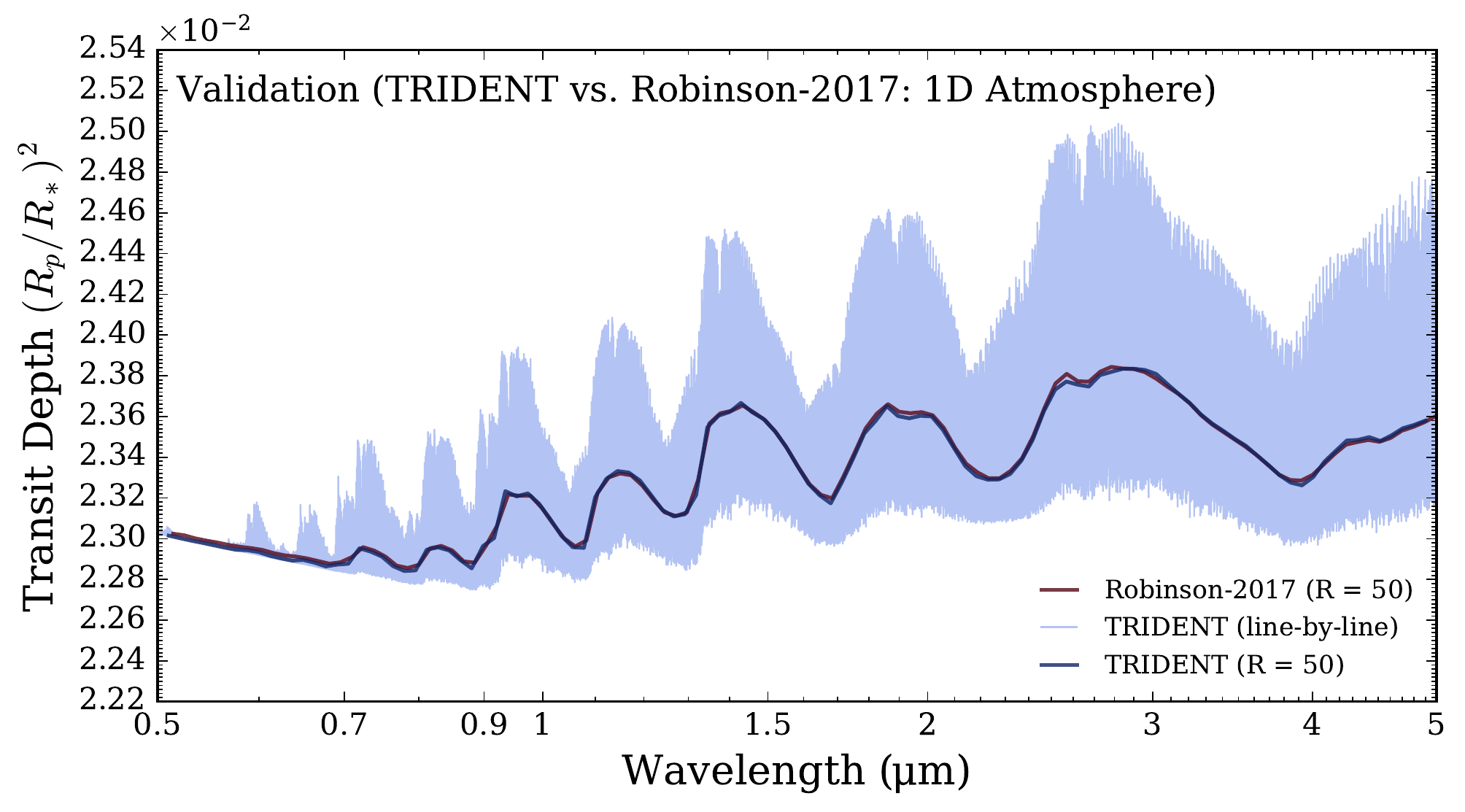}
    \includegraphics[width=\columnwidth]{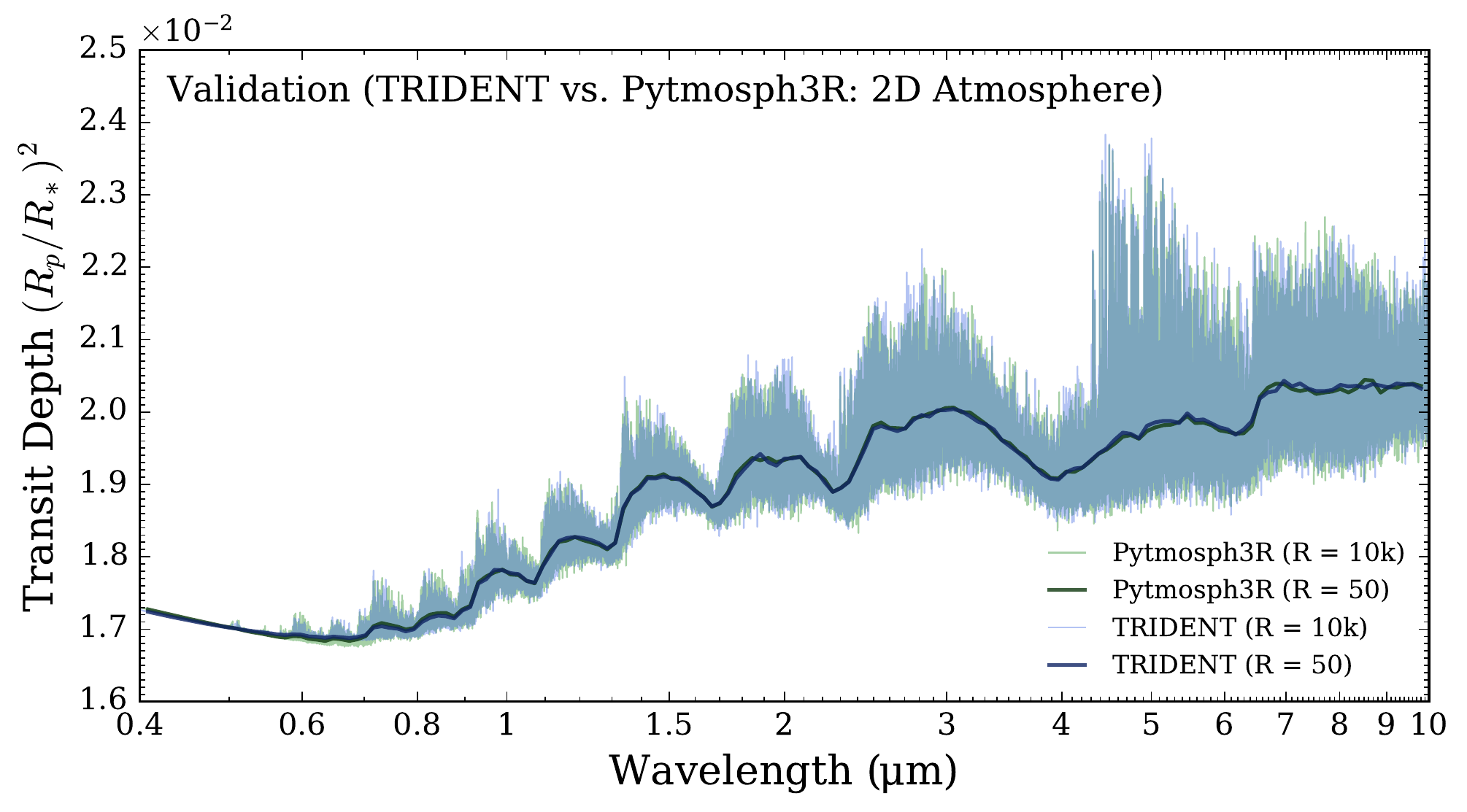}
    \caption{Validation of TRIDENT's radiative transfer against existing 1D and 3D transmission spectra models. Top: validation for a 1D hot Jupiter \citep[see][their Figure 7]{Robinson2017a}. Bottom: validation for a 2D ultra-hot Jupiter (see Figure~\ref{fig:validation_PT_geometry}) against the 3D radiative transfer code Pytmosph3R \citep{Caldas2019,Falco2021}}
\label{fig:validation_transmission_spectra}
\end{figure}

We compare TRIDENT to \citet{Robinson2017a} in Figure~\ref{fig:validation_transmission_spectra} (top panel). We generated our transmission spectrum line-by-line at the native resolution of our cross sections ($R = \frac{\lambda}{d\lambda} = 10^6$ at 1\,$\micron$), before binning down to $R = 50$ for comparison. We obtain excellent agreement between our models (mean difference = 11\,ppm), with the residual differences arising from the codes using different H$_2$O cross sections\footnote{\citet{Robinson2017a} used HITEMP2010 \citep{Rothman2010}, whilst TRIDENT uses POKAZATEL \citep{Polyansky2018} (Tyler Robinson, personal communication).}. We further verified that the path distribution technique agrees with traditional transmission spectra methods by comparing against the forward model from \citet{MacDonald2017a}. This comparison, using the same opacities, resulted in agreement to machine precision.

\subsection{Validation for 3D Transmission Spectra} \label{subsec:3D_validation}

\begin{figure*}[!htb]
    \centering
    \vspace{-0.1cm}
    \includegraphics[width=0.96\textwidth]{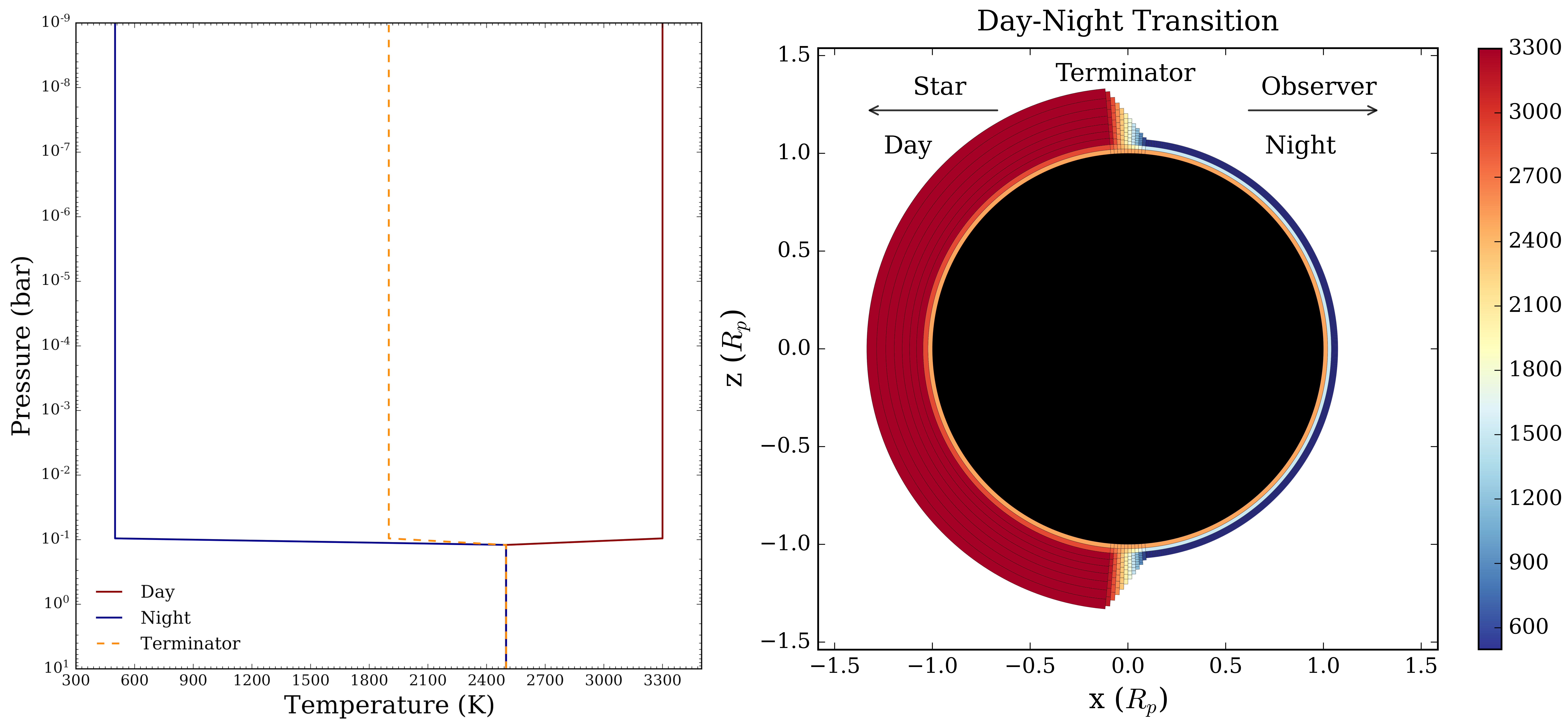}
    \vspace{-0.1cm}
    \caption{The temperature structure of an ultra-hot Jupiter model used to validate TRIDENT against Pytmosph3R. Left: P-T profiles for the dayside, nightside, and terminator plane. Right: the day-night transition temperature field and geometry (to scale). The atmospheric model is 2D, since azimuthal symmetry is assumed.}
\label{fig:validation_PT_geometry}
\end{figure*}

We next validated TRIDENT against the open source 3D radiative transfer code Pytmosph3R \citep{Caldas2019,Falco2021}. We adopt an ultra-hot Jupiter model with a day-night gradient from the Pytmosph3R documentation\footnote{\href{http://perso.astrophy.u-bordeaux.fr/\~jleconte/pytmosph3r-doc/tutorial\_pytmosph3r.html\#Parametrize-a-2D-day-night-transition}{Pytmosph3R documentation}.}. The ultra-hot Jupiter has the following properties: $R_{*} = 1.458\,\nom{R}$, $R_{\rm{p}} = 1.7670\,\nomJovianEqRadius$ (at 10\,bar), and $M_{\rm{p}} = 1.1829\,M_{\rm{J}}$ (hence $g_{\rm{p}} = 9.3897$\,ms$^{-2}$ at 10\,bar). The P-T profile consists of an isothermal dayside (3300\,K), an isothermal nightside (500\,K), and a discontinuous jump to a deep isotherm (2500\,K) for $P > 0.1$\,bar (see Figure~\ref{fig:validation_PT_geometry}, left panel). The nightside has a H$_2$O mixing ratio of $5.0119 \times 10^{-3}$, a CO mixing ratio of $4.4 \times 10^{-4}$, and the remainder is pure H$_2$. The dayside has $10 \times$ less H$_2$O (simulating photodissociation of H$_2$O), with CO and H$_2$ as on the nightside. The angular width of the day-night transition is $\beta = 10\degr$. Both codes used 100 layers spaced uniformly in log-pressure from 10$^{-9}$\,bar to 10\,bar.

We compare the spectra from Pytmosph3R and TRIDENT in Figure~\ref{fig:validation_transmission_spectra} (bottom panel). We computed both spectra on the same wavelength grid using opacity sampling with $R =$ 10,000 from 0.4--10\,$\micron$ (32,189 wavelengths). We then binned each spectrum down to $R = 50$ for closer comparison to current low-resolution transmission spectra data (and for consistency with Section~\ref{subsec:1D_validation}). We found it necessary to place a surface at 0.1\,bar for the spectra to match, so we infer that Pytmosph3R places a solid surface at the pressure where the P-T profile discontinuity occurs. Overall, we find good agreement between the models (mean difference = 29\,ppm), with the residuals dominated by differences in the H$_2$O cross sections\footnote{Pytmosph3R is not currently distributed with opacity files. We therefore used cross sections from the 2018 release of TauREx \citep{Waldmann2015} (\href{https://osf.io/hn8uk/}{https://osf.io/hn8uk/}), which we understand uses the BT2 H$_2$O line list \citep{Barber2006}.}.

\newpage

The agreement between Pytmosph3R and TRIDENT validates our coordinate system and new 3D radiative transfer algorithm. Pytmosph3R defines an atmosphere on a spherical coordinate system (by default north-pole oriented for ease of interpolating GCMs), with 49 latitudes and 64 longitudes for their reference model. TRIDENT uses a rotated coordinate system, with the pole along the observer line-of-sight (Figure~\ref{fig:coordinate_system}), to exploit the computational efficiency of the algorithm we developed (Section~\ref{subsubsec:slant_optical_depth}). We use $N_{\rm{zenith}} = 10$ and $N_{\rm{sector}} = 1$ to validate TRIDENT (Figure~\ref{fig:validation_PT_geometry}, right panel). On a single CPU, Pytmosph3R took 18.7\,s whilst TRIDENT's run time was 0.64\,s. Pytmosph3R's run time decreases to 5.9\,s when the coordinate system is rotated to align with the line-of-sight. Our coordinate system is thus a natural choice for transmission spectra, since azimuthal symmetry can be exploited to reduce a 3D radiative transfer problem to a faster 2D problem. We provide further TRIDENT model run times in Appendix~\ref{appendix_D}.

The previous test demonstrated that TRIDENT agrees with Pytmosph3R for a 2D atmosphere. Since a 3D spectrum is a linear superposition of 2D spectra for each azimuthal sector in our coordinate system (Equation~\ref{eq:transmission_spectrum_discrete}), this automatically validates TRIDENT for 3D models. Nevertheless, we conducted additional tests to verify TRIDENT for 3D atmospheres. For example, we generated 3D spectra and verified that when all morning-evening gradients ($\Delta T_{\rm term}$, $\Delta \log X_{\rm q, \, term}$) are set to zero we obtain the equivalent 2D spectrum with day-night gradients. We conclude that TRIDENT can accurately compute 3D transmission spectra.

\section{Observational Signatures of Multidimensional Atmospheres} \label{sec:3D_signatures}

We now demonstrate how transmission spectra of multidimensional atmospheres differ from their 1D counterparts. We first consider the impact of day-night and morning-evening temperature gradients, before dealing with the more general case with composition gradients. We offer several observational diagnostics to distinguish between transmission spectra of 1D and multidimensional atmospheres. Finally, we conclude by examining how 3D clouds influence transmission spectra.

\subsection{Temperature Gradients} \label{subsec:temperature_gradients}

Here we investigate how the spectrum of a typical hot Jupiter changes when its atmospheric properties become spatially non-uniform. We consider a planet with physical properties based on HD~209458b: $R_{*} = 1.155\,\nom{R}$, $R_{\rm{p}} = 1.30464\,\nomJovianEqRadius$ (at 10\,bar), and $M_{\rm{p}} = 0.6845\,M_{\rm{J}}$ ($g_{\rm{p}} = 9.186$\,ms$^{-2}$ at 10\,bar). Our reference 1D atmosphere has 500 layers spaced uniformly in log-pressure from 10$^{-9}$--100\,bar. The P-T profile has $T_{\rm{high}} = 1400\,$K, $T_{\rm{deep}} = 2200\,$K, and a temperature gradient linear in log-pressure for $10^{-5} < P < 10\,$bar (see Section~\ref{subsubsec:P-T_profiles}). We assume mixing ratios representative of a solar composition hot Jupiter at 1400\,K \citep[e.g.][]{Woitke2018}: $X_{\rm{Na}} = 10^{-6}$, $X_{\rm{Na}} = 10^{-7}$, $X_{\rm{H_2 O}} = 5 \times 10^{-4}$, and $X_{\rm{CO_2}} = 10^{-6}$. The bulk composition is H$_2$ and He with $X_{\rm{He}}/X_{\rm{H_2}} = 0.17$. We assume a clear atmosphere, with the impact of clouds considered in Section~\ref{subsec:multidimensional_clouds}.

\newpage

We generated a series of multidimensional transmission spectra to explore the impact of day-night and morning-evening temperature gradients. We consider day-night temperature differences of $\Delta T_{\rm{DN}} =$ 200--1,000\,K over opening angles from $\beta =$ 0--40$\degr$, with all models having the same terminator plane P-T profile as the 1D reference model. Similarly, we consider morning-evening temperature differences of $\Delta T_{\rm{term}} =$ 200--1,000\,K over opening angles from $\alpha = $ 0--160$\degr$, with all models having the same average terminator P-T profile as the 1D reference model. We computed all model spectra at $R =$ 10,000 from 0.4--5\,$\micron$, before binning to $R = 100$ to resemble the typical resolution of binned JWST transmission spectra \citep[e.g.][]{Greene2016}.

\subsubsection{Day-night Temperature Gradients} \label{subsubsec:day-night_temperature_gradients}

\begin{figure*}[htb!]
    \centering
    \includegraphics[width=0.49\textwidth]{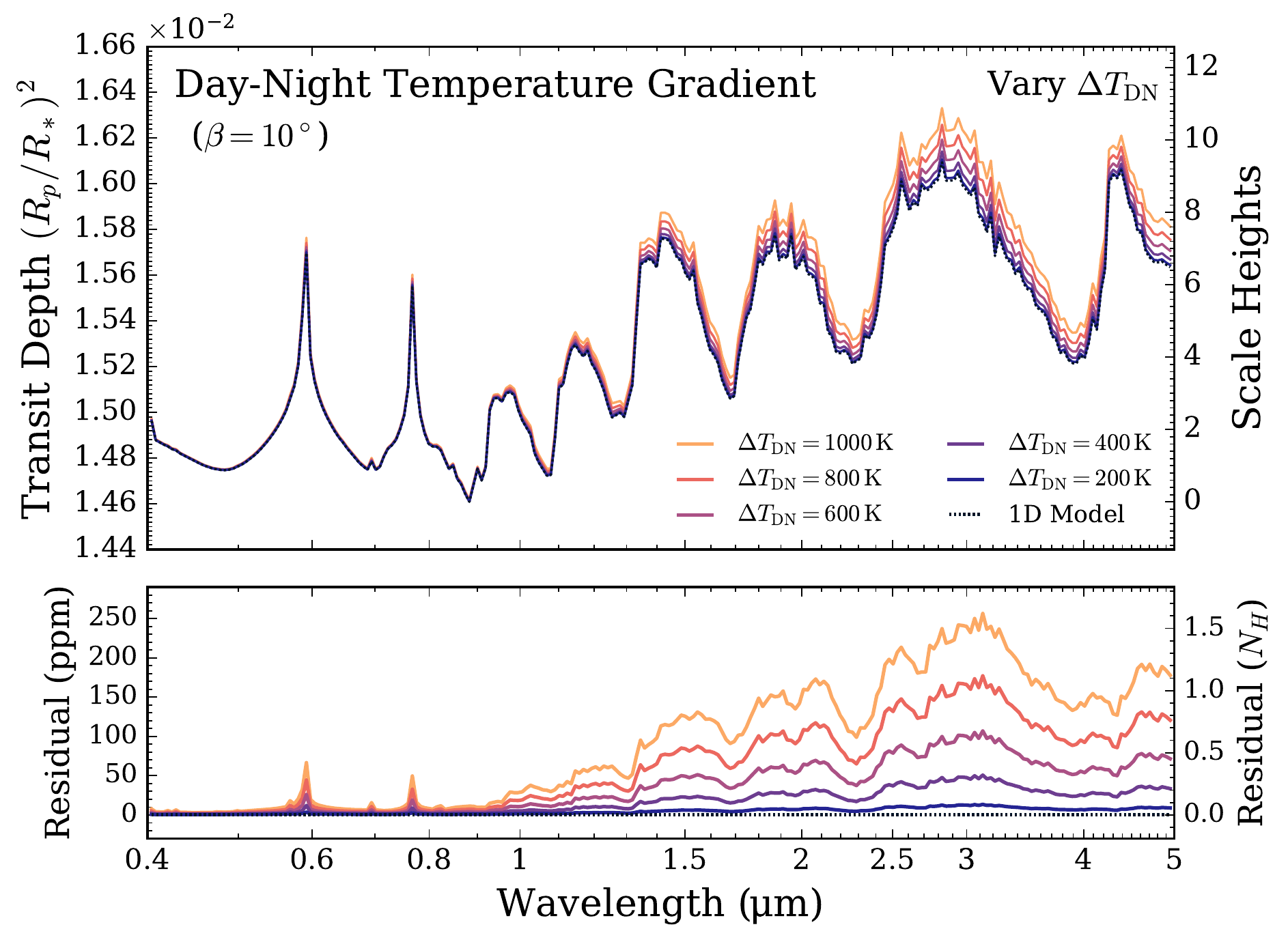}
    \includegraphics[width=0.49\textwidth]{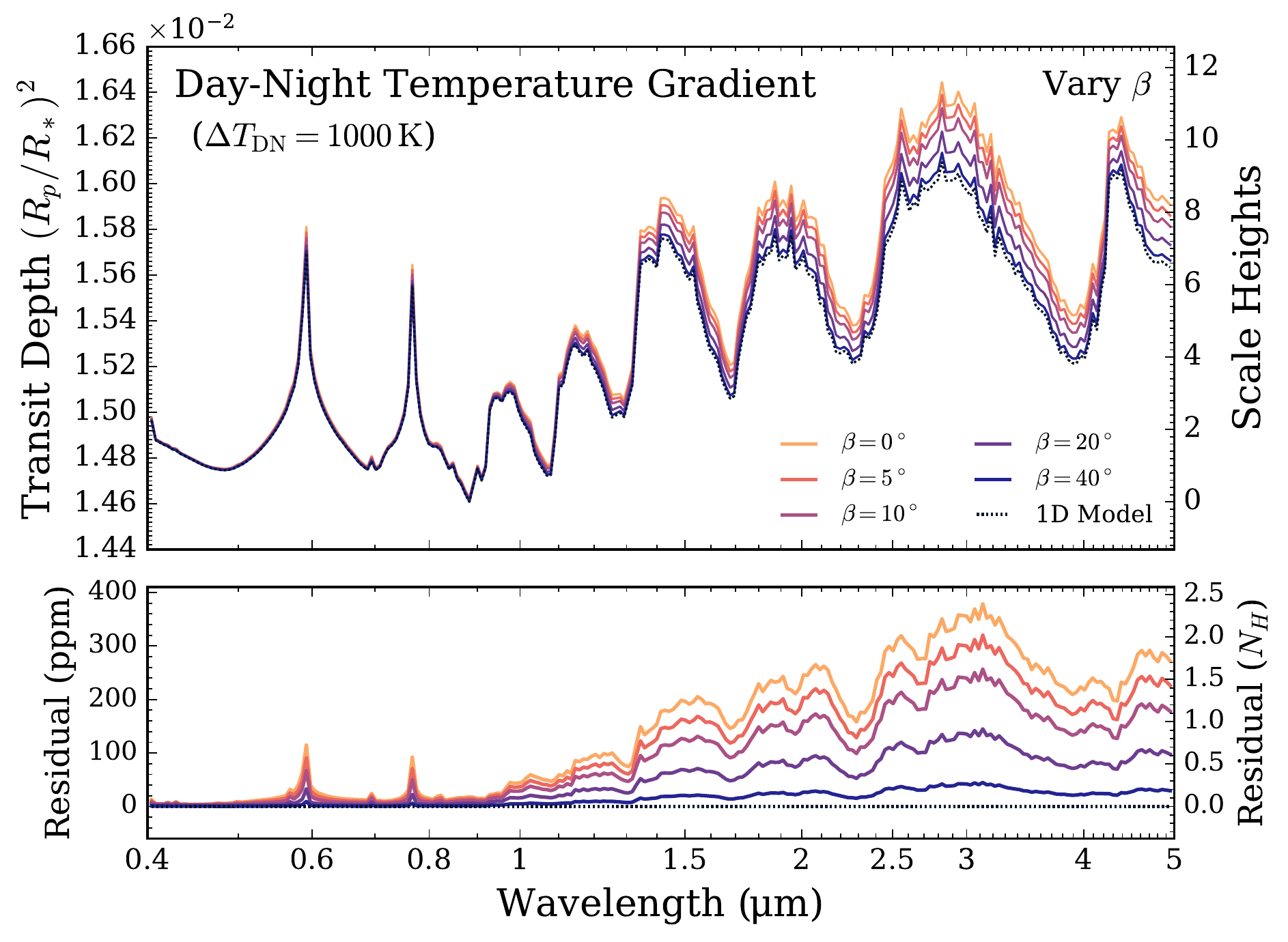}
    \includegraphics[width=0.49\textwidth]{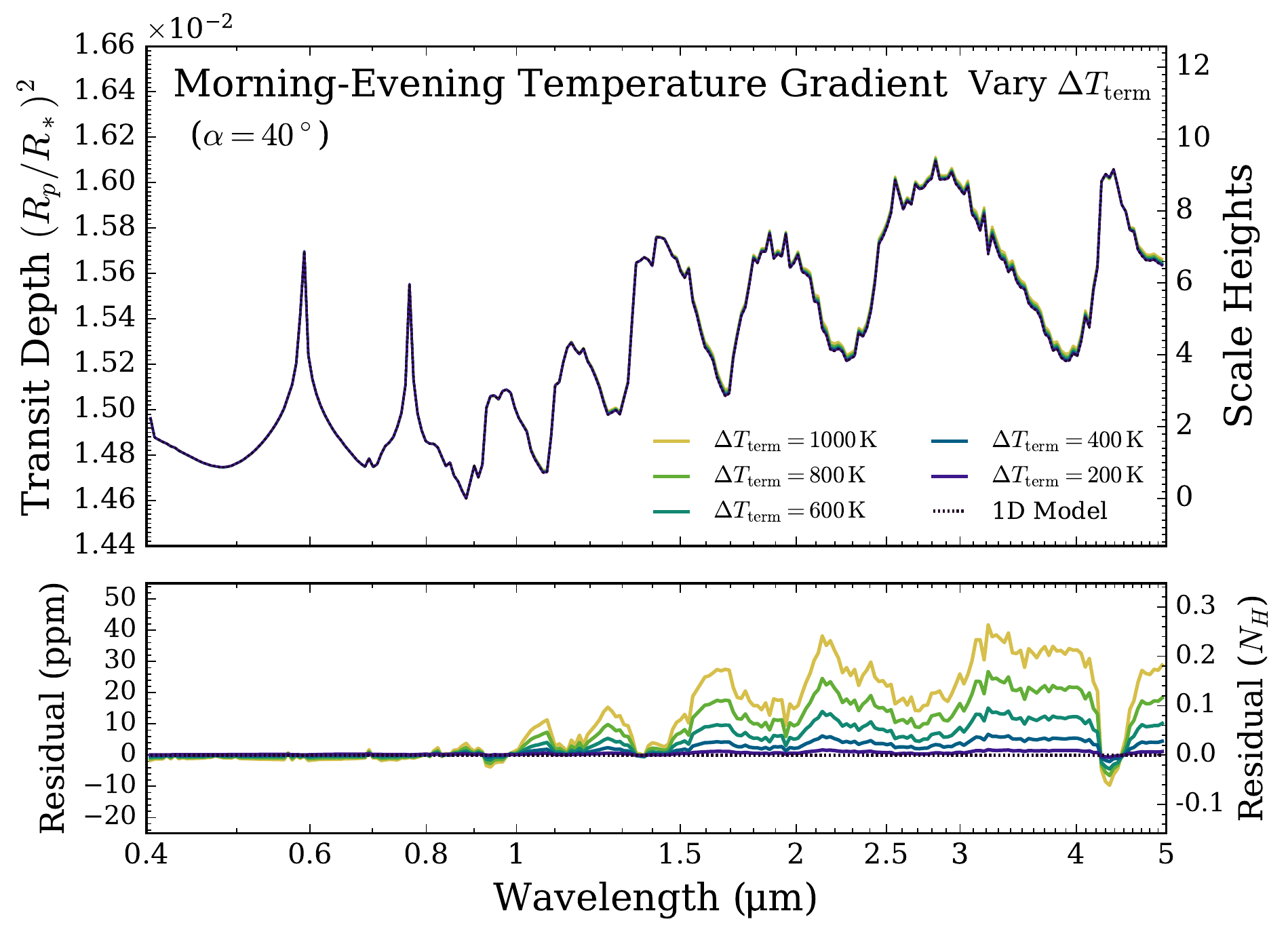}
    \includegraphics[width=0.49\textwidth]{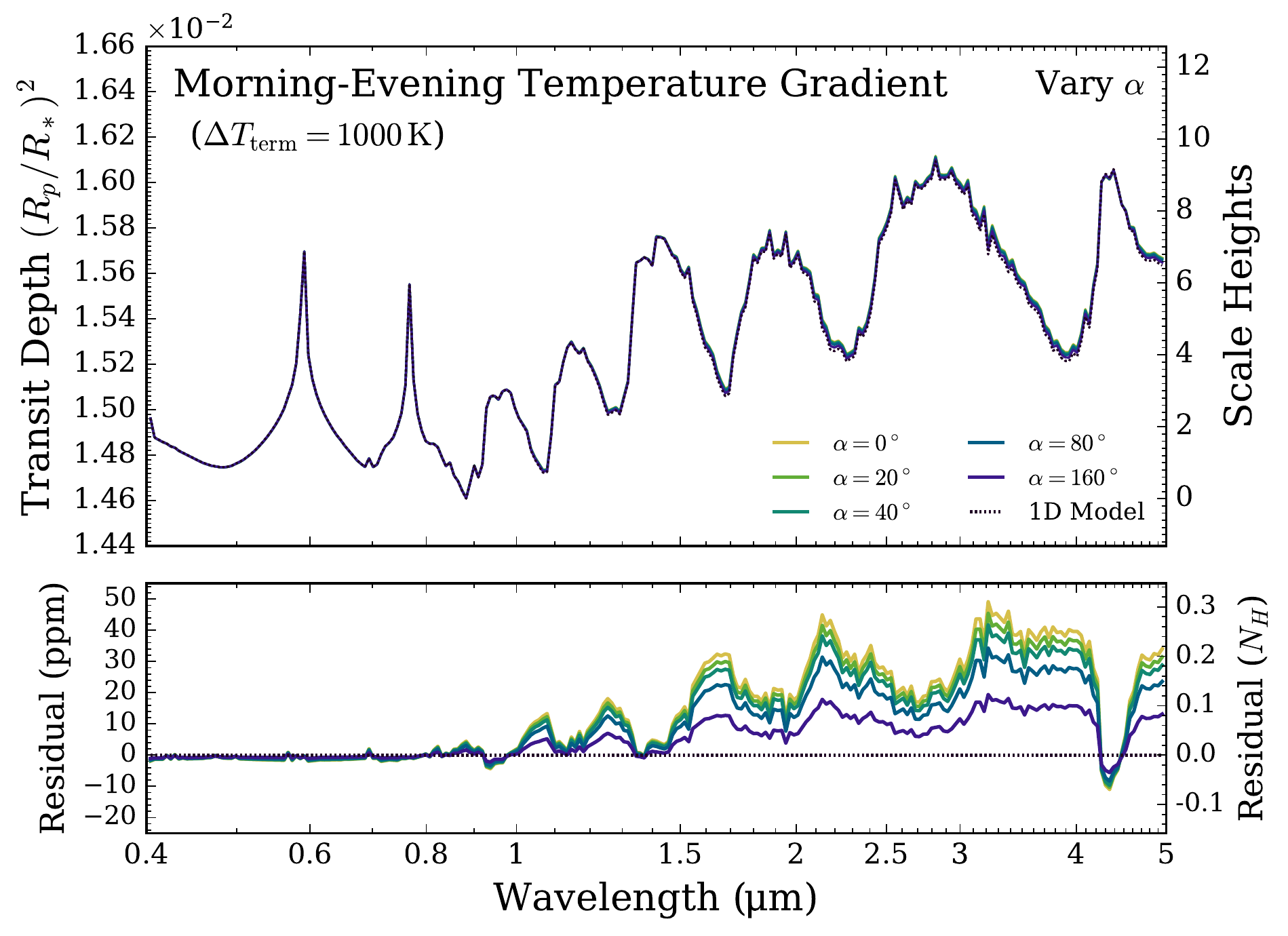}
    \caption{Signatures of multidimensional temperature gradients on hot Jupiter transmission spectra. Top row: the impact of day-night temperature gradients for models with the same P-T profile at the terminator plane. Bottom row: the impact of morning-evening temperature gradients for models with the same average terminator P-T profile. Left column: variation with the day-night (top row) or the morning-evening (bottom row) temperature gradient for a fixed opening angle. Right column: variation with opening angle for a fixed temperature difference. The residual panels show the difference (in parts per million and number of scale heights) compared to a 1D spectrum with the same atmospheric properties as the terminator plane (top row) or the average of the morning and evening terminators (bottom row). Day-night temperature gradients manifest as in increase in the strength of all absorption features. Morning-evening temperature gradients can strengthen or weaken absorption features, depending on the cross section temperature dependence of each species.}
\label{fig:results_temperature_gradients}
\end{figure*}

A day-night temperature gradient strengthens the amplitude of transmission spectra absorption features (Figure~\ref{fig:results_temperature_gradients}, top panels). The increased band strengths arise from longer ray paths through the dayside (see Figure~\ref{fig:2D_atm_structure}), which effectively increases the equivalent scale height of the atmosphere. This agrees with the analytic derivation by \citet{Caldas2019}, who showed that day-night temperature gradients induce a higher slant optical depth than a 1D atmosphere with the same terminator temperature. For a given temperature difference, increasing the opening angle ($\beta$) counteracts the increased absorption from the dayside. This effect stems from the finite angular range a ray can sample about the terminator plane. Since the maximum zenith angle a ray samples upon entering the dayside is $\theta_{\rm{max}} = 2 \cos^{-1} (b / R_{\mathrm{p, \, top}})$, we estimate that $\theta_{\rm{max}} \approx 67\degr$ for our HD~209458b model (assuming the deepest ray samples down to 0.1\,bar). Therefore, as $\beta \rightarrow \theta_{\rm{max}}$ the temperature gradient becomes sufficiently shallow along the line of sight that the 1D limit is recovered (Figure~\ref{fig:results_temperature_gradients}, top right).

The detectability of day-night temperature gradients from transmission spectra has mixed prospects. Though the residuals from even moderate temperature gradients ($\Delta T_{\rm{DN}} \gtrsim 600$\,K) can exceed 100\,ppm, the detection of a 2D day-night temperature difference relies on the inability of a 1D model to compensate for the residuals. Since the residuals for a day-night temperature gradient manifest as a scale height increase, a 1D model can simply increase the planet temperature to compensate. Indeed, \citet{Caldas2019} showed that a 1D retrieval of a spectrum with such a day-night gradient infers a biased temperature warmer than the true terminator temperature. We show in Section~\ref{subsec:composition_gradients} that this holds only when one assumes uniform abundances, with the presence of a composition gradient substantially improving the detection prospects for day-night temperature gradients.     

\subsubsection{Morning-evening Temperature Gradients} \label{subsubsec:morning-evening_temperature_gradients}

A morning-evening temperature gradient can strengthen or weaken absorption features (Figure~\ref{fig:results_temperature_gradients}, bottom panels). This effect is distinct from day-night temperature gradients, albeit yielding residuals $\sim 10 \times$ less pronounced. The small morning-evening residuals arise from rays sampling both the morning and evening terminators (via the $\phi$ integral in Equation~\ref{eq:transmission_spectrum_cyclindrical}). The greater effective area of the warmer evening terminator therefore largely cancels with the lower effective area of the colder morning terminator. This agrees with an analytic result from \citet{MacDonald2020}, where we showed that morning-evening temperature gradients negligibly alter 2D transmission spectra from their 1D counterpart. However, the superposition of the morning and evening terminators is not exactly identical to a 1D spectrum with the average terminator temperature. The residuals are caused by the different cross section temperature dependencies of each chemical species in the atmosphere. For example, the residuals in H$_2$O bands have a different magnitude (and sign) than the 4.3\,$\micron$ CO$_2$ feature (see Figure~\ref{fig:results_temperature_gradients}). Increasing the opening angle ($\alpha$) reduces the residuals, since a greater fraction of the azimuthal integral covers sectors with a similar temperature to the average terminator temperature.

A morning-evening temperature gradient may prove detectable in high-precision ($\sim$ 20\,ppm) transmission spectra. A morning-evening temperature gradient distorts the shape of absorption bands (Figure~\ref{fig:results_temperature_gradients}). For example, the wings of the 3\,$\micron$ H$_2$O band strengthen more than the corresponding absorption peak. Since this residual structure is distinct for different molecular bands and for different molecules, a multidimensional retrieval of a sufficiently precise dataset, covering a wide wavelength range, may provide a better fit than a 1D model. While the  magnitude of morning-evening temperature gradient residuals is comparable to the $\sim$ 20\,ppm JWST noise floor \citep{Greene2016}, we show in the next section that the addition of a composition gradient significantly amplifies the residuals.

\subsection{Composition Gradients} \label{subsec:composition_gradients}

We now investigate how day-night and morning-evening compositional gradients influence transmission spectra. To isolate the impact of compositional gradients, we consider two cases: (i) `pure' composition gradients ($\Delta T_{\rm{DN}} = \Delta T_{\rm{term}} = 0$), and (ii) combined composition and temperature gradients ($\Delta T_{\rm{DN}} =$ 1,000\,K or $\Delta T_{\rm{term}} =$ 1,000\,K). The latter case is more representative of real hot Jupiters, since temperature gradients induce compositional gradients (e.g. equilibrium abundance variations with temperature). We consider H$_2$O abundance variations of up to 2 dex between the dayside and nightside ($\Delta \log X_{\rm{H_{2}O, \, DN}} =$ -2 to +2) or morning and evening terminators ($\Delta \log X_{\rm{H_{2}O, \, term}} =$ -2 to +2). We assume the H$_2$O log-mixing ratio varies linearly over the transition regions, with fixed opening angles of $\beta = 10\degr$ or $\alpha = 40\degr$. Our multidimensional models thus have the same terminator plane H$_2$O abundance (for day-night gradients) or terminator-averaged H$_2$O abundance (for morning-evening gradients) as our 1D reference atmosphere. We maintain uniform abundances for Na, K, and CO$_2$, so composition gradients manifest at wavelengths dominated by H$_2$O opacity. Although we focus here on H$_2$O gradients, similar signatures would be expected for gradients of other species.

\subsubsection{Day-night Composition Gradients} \label{subsubsec:day-night_composition_gradients}

\paragraph{Pure Composition Gradients} \label{paragraph:pure_day-night_composition_gradient}

\begin{figure*}[htb!]
    \centering
    \includegraphics[width=0.49\textwidth]{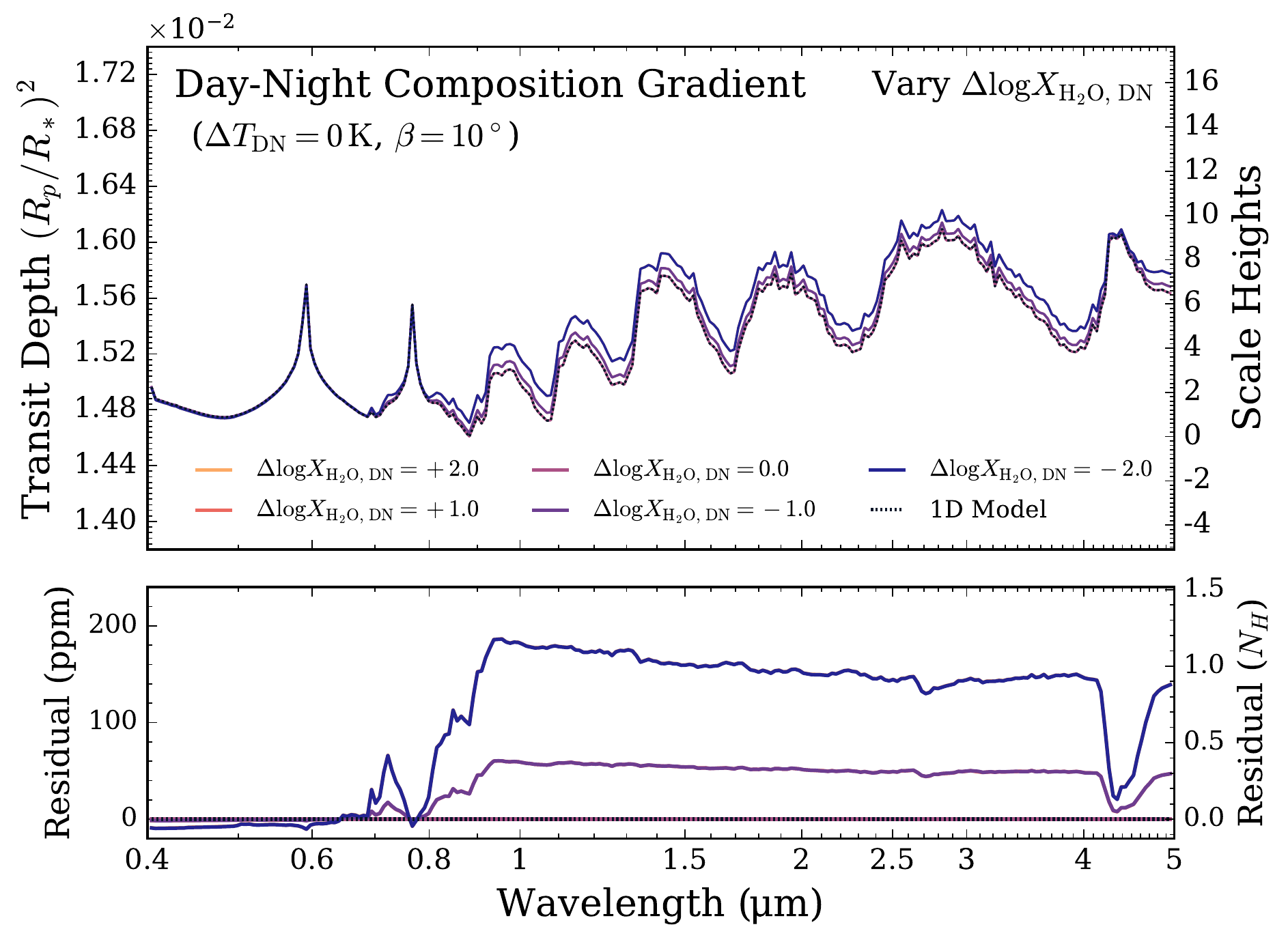}
    \includegraphics[width=0.49\textwidth]{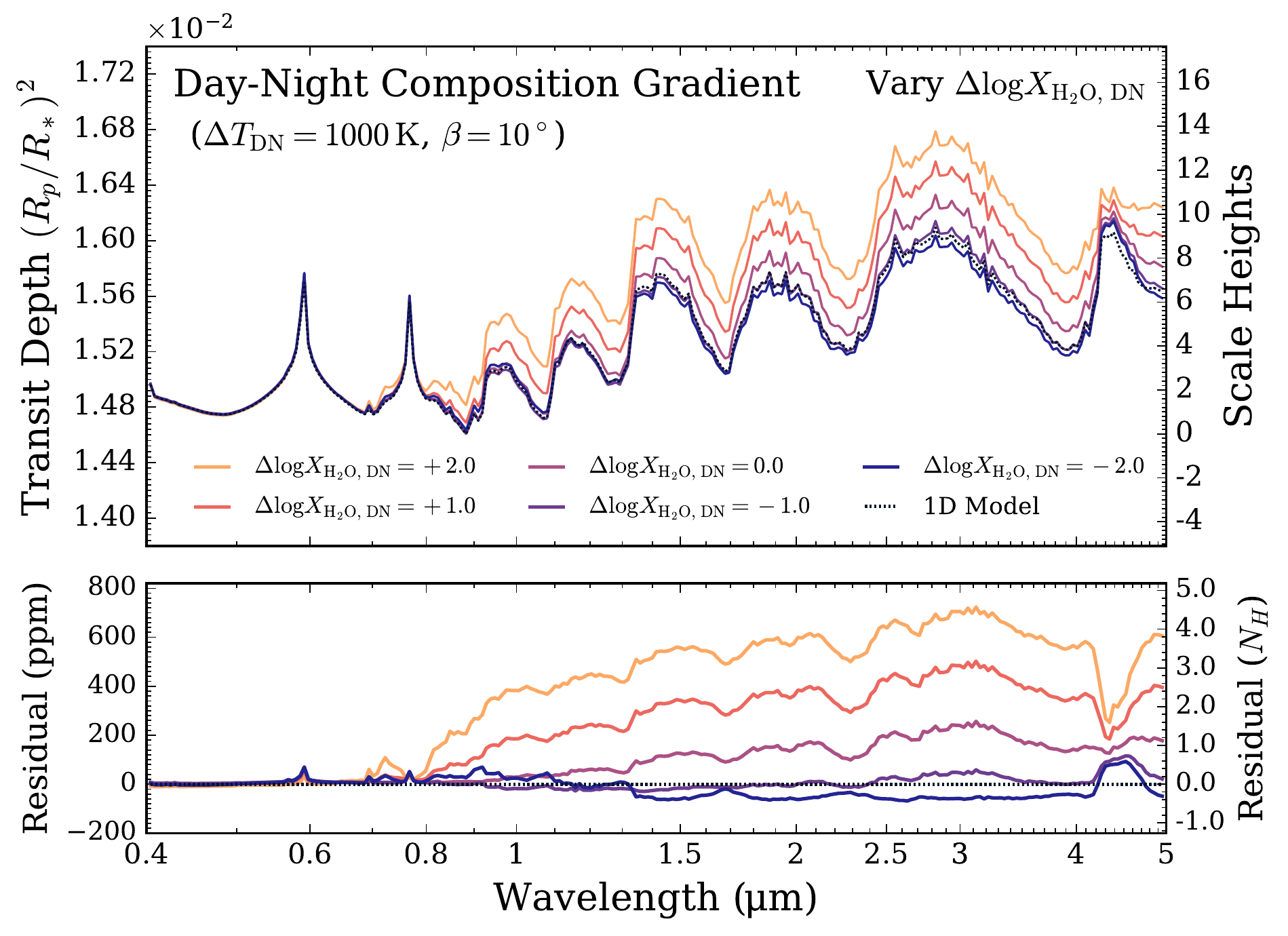}
    \includegraphics[width=0.49\textwidth]{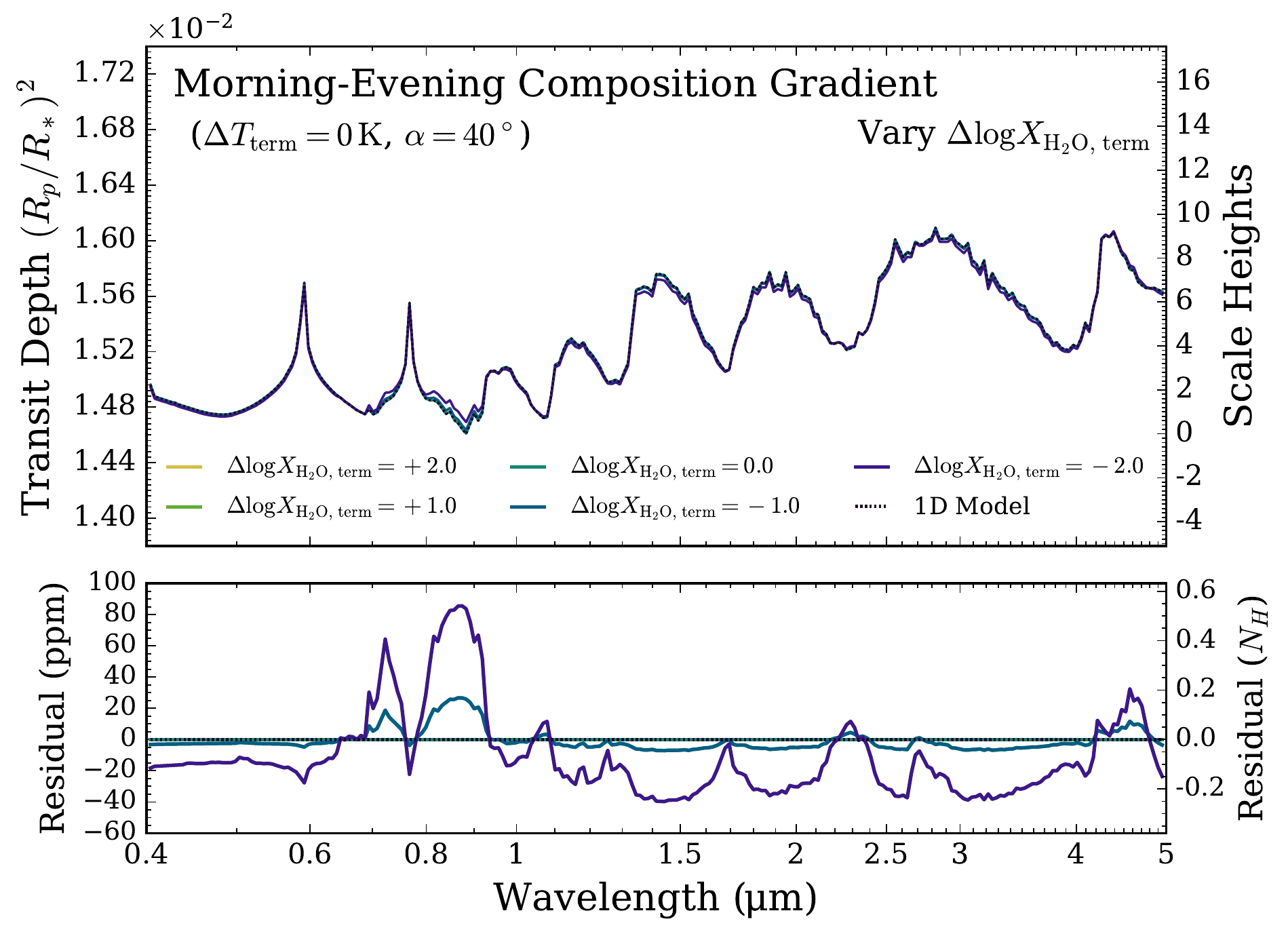}
    \includegraphics[width=0.49\textwidth]{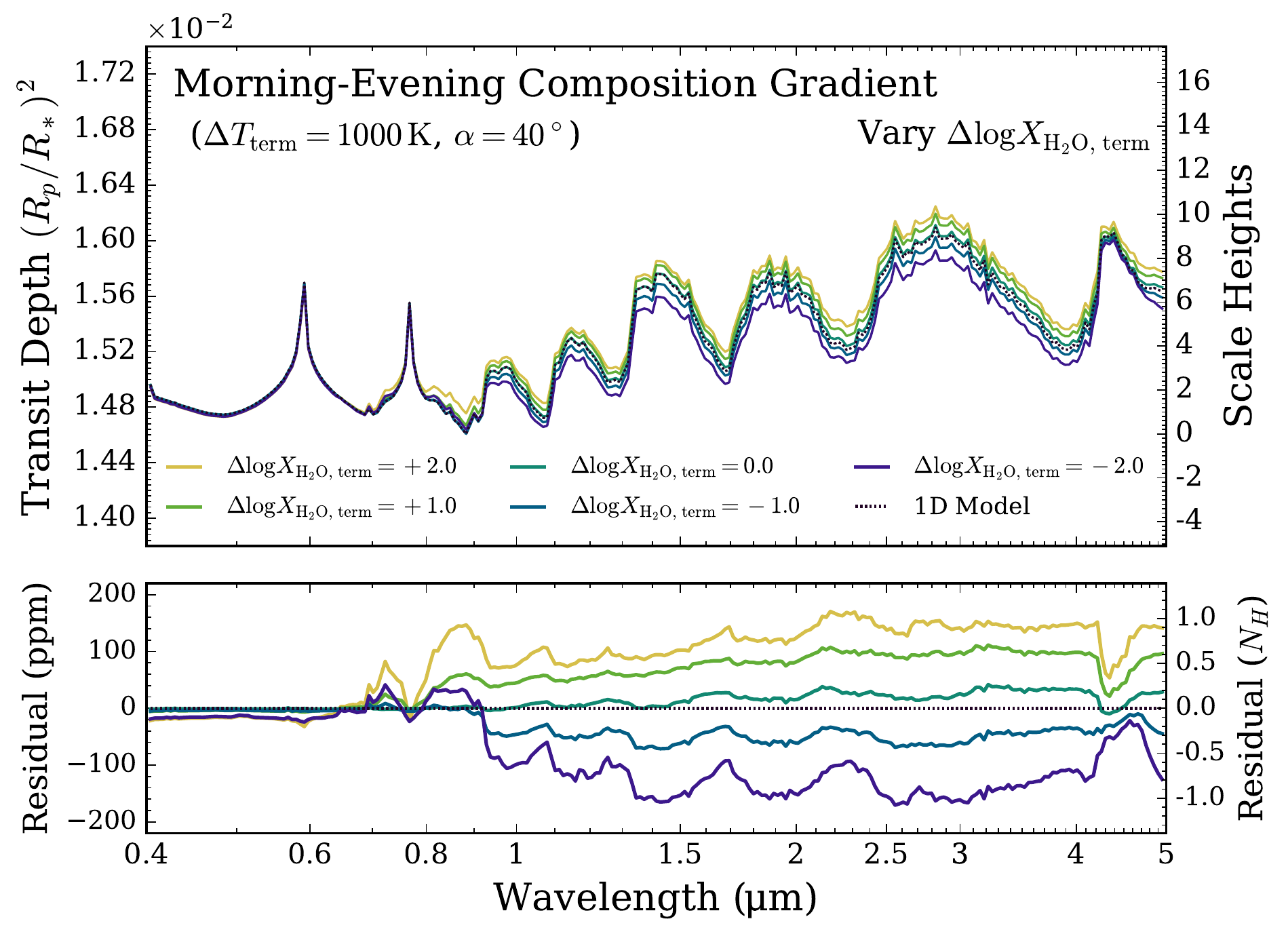}
    \caption{Signatures of multidimensional composition gradients on hot Jupiter transmission spectra. Top row: the impact of day-night H$_2$O abundance gradients for models with the same composition at the terminator plane. Bottom row: the impact of morning-evening H$_2$O abundance gradients for models with the same average terminator composition. Left column: variation with pure composition gradients. Right column: variation with a combined composition gradient and a 1000\,K temperature difference. The residual panels show the difference (in parts per million and number of scale heights) compared to a 1D spectrum with the same atmospheric properties as the terminator plane (top row) or the average of the morning and evening terminators (bottom row). Day-night composition gradients change the relative amplitudes of absorption features for the species exhibiting the gradient (i.e. the H$_2$O band strengths here). Morning-evening composition gradients alter the peak to wing shape of absorption features. Both effects alter spectra by $> 100$\,ppm, rendering these diagnostics readily detectable with JWST.}
\label{fig:results_composition_gradients}
\end{figure*}

Day-night composition gradients produce three key differences compared to our 1D reference model (Figure~\ref{fig:results_composition_gradients}, top left):

\begin{enumerate}
    \item A transit depth increase in bands of the species exhibiting the composition gradient (here, H$_2$O).
    \item Weaker H$_2$O bands (e.g. 1\,$\micron$) are strengthened more than stronger bands (e.g. 3\,$\micron$).
    \item Large composition gradients lower the short-wavelength continuum transit depth.
\end{enumerate}

To understand the first effect, consider a model with $100 \times$ less H$_2$O on the dayside than on the nightside ($\Delta \log X_{\rm{H_{2}O, \, DN}} = -2$). Even though a stellar ray traverses the same distance through the dayside and nightside (for pure composition gradients), the nightside dominates the slant optical depth. Consequently, all H$_2$O absorption bands are stronger than the 1D reference model with a uniform H$_2$O abundance equal to that at the terminator plane ($10 \times$ less than the nightside).

The second effect, unequal band strengthening, comes from geometry: wavelengths with low H$_2$O opacity allow the transmission of rays with deeper impact parameters. Since low-impact parameter rays traverse a greater distance in the dayside/nightside before encountering the transition region (see Figure~\ref{fig:2D_atm_structure}, right panel), the high-H$_2$O hemisphere has a more pronounced effect than for rays only sampling the upper atmosphere.

The third effect is an indirect consequence of a high H$_2$O abundance gradient. Despite these optical wavelengths having negligible H$_2$O opacity, the H$_2$O abundance dichotomy slightly alters the mean molecular weight of the dayside and nightside. The smaller scale height in the high-H$_2$O hemisphere then weakens the alkali wings and H$_2$ Rayleigh opacities.   

Finally, we note that models with more H$_2$O on the dayside (positive gradient) produce identical spectra to those with the same H$_2$O enhancement on the nightside (negative gradient). This symmetry holds only for pure composition gradients, since the dayside and nightside---with the same physical size and temperature in this case---can be freely interchanged without altering the slant optical depths through the atmosphere. 

\paragraph{Composition \& Temperature Gradients} \label{paragraph:day-night_combined_gradient}

Combined, day-night composition and temperature gradients change the relative amplitudes of absorption bands (Figure~\ref{fig:results_composition_gradients}, top right). Compared to pure composition gradients, a temperature gradient breaks the symmetry between models with positive or negative H$_2$O gradients. Models with a higher dayside H$_2$O abundance have significantly amplified H$_2$O bands due to the extended ray paths through the hot dayside. Conversely, models with a higher nightside H$_2$O abundance have much weaker H$_2$O bands due to the shortened ray paths through the cool nightside. In concert, the effective scale height increase (from temperature gradients) and band strengthening (from composition gradients) alters the relative band strengths of the species with a day-night abundance gradient. 

One can detect day-night composition and temperature gradients via the distortion of band strengths compared to 1D models. Atmospheres with lower dayside abundances can have lower band amplitudes at wavelengths with strong absorption---alongside higher band amplitudes at wavelengths with weak absorption---than a 1D model with the same abundance as the terminator plane (e.g. compare the 3\,$\micron$ and 1\,$\micron$ H$_2$O bands in Figure~\ref{fig:results_composition_gradients}, top right). Alternatively, atmospheres with a higher dayside abundance display strongly amplified absorption features. The residuals from day-night composition and temperature gradients are often greater than 100\,ppm, while exhibiting a structure different from the scale height enhancement seen for pure temperature gradients. Consequently, 1D retrievals attempting to fit a spectrum with day-night composition gradients experience significant biases in retrieved abundances, temperatures, and planetary radii \citep{Pluriel2021}. However, the same phenomenon causing these biases may offer the solution to reliably identify day-night gradients with multidimensional retrievals. Our results suggest that transmission spectra datasets covering multiple absorption bands of species expected to exhibit composition gradients is the key to detect day-night gradients.

\subsubsection{Morning-evening Composition Gradients} \label{subsubsec:morning-evening_composition_gradients}

\paragraph{Pure Composition Gradients} \label{paragraph:pure_morning-evening_composition_gradient}

Morning-evening composition gradients produce three differences compared to our 1D reference model (Figure~\ref{fig:results_composition_gradients}, bottom left):

\begin{enumerate}
    \item The bands of the species exhibiting the composition gradient (here, H$_2$O) weaken at wavelengths where it dominates the opacity.
    \item Conversely, H$_2$O bands strengthen at wavelengths where another overlapping species dominates the opacity (e.g. in the K line wings).
    \item Composition gradients lower the short-wavelength continuum transit depth.
\end{enumerate}

The first effect arises from the superposition of the two terminators (and transition region) not being equivalent to a 1D atmosphere with the average terminator H$_2$O abundance. Specifically, the weighted sum of regions with lower H$_2$O opacity and regions with higher H$_2$O opacity produces a weaker H$_2$O band than our 1D reference atmosphere (with the same log mixing ratio of H$_2$O as the average of the morning and evening terminators). We analytically demonstrated this effect in \citet{MacDonald2020}, where we showed that the equivalent 1D temperature of a 2D atmosphere with a morning-evening composition gradient is colder than the average terminator temperature. We note that our derivation relied on the assumption that \emph{only} the species exhibiting the composition gradient dominates the opacity at a given wavelength. Such an assumption holds for strong H$_2$O bands, hence explaining the negative residuals in most of the infrared seen in Figure~\ref{fig:results_composition_gradients} (e.g. for H$_2$O at $3\,\micron$). Consequently, a 1D retrieval would be biased by retrieving a lower temperature than the true terminator average temperature \citep{MacDonald2020}.

The second effect occurs at wavelengths where absorption features from multiple species overlap. Such an opacity overlap amplifies weak H$_2$O features relative to the 1D case, resulting in positive instead of negative residuals. We see this effect most clearly in the K line wings (near $0.7\,\micron$ and $0.9\,\micron$) and the $4.3\,\micron$ CO$_2$ band. Effects (i) and (ii) have opposing influences on the absorption bands of the species displaying the composition gradient, underscoring the importance of observing both weak and strong molecular bands to detect morning-evening composition gradients. 

The third effect is caused by mean molecular weight differences (as with day-night composition gradients). For morning-evening composition gradients, the terminator with the higher mean molecular weight (higher H$_2$O abundance) has a lower scale height and hence presents a decreased effective area to incoming rays. The terminator with the lower H$_2$O abundance has a relatively unchanged mean molecular weight (since it is dominated by H$_2$ and He), so presents a similar effective area. The net effect of the terminator superposition is thus a lower continuum transit depth.

Finally, we see symmetry between models with more H$_2$O on one terminator and the mirror model with the H$_2$O gradient reversed. This results from the morning and evening terminators being freely interchangeable for pure composition gradients, since stellar rays sample the full terminator annulus during a transit.

\paragraph{Composition \& Temperature Gradients} \label{paragraph:morning-evening_combined_gradient}

Combined, morning-evening composition and temperature gradients modulate the strength of absorption bands and alter the peak to wing shape (Figure~\ref{fig:results_composition_gradients}, bottom right). As with day-night combined gradients, the symmetry between positive and negative H$_2$O gradients breaks with the addition of an inter-terminator temperature gradient. The first order effect of a positive H$_2$O gradient is stronger H$_2$O bands, arising from the warmer evening terminator presenting both a greater effective area and a higher H$_2$O opacity. Conversely, a negative H$_2$O gradient significantly weakens H$_2$O bands due to the colder morning terminator (with more H$_2$O) presenting a lower effective area than the warmer evening terminator (with less H$_2$O). The residuals for morning-evening composition and temperature gradients have a unique signature: they are \emph{anti-correlated} with the shape of H$_2$O bands. For positive H$_2$O gradients, the wings of absorption features strengthen more than the band peak. For negative H$_2$O gradients, the wings weaken less than the band peak. Consequently, the wing to peak slope of absorption bands becomes shallower as the magnitude of morning-evening composition gradients increases.

One can detect morning-evening composition and temperature gradients via the sensitivity of peak to wing band shapes to composition gradients. We stress that morning-evening composition gradients are far more detectable than pure temperature gradients. Comparing Figures~\ref{fig:results_temperature_gradients} and \ref{fig:results_composition_gradients}, we see that the residuals for combined morning-evening composition and temperature gradients often exceed 100\,ppm---or 3--5$\times$ more prominent than for pure temperature gradients. Given the magnitude and complex structure of the residuals, multidimensional retrievals are required to correctly interpret transmission spectra shaped by morning-evening gradients \citep{MacDonald2020,Espinoza2021}. Our results indicate two requirements to detect morning-evening gradients: (i) data with sufficient spectral resolution to resolve band wing shapes ($R \sim 100$ is sufficient); and (ii) data with a wide wavelength baseline covering both strong and weak absorption bands for key chemical species. We caution that patchy clouds can also alter band wing shapes in the infrared (see \citealt{Line2016,MacDonald2017a} and Section~\ref{subsec:multidimensional_clouds}), so uniquely identifying morning-evening composition gradients requires both visible wavelength and infrared observations. Finally, the distinct ways in which day-night and morning-evening gradients influence multidimensional transmission spectra strongly implies that a 3D retrieval code can \emph{simultaneously} detect gradients around and across exoplanet terminators. 

\subsection{Multidimensional Clouds} \label{subsec:multidimensional_clouds}

\begin{figure*}[htb!]
    \centering
    \includegraphics[width=0.497\textwidth]{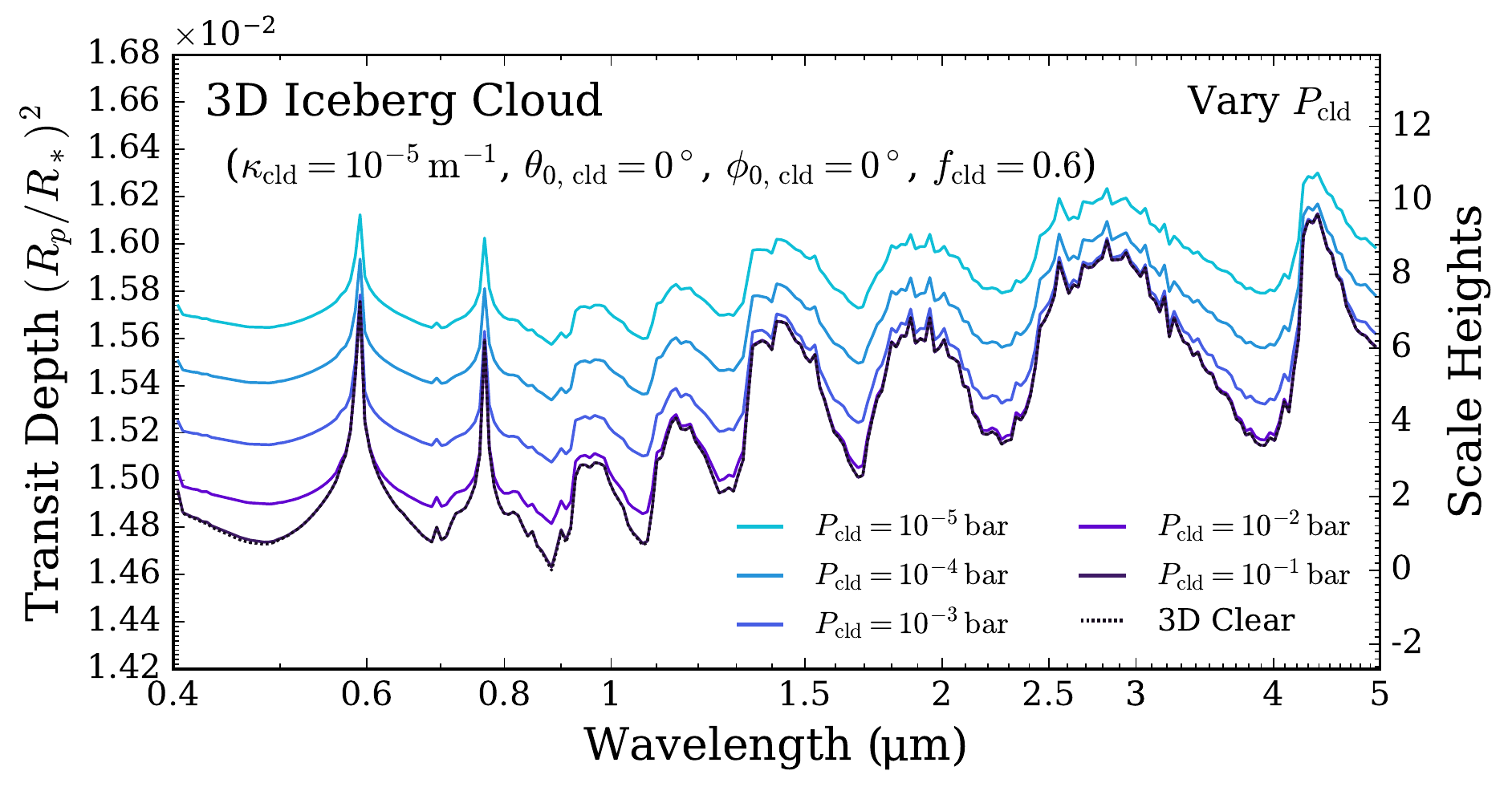}
    \includegraphics[width=0.497\textwidth]{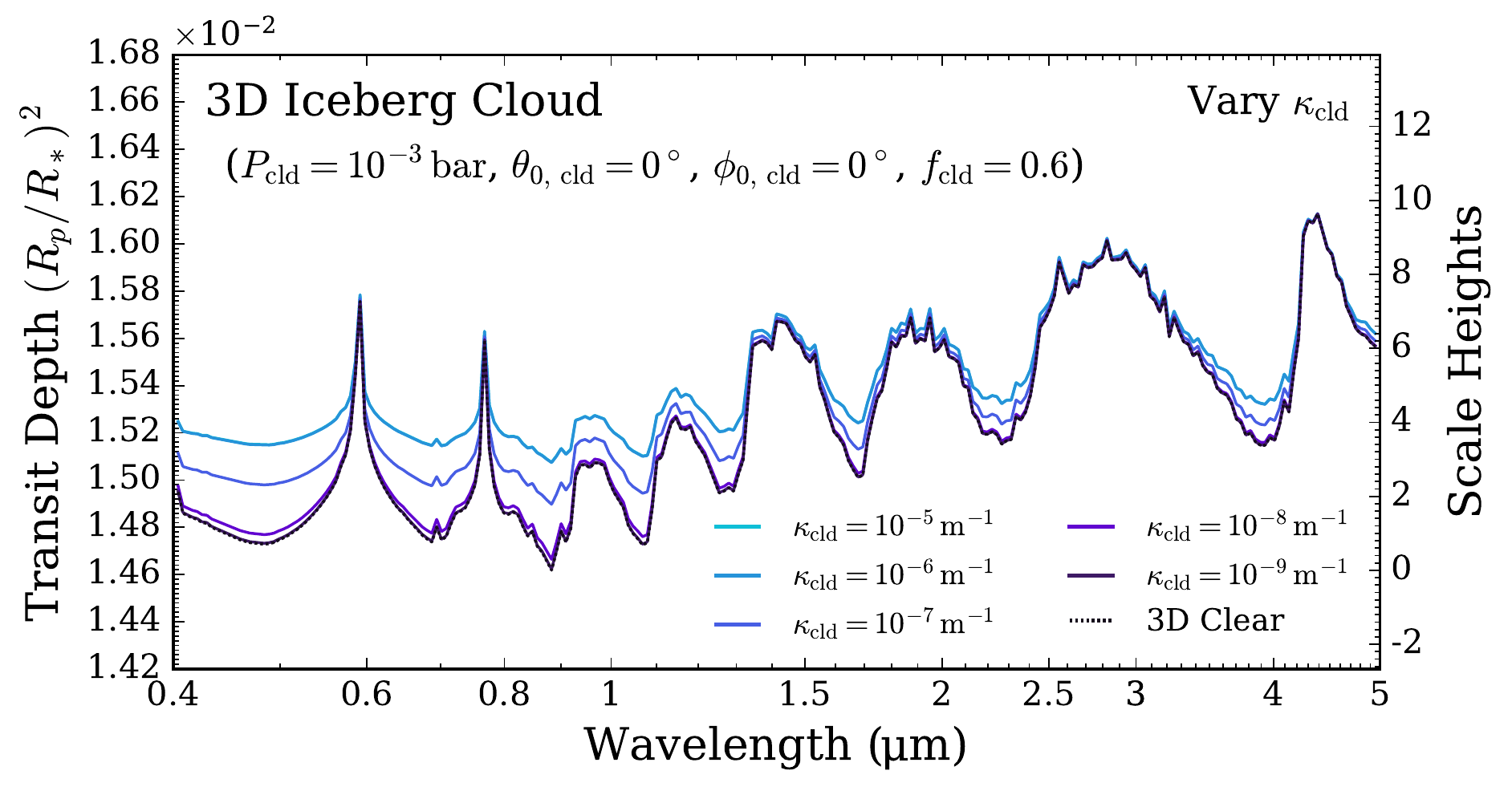}
    \includegraphics[width=0.497\textwidth]{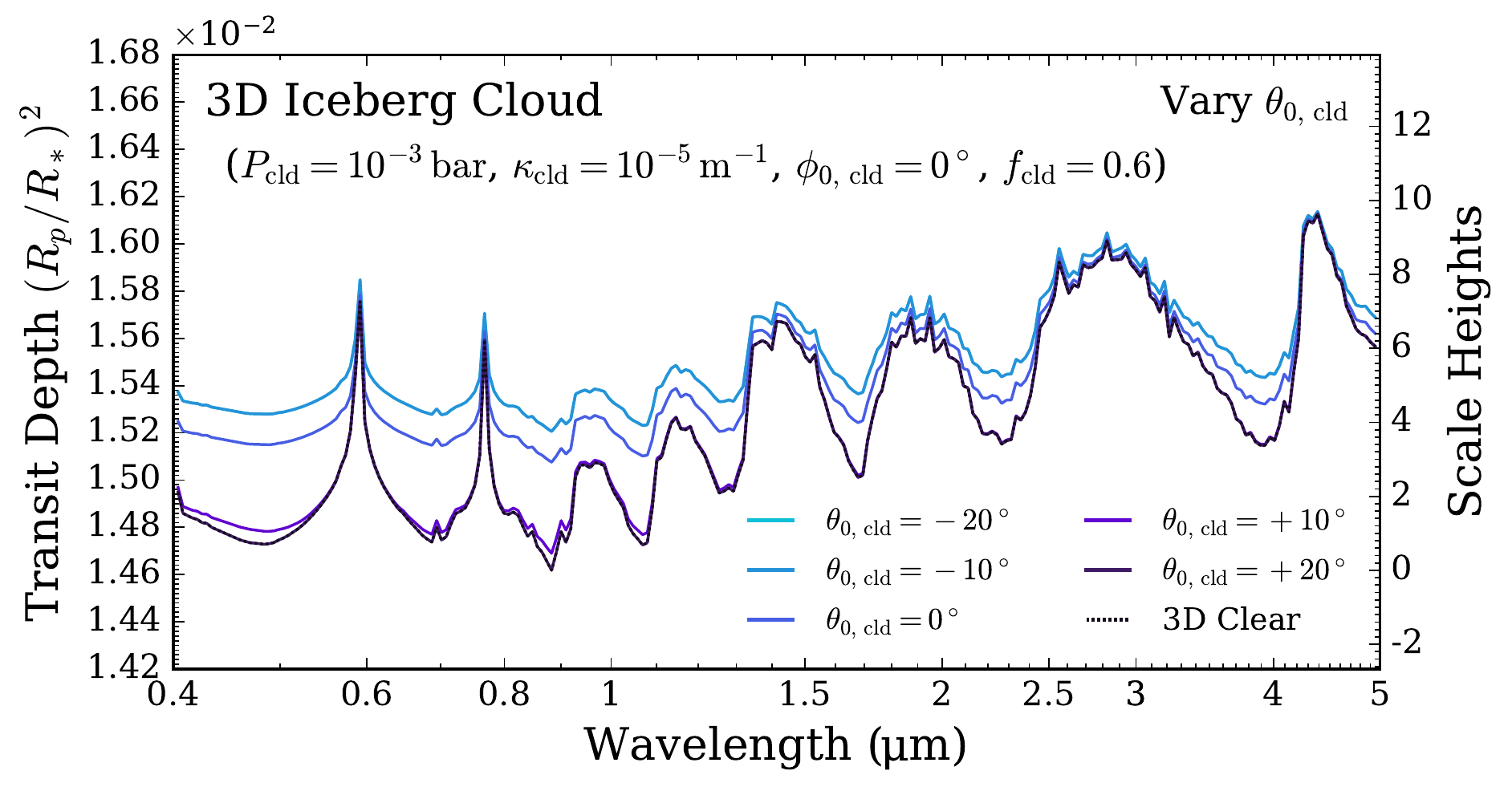}
    \includegraphics[width=0.497\textwidth]{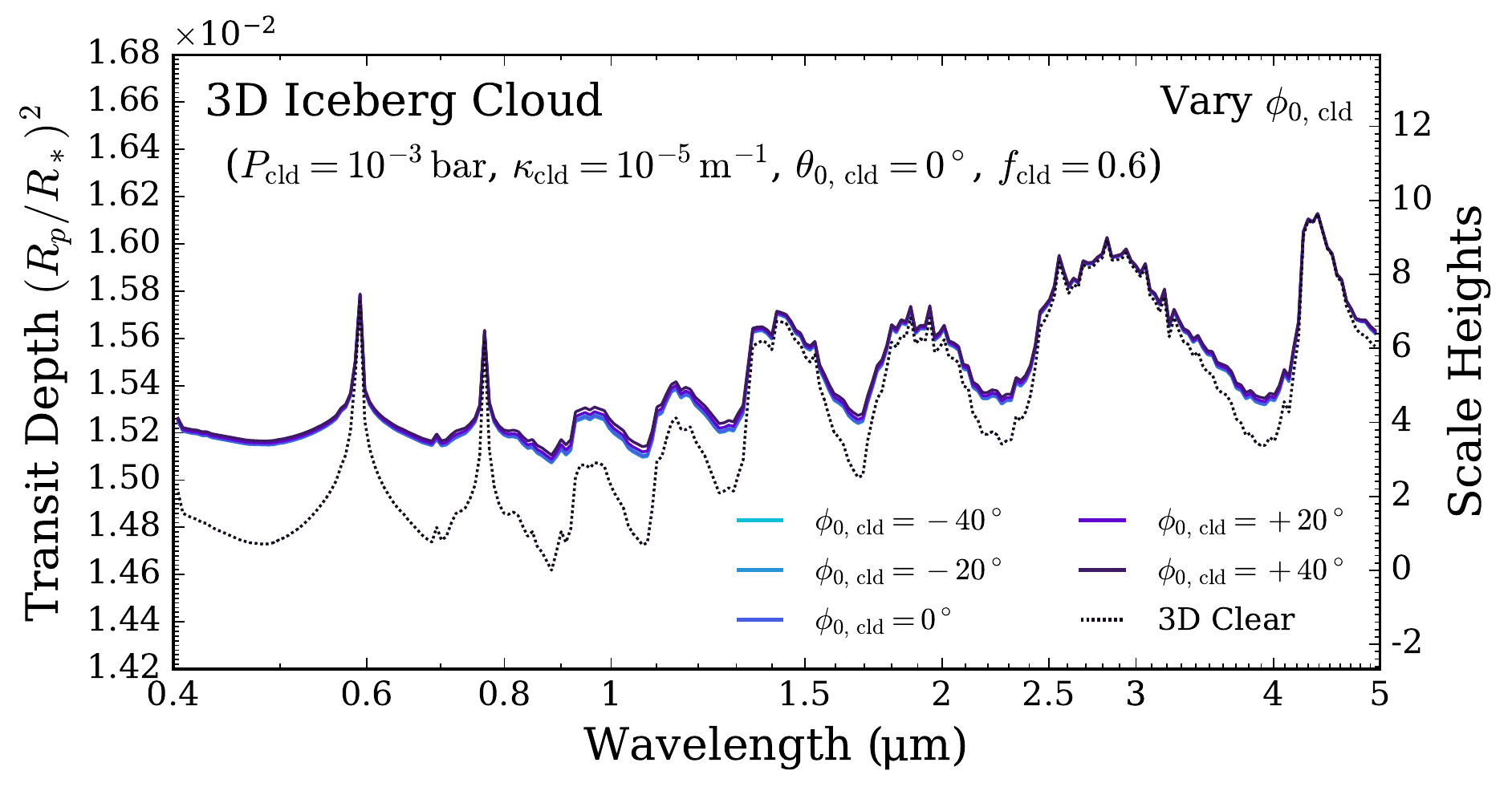}
    \includegraphics[width=0.497\textwidth]{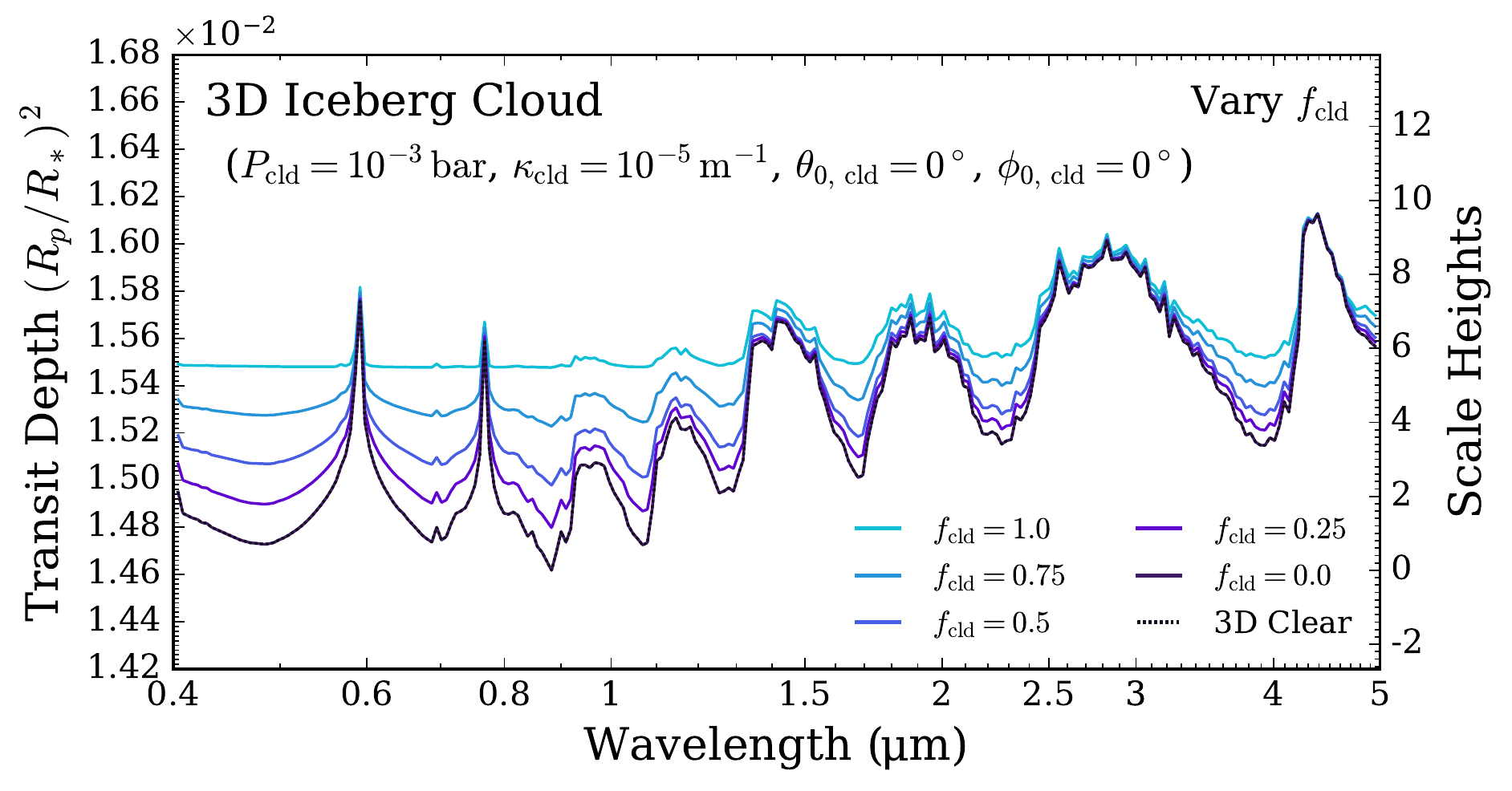}
    \caption{Influence of a 3D Iceberg cloud on the transmission spectrum of a 3D hot Jupiter atmosphere. Each panel shows the variation with the five free parameters defining an Iceberg cloud (see Section~\ref{subsec:clouds} and Figure~\ref{fig:iceberg_cloud_model}): cloud-top pressure (top left), extinction (top right), starting zenith angle (middle left), starting azimuthal angle (middle right), and azimuthal coverage fraction (bottom). The spatial geometry of a 3D cloud alters transmission spectra in ways distinct from a 1D cloud deck.}
\label{fig:results_clouds}
\end{figure*}

We now explore how multidimensional Iceberg clouds affect transmission spectra of 3D atmospheres. Though we focus on clouds, our framework is extendable to general 3D aerosol distributions (see Section~\ref{subsec:clouds}). Our goal is to demonstrate that the spatial extent of non-uniform clouds may be inferred from transmission spectra.

We consider a hot Jupiter with the same planetary properties as Section~\ref{subsec:temperature_gradients}. Our model has a 3D atmosphere exhibiting day-night and morning-evening gradients in the temperature ($\overline{T}_{\rm term} = 1400$\,K, $\Delta T_{\rm term} = 500$\,K, $\Delta T_{\rm DN} = 1000$\,K, and $T_{\rm deep} = 2000$\,K) and H$_2$O abundance ($\overline{\log X}_{\rm{H_{2}O, \, term}} = -3.3$, $\Delta \log X_{\rm{H_{2}O, \, term}} = -1.0$, $\Delta \log X_{\rm{H_{2}O, \, DN}} = -2.0$, and $\log X_{\rm{H_{2}O, \, deep}} = \log X_{\rm{H_{2}O, \, high}} (\theta, \phi)$), with $\alpha = 40\degr$ and $\beta = 10\degr$. We add a reference Iceberg cloud (see Figure~\ref{fig:iceberg_cloud_model}) to this 3D atmosphere with a cloud-top at $P_{\rm cld} = 1$\,mbar and an extinction of $\kappa_{\rm cld} = 10^{-5}\, \rm{m}^{-1}$. The reference cloud starts at the north pole ($\phi_{0, \, \rm{cld}} = 0\degr$) in the terminator plane ($\theta_{0, \, \rm{cld}} = 0\degr$) and extends clockwise into the morning terminator with 60\% cloud coverage ($f_{\rm cld} = 0.6$). 

We computed a grid of transmission spectra by varying the five parameters defining the Iceberg cloud. We consider cloud-top pressures from $10^{-5}$ to $10^{-1}$\,bar, extinctions from $10^{-9}$ to $10^{-5}\, \rm{m}^{-1}$, zenith angles from $-20\degr$ to $+20\degr$ (negative angles start in the dayside, positive in the nightside), azimuthal angles from $-40\degr$ to $+40\degr$ (negative angles start in the evening terminator, positive in the morning terminator), and terminator cloud fractions from 0 to 1. We computed all model spectra at $R =$ 10,000 from 0.4--5\,$\micron$ before binning to $R = 100$. We show our grid of 3D transmission spectra with multidimensional clouds in Figure~\ref{fig:results_clouds}.

\subsubsection{Cloud-top Pressure} \label{subsubsec:cloud-top_pressure}

The cloud-top pressure of an Iceberg cloud demarks the atmospheric layers where an additional continuum opacity source is present. Low-pressure clouds elevate the impact parameter for which the slant optical depth equals unity, resulting in a higher transit depth at all wavelengths (Figure~\ref{fig:results_clouds}, top left). Iceberg clouds do not act as a hard surface, unlike the classical 1D cloud deck, since the non-uniform terminator cloud coverage allows rays with certain azimuthal angles to sample clear atmospheric regions (see Figure~\ref{fig:iceberg_cloud_model}).

\subsubsection{Cloud Extinction} \label{subsubsec:cloud_extinction}

The cloud extinction controls the overall cloud transparency. Our reference extinction, $\kappa_{\rm cld} = 10^{-5}\, \rm{m}^{-1}$, corresponds to an optically thick cloud at all wavelengths. Partially transparent clouds exist over a narrow range from $\kappa_{\rm cld} \sim 10^{-7}$--$10^{-8}\, \rm{m}^{-1}$, with the cloud presenting negligible opacity for $\kappa_{\rm cld} \leq 10^{-9}\, \rm{m}^{-1}$ (Figure~\ref{fig:results_clouds}, top right). The long slant paths in transmission geometry cause even moderate cloud extinction to rapidly behave as an opaque cloud \citep{Fortney2005}, explaining the limited range of $\kappa_{\rm cld}$ affording partial transparency. The cloud extinction strongly affects spectral regions probing deep layers in the atmosphere, such as the alkali wings and Rayleigh slope in the optical, with little impact on strong absorption bands forming mostly above the cloud-top pressure (e.g. the $3\,\micron$ H$_2$O band).

\subsubsection{Cloud Zenith Angle} \label{subsubsec:cloud_zenith}

The cloud zenith angle dependence reveals how far from the terminator plane rays are sensitive to cloud opacity. Clouds commencing in the dayside ($\theta_{0, \, \rm{cld}} < 0\degr$) are encountered by rays at an earlier stage, resulting in higher transit depths and shallower spectral features (Figure~\ref{fig:results_clouds}, middle left). We see no change in the spectrum between a cloud with $\theta_{0, \, \rm{cld}} = -10\degr$ and $\theta_{0, \, \rm{cld}} = -20\degr$. Conversely, a cloud extending $10\degr$ into the nightside ($\theta_{0, \, \rm{cld}} = 10\degr$) produces almost the same spectrum as a clear atmosphere. Therefore, our 3D atmosphere has an effective cloud horizon constrained to $\mod{\theta_{0, \, \rm{cld}}} =\lesssim 10\degr$. The sensitivity to slightly larger nightside zenith angles is due to the smaller nightside scale height (and hence a shorter slant path traversed per unit zenith angle). From Figure~\ref{fig:results_clouds}, we see that the cloud zenith angle has a similar influence on 3D transmission spectra to the cloud extinction. Therefore, we expect these two parameters to be somewhat degenerate in multidimensional retrievals.

\subsubsection{Cloud Azimuth Angle} \label{subsubsec:cloud_azimuth}

Sensitivity to the cloud azimuthal starting angle is a unique feature of multidimensional atmospheres. A 1D background atmosphere with a 2D patchy cloud \citep[e.g.][]{Line2016} is insensitive to $\phi_{0, \, \rm{cld}}$, since one can azimuthally rotate the cloud's starting location and obtain an identical transmission spectrum. However, a non-uniform background temperature or composition breaks this symmetry. Consequently, multidimensional transmission spectra are weakly sensitive to $\phi_{0, \, \rm{cld}}$ (Figure~\ref{fig:results_clouds}, middle right). We see that the visible wavelength continuum changes as the cloud rotates, due to the cloud obscuring sectors with different temperatures as $\phi_{0, \, \rm{cld}}$ varies. Further, absorption bands forming below the cloud-top are modulated for species exhibiting a morning-evening gradient. For example, the $1\,\micron$ H$_2$O band is strong when $\phi_{0, \, \rm{cld}} = 40\degr$, since $40\degr/180\degr = 22\%$ of the morning half-annulus (containing a higher H$_2$O abundance than the evening half-annulus) is unobscured by clouds. Conversely, the same H$_2$O band is weaker when $\phi_{0, \, \rm{cld}} = -40\degr$, since only $(40 - 0.1*180)\degr/180\degr = 12\%$ of the morning half-annulus is unobscured. Despite the lower sensitivity to $\phi_{0, \, \rm{cld}}$ than for the other Iceberg cloud parameters, our results demonstrate a key point: morning-evening temperature and composition gradients break azimuthal symmetry, allowing one to infer the region of the terminator annulus containing 3D clouds.

\subsubsection{Terminator Cloud Coverage} \label{subsubsec:cloud_fraction}

The terminator cloud coverage controls the fraction of terminator containing cloud opacity. Higher cloud fractions increase the transit depth, decrease the amplitudes of spectral features, and alter absorption band wing shapes (Figure~\ref{fig:results_clouds}, bottom). When $f_{\rm cld} = 1$, one recovers the limit of a uniform cloud deck which emulates a hard surface. When $f_{\rm cld} = 0$, one obtains a clear atmosphere. We also note that the $f_{\rm cld}$ and $\phi_{0, \, \rm{cld}}$ parameters interface in several ways, such as: (i) $f_{\rm cld} = 1$ means $\phi_{0, \, \rm{cld}}$ has no effect; and (ii) $f_{\rm cld} = 0.5$ yields the same spectrum for $\phi_{0, \, \rm{cld}}$ and $-\phi_{0, \, \rm{cld}}$ (due to north-south symmetry of the background atmosphere). Given the strong impact of patchy clouds on transmission spectra---and the unique way $f_{\rm cld}$ alters spectral features---evidence of patchy clouds has already emerged from \emph{Hubble} spectra of hot Jupiters \citep{MacDonald2017a,Pinhas2019,Barstow2020}. Our analysis builds on these inferences by demonstrating that multidimensional transmission spectra encode not only the terminator cloud fraction, but also the \emph{spatial} location of a 3D cloud.

\newpage

\section{Summary and Discussion} \label{sec:discussion}

Exoplanet transmission spectra probe the highly non-uniform terminator region, encoding multidimensional atmospheric properties in transmission spectra. The improvement in spectral data quality following the launch of JWST will necessitate incorporating multidimensional models inside atmospheric retrieval frameworks. Here we introduced TRIDENT, a new 3D radiative transfer model designed to meet the computational requirements for 3D atmospheric retrievals in the era of JWST. Our study offers several significant advancements to the theory and analysis of  transmission spectra:

\begin{itemize}
    \item We presented a unified theoretical framework for exoplanet transmission spectra. Our model, summarised in Equation~\ref{eq:transmission_spectrum_general}, combines a wide range of processes important for transmission spectroscopy: non-uniform 3D atmospheres, refraction, multiple scattering, stellar heterogeneities, and nightside thermal emission. The model also covers general partial transit geometries, such as ingress/egress and grazing transits.
    \item We developed a linear algebra-based approach to rapidly compute 3D transmission spectra. This new 3D radiative transfer technique is sufficiently fast to enable 3D atmospheric retrievals.
    \item We introduced TRIDENT, our new 3D radiative transfer model implementing the above methods. We validated TRIDENT against two codes in the literature, demonstrating its efficacy.
    \item In anticipation of 3D retrievals, we proposed parametric prescriptions for 3D temperature and abundance fields. Our new parametrisations describe 3D atmospheres with a minimal free parameter increase, solving the problem of model complexity balance for multidimensional retrievals.
    \item We further proposed a 3D `Iceberg' cloud model for spectral analyses of multidimensional atmospheres. We demonstrated that the spatial location of a 3D cloud can imprint unique signatures in transmission spectra, necessitating a full 3D treatment instead of previous patchy cloud models.
    \item Finally, we showed that day-night and morning-evening composition gradients imprint significant ($> 100$\,ppm) signatures in transmission spectra. Multidimensional atmospheres alter the relative band strengths and peak-to-wing shapes of absorption features, offering the promise to detect temperature and composition gradients with JWST. 
\end{itemize}

We proceed to discuss the implications of our study.

\subsection{A Unified Model for Transmission Spectra} \label{subsec:discussion_unified_equation}

Transmission spectra encode a rich array of information about transiting exoplanet atmospheres and their host star. We have introduced a general equation that unifies many important phenomena influencing transmission spectra (Equation~\ref{eq:transmission_spectrum_general}). By presenting a first-principles derivation of this general equation (Appendix~\ref{appendix_A}), we elucidated the key principles underlying theoretical models of transmission spectra. Equation~\ref{eq:transmission_spectrum_general} is a straightforward generalisation of commonly used equations for 1D transmission spectra \citep[e.g.][]{Brown2001,Tinetti2012,MacDonald2017a}, while holding for 3D atmospheres including refraction, multiple scattering, stellar heterogeneities, and nightside thermal emission. Our new equation for transmission spectra can be readily integrated into other radiative transfer codes and retrieval forward models.

Our theoretical framework opens multiple directions for future studies. By relaxing the assumption of a full transit, one can pursue many applications of \emph{partial transit spectroscopy} (see Section~\ref{subsec:discussion_TRIDENT}). The inclusion of refraction and multiple scattering allows increased realism for model transmission spectra of terrestrial planets or atmospheres with highly scattering aerosols. Additionally, one can investigate scenarios where the two contamination sources --- nightside thermal flux \citep{Kipping2010} and the transit light source effect \citep{Rackham2018} --- are simultaneously prominent (e.g. spectra of ultra-hot Jupiters transiting active stars).  

\subsection{Further Applications of TRIDENT} \label{subsec:discussion_TRIDENT}

TRIDENT can model transmission spectra for many scenarios beyond those explored in this paper. While this study has focused on low-resolution ($R \sim 100$) 3D transmission spectra of hot Jupiters --- of key importance for early JWST observations --- here we mention several ancillary applications of TRIDENT. 

\subsubsection{Ingress and Egress Spectroscopy} \label{subsubsec:discussion_ingress_egress}

Ingress and egress spectroscopy is one key application of TRIDENT. When a planet partially transits its host star, the fraction of the planet and its atmosphere occulting the stellar disc changes with time. Consequently, transmission spectra are time-dependent during partial transits. All transiting planets undergo partial transits during ingress and egress, producing distinct time-averaged spectra for the ingress and egress if the morning and evening terminators differ in their temperature, composition, or aerosol properties \citep[e.g.][]{Fortney2010,Kempton2017}. We show an ingress and egress spectrum for a typical hot Jupiter in Figure~\ref{fig:TRIDENT_applications} (top panel), computed with TRIDENT using Equations~\ref{eq:atmosphere_factor} and \ref{eq:transit_depth_intensity_form_4}. With sufficient time resolution and data precision, one could in principle extract multiple spectra as a function of time from ingress/egress observations. Fitting such time-resolved transmission spectra would open the field of \emph{transmission mapping}, analogous to secondary eclipse mapping \citep[e.g.][]{Majeau2012,deWit2012,Challener2021}. Transmission mapping is a rich area for future theoretical and observational studies.

\subsubsection{Grazing Transit Spectroscopy} \label{subsubsec:discussion_grazing}

\begin{figure*}[ht!]
    \centering
    \includegraphics[width=0.96\textwidth, trim={0.0cm 0.0cm 0.0cm 0.0cm}]{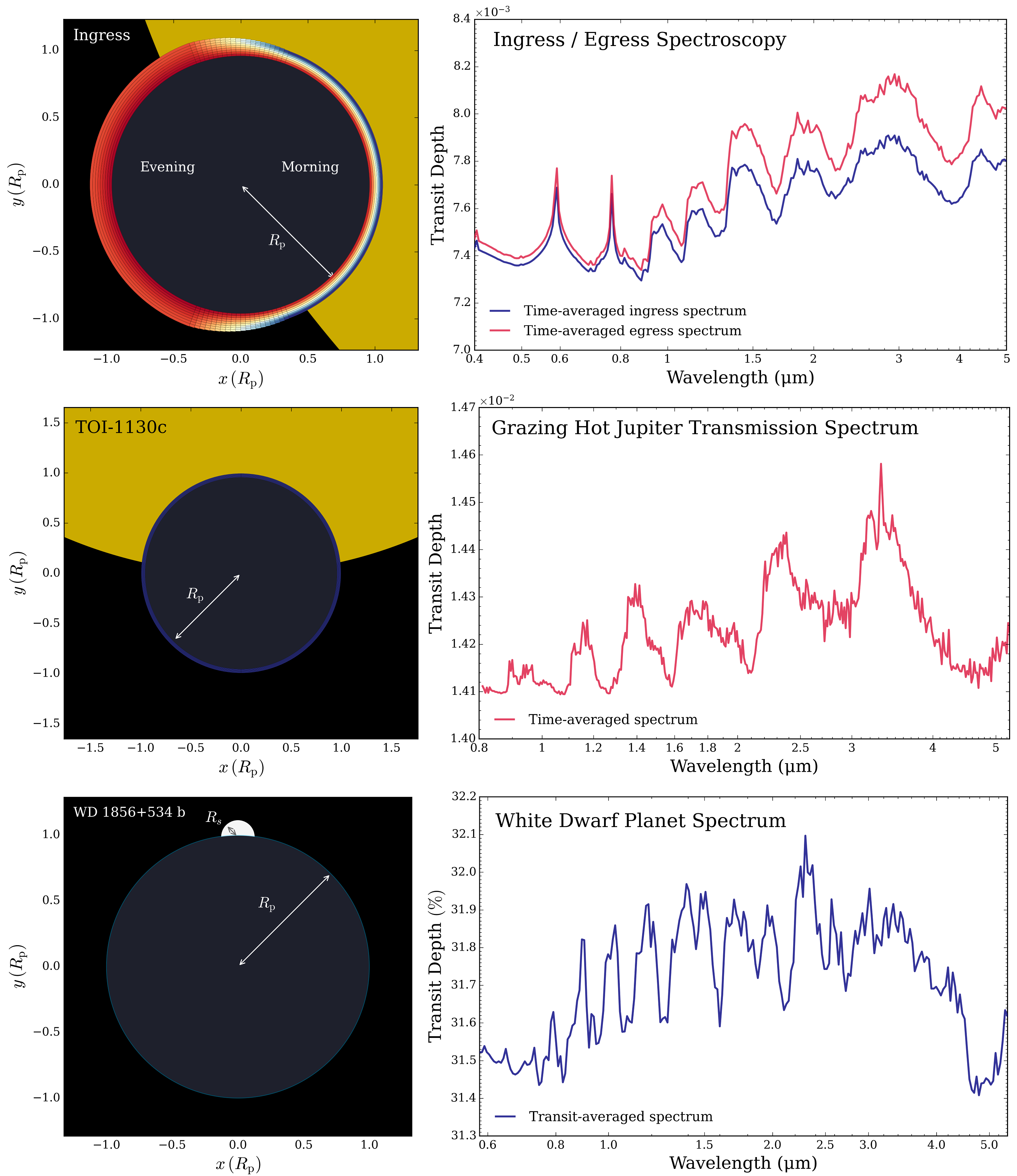}
    \caption{Applications of TRIDENT to transmission spectra with partial transit geometries. Top: measuring multidimensional atmospheric properties via ingress/egress spectroscopy. The time-averaged spectrum during ingress preferentially samples the cooler morning terminator, resulting in a lower average transit depth than the egress spectrum. Middle: transmission spectra of planets with grazing transits. We show the average spectrum of an illustrative grazing hot Jupiter, TOI-1130c \citep{Huang2020}, assuming a solar composition atmosphere. Bottom: transmission spectroscopy of planets orbiting white dwarfs. We show the average spectrum of WD~1856+534b \citep{Vanderburg2020}, assuming a Jupiter-like composition. }
\label{fig:TRIDENT_applications}
\end{figure*}

Grazing transit spectroscopy is another application of TRIDENT. A planet with a grazing transit (satisfying $b_{\mathrm{p}} R_{*} > R_{*} - R_{\mathrm{p}}$) has an ingress/egress lasting the entire transit. To date, over 20 confirmed exoplanets have grazing transits \citep{Davis2020}. We show an example grazing transit spectrum for the hot Jupiter TOI-1130c \citep{Huang2020}, assuming a solar composition atmosphere at 650\,K, in Figure~\ref{fig:TRIDENT_applications} (middle panel). For grazing transits, TRIDENT computes spectra at a series of time steps, with the time-averaged spectrum shown in Figure~\ref{fig:TRIDENT_applications}. An extreme grazing transit is WD~1856+534b \citep{Vanderburg2020} --- a planet transiting a white dwarf with $R_{\mathrm{p}}/R_{*} \approx 8$ --- for which we show a model, assuming a Jupiter-like composition, in Figure~\ref{fig:TRIDENT_applications} (bottom panel). Recently, we demonstrated the principles of TRIDENT's grazing transit modelling in an analysis of Gemini/GMOS observations of WD~1856+534b \citep{Xu2021}. With the success of the Transiting Exoplanet Survey Satellite (TESS), the number of planets with grazing transits is steadily increasing. Since TESS planets orbit nearby bright stars, TRIDENT offers a new capability to model grazing transit spectra for the interpretation of JWST observations.

\subsubsection{High-Resolution Transmission Spectroscopy} \label{subsubsec:high_res}

TRIDENT can conduct line-by-line radiative transfer, enabling high spectral resolution model applications. We demonstrated this capability by showing a line-by-line transmission spectrum of a hot Jupiter in Figure~\ref{fig:validation_transmission_spectra}. Since our cross section database (Appendix~\ref{appendix_C}) is pre-computed at $R \sim 10^6$ (see Section~\ref{subsec:opacity}), TRIDENT can therefore generate template spectra for high-resolution cross correlation analyses \citep[e.g.][]{Snellen2010,Hoeijmakers2018,Sedaghati2021}. Future improvements to TRIDENT can focus on physical effects that manifest in high-resolution transmission spectra, including planetary rotation and winds \citep[e.g.][]{Brogi2016,Flowers2019,Seidel2020}. Another promising avenue is integrating TRIDENT within a high-resolution retrieval framework \citep[e.g.][]{Brogi2019,Gibson2020,Pelletier2021}.

\subsubsection{Terrestrial Exoplanet Transmission Spectra} \label{subsubsec:terrestrial_spectra}

Transmission spectra of terrestrial exoplanets is one of the most significant advances enabled by JWST. While we have focused in this study on giant planet spectra, TRIDENT also contains a growing database of cross sections applicable for radiative transfer through temperate terrestrial exoplanet atmospheres (see Table~\ref{table:molecular_line_lists}). Our cross section database has already been used in the literature to simulate transmission spectra of an Earth-like planet transiting a white dwarf \linktocite{KalteneggerMacDonald2020}{(Kaltenegger \& MacDonald et al.} \citeyear{KalteneggerMacDonald2020}) and TRAPPIST-1e \citep{Lin2021}. Future extensions to TRIDENT can add a refractive ray tracing algorithm, which can be important to accurately simulate transmission spectra of terrestrial atmospheres \citep[e.g.][]{Betremieux2014,Misra2014}. The inclusion of refraction requires one to numerically compute the path distribution tensor \citep{Robinson2017a}, but is otherwise covered by our theoretical framework.

\subsection{Prospects for Multidimensional Retrievals of Exoplanet Transmission Spectra} \label{subsec:discussion_3D_retrievals}

The central goal of this study was to overcome two significant obstacles preventing 3D retrievals of exoplanet transmission spectra: 

\begin{enumerate}
    \item The prohibitive computational demands of existing 3D radiative transfer techniques.
    \item A lack of parametrisations for 3D atmospheres with day-night and morning-evening gradients.
\end{enumerate}

We have offered solutions to these problems. First, our new 3D radiative transfer approach (Section~\ref{subsec:radiative_transfer}) decouples the wavelength-dependent and geometric calculations in transmission spectra models (generalising the path distribution theory introduced by \citealt{Robinson2017a}). By further expressing 3D transmission spectra as a linear algebra operation (Section~\ref{subsubsec:TRIDENT_radiative_transfer}), TRIDENT can rapidly compute 3D transmission spectra (see Appendix~\ref{appendix_D}). Second, we introduced simple parametrisations for temperatures, chemical abundances, and clouds that can vary vertically, across the day-night transition, and between the morning and evening terminators. We do not claim these to be the only parametrisations applicable for 3D retrievals; our parametrisations represent a foundation from which more complex models can be developed as required by improving data quality.  

Our results demonstrate that multidimensional atmospheric properties induce significant discernible features in transmission spectra (see Section~\ref{sec:3D_signatures}). However, these multidimensional features only guarantee detectability if a 1D model cannot compensate for their signatures. \citet{Caldas2019} demonstrated this point for 2D spectra with day-night temperature gradients, which they showed a 1D retrieval can readily fit without significant residuals. However, \citet{Pluriel2020} and \citet{Pluriel2021} subsequently showed that day-night composition gradients are far more challenging for 1D retrievals. Our results offer an intuitive explanation for \emph{why} 1D models struggle to fit day-night composition gradients: the relative band strengths change for a molecule with variable abundance along the line of sight (e.g. day-night H$_2$O dissociation for ultra-hot Jupiters). Similarly, we find that morning-evening composition gradients distort the peak-to-wing shape of absorption features in transmission spectra. These prominent signatures are the key to detecting multidimensional atmospheric properties from transmission spectra.

Ultimately, the detection of 2D and 3D effects in exoplanet atmospheres will be enabled by multidimensional atmospheric retrievals. Recently, several studies have developed 2D retrieval frameworks to study the detectability of either day-night or morning-evening gradients \citep{Lacy2020a,Espinoza2021}. These studies have generally relied on the assumption of chemical equilibrium to render the parameter space tractable for retrievals. Our present study is intended to enable the next logical step --- \emph{3D atmospheric retrievals} --- while maintaining the general flexibility of `free retrieval' parametrisations. We have already integrated TRIDENT within the POSEIDON atmospheric retrieval framework \citep{MacDonald2017a} for these purposes. We will describe the nuances and applications of 2D and 3D retrievals in a forthcoming paper. Transmission spectroscopy has a bright future, with the observation of multidimensional atmospheric properties resting one small step over the horizon.

\begin{acknowledgments}
R.J.M. thanks Tyler Robinson for providing a model spectrum for intercomparison with TRIDENT. R.J.M. acknowledges conference travel support from the Royal Astronomical Society (RAS). R.J.M. and N.K.L acknowledge support from NASA Grant 80NSSC20K0586 issued through the James Webb Space Telescope Guaranteed Time Observer Program. R.J.M. thanks Jake Turner, Jayesh Goyal, Jonathan Gomez Barrientos, Ishan Mishra, and Aurélien Falco for helpful discussions. We thank the anonymous referee for their suggestions.
\end{acknowledgments}

\software{Numpy \citep{Harris2020}, Matplotlib \citep{Hunter2007}, Numba \citep{Lam2015}, SciPy \citep{Virtanen2020}, Astropy \citep{AstropyCollaboration2018}, pandas \citep{Mckinney2010}, SpectRes \citep{Carnall2017}, CMasher \citep{vanderVelden2020}, Mayavi \citep{Ramachandran2011}.}

\appendix

\section{Derivation of a General Equation for Exoplanet Transmission Spectra} \label{appendix_A}

Here we derive the general equation for transmission spectra presented in Equation~\ref{eq:transmission_spectrum_general}. We seek an equation valid for 3D exoplanet atmospheres, during both full and partial transits (ingress/egress or grazing transits) of a star with unocculted stellar heterogeneities, and including the effects of refraction, multiple scattering, and nightside thermal emission. We consider in this derivation the transiting exoplanet geometry shown in Figure~\ref{fig:schematic_diagram}.

A transmission spectrum is defined as the fractional spectral flux difference between a time outside of transit and a time during transit
\begin{equation}
    \Delta_{\lambda} \equiv \frac{F_{\lambda, \, \rm{out}} - F_{\lambda, \, \rm{in}}}{F_{\lambda, \, \rm{out}}} 
\label{eq:transit_depth_definition}
\end{equation}
The definition of flux is
\begin{equation}
    F_{\lambda} = \int_{\Omega} I_{\lambda} \; \hat{\bm{n}} \cdot \hat{\bm{k}} \,  d\Omega
\label{eq:flux_definition}
\end{equation}
where $I_{\lambda}$ is the spectral intensity, $\hat{\bm{n}}$ is a unit vector in the direction of beam propagation, $\hat{\bm{k}}$ is a unit vector in the direction of the observer, and $\Omega$ is the solid angle subtended by the source at the observer. In transit geometry, stellar rays satisfy $\hat{\bm{n}} \cdot \hat{\bm{k}} = 1$. The flux outside transit arises from the stellar flux and the nightside planetary flux
\begin{equation}
    F_{\lambda, \, \rm{out}} = F_{\lambda, \, \rm{* \, (out)}} + F_{\lambda, \, \rm{p \, (night)}}
\label{eq:flux_out}
\end{equation}
whilst the flux inside transit\footnote{$F_{\lambda, \, \rm{in}}$ need not only refer to times when the planet fully transits its host star. During \emph{partial transits} (e.g. ingress/egress), $F_{\lambda, \, \rm{in}}$ continually changes due to the changing fraction of the planet occulting its host star.} also contains a contribution from light transmitted through the planetary atmosphere
\begin{equation}
    F_{\lambda, \, \rm{in}} = F_{\lambda, \, \rm{* \, (in)}} + F_{\lambda, \, \rm{* \, (trans)}} + F_{\lambda, \, \rm{p \, (night)}}
\label{eq:flux_in}
\end{equation}
Substituting the in-transit and out-of-transit fluxes into Equation~\ref{eq:transit_depth_definition}, we have
\begin{equation}
    \Delta_{\lambda} = \frac{F_{\lambda, \, \rm{* \, (out)}} - F_{\lambda, \, \rm{* \, (in)}} - F_{\lambda, \, \rm{* \, (trans)}}}{F_{\lambda, \, \rm{* \, (out)}} + F_{\lambda, \, \rm{p \, (night)}}} = \frac{F_{\lambda, \, \rm{* \, (out)}} - F_{\lambda, \, \rm{* \, (in)}} - F_{\lambda, \, \rm{* \, (trans)}}}{F_{\lambda, \, \rm{* \, (out)}}} \left(\frac{1}{1 + \frac{F_{\lambda, \, \rm{p \, (night)}}}{F_{\lambda, \, \rm{* \, (out)}}} } \right)
\label{eq:transit_depth_nightside_factored}
\end{equation}
where in the last equality we factored out the nightside planetary flux. The factor in parentheses can thus be considered a multiplicative `contamination factor' that differs from unity only when the nightside flux is non-negligible \citep{Kipping2010}. 

To unpack the pre-factor in Equation~\ref{eq:transit_depth_nightside_factored}, we can write the stellar fluxes inside and outside transit as
\begin{equation}
    F_{\lambda, \, \rm{* \, (out)}} = \int_{\Omega_{\rm{* \, (full)}}} I_{\lambda, \, *} \,  d\Omega
\label{eq:stellar_flux_out}
\end{equation}
\begin{equation}
    F_{\lambda, \, \rm{* \, (in)}} = \int_{\Omega_{\rm{* \, (unobscured)}}} I_{\lambda, \, *} \,  d\Omega
\label{eq:stellar_flux_in}
\end{equation}
\begin{equation}
    F_{\lambda, \, \rm{* \, (trans)}} = \int_{\Omega_{\rm{p}}} I_{\lambda, \, \rm{* \, (trans)}} \,  d\Omega
\label{eq:stellar_flux_trans}
\end{equation}
The out-of-transit stellar flux is simply the integrated stellar intensity over the full stellar disc. In Equation~\ref{eq:stellar_flux_in}, we define the stellar flux inside transit as the stellar flux reaching the observer \emph{without} interacting with the planetary atmosphere (i.e. the flux from the portion of the stellar disc unobscured by the planet and its surrounding atmosphere). Finally, Equation~\ref{eq:stellar_flux_trans} contains the transmitted stellar intensity, $I_{\lambda, \, \rm{* \, (trans)}}$, through the planetary atmosphere. Given these definitions, the difference between the stellar fluxes outside and inside transit can be expressed as
\begin{equation}
    F_{\lambda, \, \rm{* \, (out)}} - F_{\lambda, \, \rm{* \, (in)}} = \int_{\Omega_{\rm{* \, (obscured)}}} I_{\lambda, \, *} \,  d\Omega = \int_{\Omega_{\rm{p \, (overlap)}}} I_{\lambda, \, *} \,  d\Omega
\label{eq:stellar_flux_difference}
\end{equation}
where the last equality denotes that the solid angle subtended by the obscured portion of the stellar disc is identical to the solid angle subtended by the portion of the planet disc overlapping the star. We stress that the integration region here only covers the full planet disc, $\Omega_{\rm{p}}$, for the case of full transits ($\Omega_{\rm{p \, (overlap)}} < \Omega_{\rm{p}}$ during ingress/egress or for the entire transit for planets exhibiting grazing transits). Next, we introduce the mean stellar intensity
\begin{equation}
    \overline{I}_{\lambda, \, \rm{*}} = \frac{\displaystyle\int_{\Omega_{\rm{* \, (full)}}} I_{\lambda, \, *} \,  d\Omega}{\Omega_{\rm{* \, (full)}}}
\label{eq:mean_intensity}
\end{equation}
allowing the out of transit stellar flux to be written as
\begin{equation}
    F_{\lambda, \, \rm{* \, (out)}} = F_{\lambda, \, *} = \overline{I}_{\lambda, \, \rm{*}} \, \Omega_{\rm{* \, (full)}}
\label{eq:stellar_flux_out_2}
\end{equation}
where in the first equality we dropped the outside transit distinction, since $F_{\lambda, \, \rm{* \, (out)}}$ is what is usually considered as the `stellar flux', $F_{\lambda, \, *}$. Substituting Equations~\ref{eq:stellar_flux_trans}, \ref{eq:stellar_flux_difference}, and \ref{eq:stellar_flux_out_2} into Equation~\ref{eq:transit_depth_nightside_factored}, we have
\begin{equation}
    \Delta_{\lambda} = \frac{ \displaystyle\int_{\Omega_{\rm{p \, (overlap)}}} \left( \frac{I_{\lambda, \, *} }{\overline{I}_{\lambda, \, \rm{*}}} \right) \, d\Omega - \displaystyle\int_{\Omega_{\rm{p}}} \left( \frac{I_{\lambda, \, \rm{* \, (trans)}}}{\overline{I}_{\lambda, \, \rm{*}}} \right) \, d\Omega }{\Omega_{\rm{* \, (full)}}}
    \left( \frac{1}{1 + \frac{F_{\lambda, \, \rm{p \, (night)}}}{F_{\lambda, \, *}} } \right)
\label{eq:transit_depth_intensity_form_1}
\end{equation}
In transit geometry, a distant observer sees plane-parallel stellar rays. The subtended solid angle of the star is therefore $\Omega_{\rm{* \, (full)}} = A_{*} / D^2$ (where $A_{*} = \pi \, R_{*}^2$ and $D$ is the distance to the system). Similarly, $d\Omega = dA / D^2$. We can thus replace the solid angle integrals with integrals over the projected area of the planetary disc
\begin{equation}
    \Delta_{\lambda} = \frac{ \displaystyle\int_{A_{\rm{p \, (overlap)}}} \left( \frac{I_{\lambda, \, *} }{\overline{I}_{\lambda, \, \rm{*}}} \right) \, dA - \displaystyle\int_{A_{\rm{p}}} \left( \frac{I_{\lambda, \, \rm{* \, (trans)}}}{\overline{I}_{\lambda, \, \rm{*}}} \right) \, dA }{\pi \, R_{*}^2}
    \left( \frac{1}{1 + \frac{F_{\lambda, \, \rm{p \, (night)}}}{F_{\lambda, \, *}} } \right)
\label{eq:transit_depth_intensity_form_2}
\end{equation}
An important subtlety here is that the second integral covers the full area of the planetary disc, whilst the first integral covers only the portion of the planet overlapping the host star. If rays travel on straight line paths this distinction does not matter, since $I_{\lambda, \, \rm{* \, (trans)}} = 0$ for any area elements off the stellar disc. However, in general, processes like refraction and scattering can deflect rays as they traverse the planetary atmosphere. Therefore, lines of sight without the stellar disc in the background may still intersect the star ($I_{\lambda, \, \rm{* \, (trans)}} \neq 0$) due to refraction or scattering. An example of this effect can be seen in \citet{Robinson2017a} (their Figure 4), where they show a halo of scattered light during ingress.

Model spectroscopic light curves can be made by including a stellar limb darkening law in Equation~\ref{eq:transit_depth_intensity_form_2}, such that $I_{\lambda, \, *} = I_{\lambda, \, *} (\mu_*)$ (where $\mu_* \equiv \hat{\bm{n}}_* \cdot \hat{\bm{k}}$ is the cosine of the angle between the observer's direction and the local normal on the stellar surface where the ray originates). However, the usual procedure in processing transit data is to first fit an analytic transit model, including limb darkening, to extract a spectrum of $\frac{R_{\rm{p, eff}}(\lambda)}{R_*}$ \citep[e.g.][]{Mandel2002,Kreidberg2015}. If the star had a uniform intensity distribution over the transit chord, its observed transmission spectrum would be $\Delta_{\lambda, \, \rm{obs}} = \left(\frac{R_{\rm{p, eff}}(\lambda)}{R_*}\right)^2$. These transit depths are quoted in the literature and can thus be considered `limb darkening corrected'. Therefore, to compare a model transit depth, $\Delta_{\lambda, \, \rm{model}}$, to an `observed' transit depth, $\Delta_{\lambda, \, \rm{obs}}$, one can use a uniform stellar intensity\footnote{Only the transit chord intensity need be uniform. We will see later how unocculted stellar heterogeneities enter the equation.} to compute model transmission spectra. We hence assume, without loss of generality, uniform stellar intensities, allowing $I_{\lambda, \, *}$ to be factored out of the integrals in Equation~\ref{eq:transit_depth_intensity_form_2}.

The transmitted stellar intensity, $I_{\lambda, \, \rm{* \, (trans)}}$, encodes how the planetary atmosphere interacts with starlight. This intensity can be considered in two ways: (i) as an inherent property of a stellar ray \citep[e.g.][]{Seager2010} or (ii) an ensemble average over stellar photons \citep[e.g.][]{Robinson2017a}. The ray treatment is generally appropriate when a beam of light incident on the atmosphere at a given location always emerges at a deterministic exit location (e.g. models including refraction). The photon treatment is more apt when a packet of photons with the same incident location can have many exit locations (e.g. models including multiple scattering, causing random photon deflection). A statistical average over many photons should reproduce the ray treatment, but with the downside of being far more computationally intensive. We consider both the ray and photon treatments in turn.

In the ray treatment, the transmitted stellar intensity can be written as
\begin{equation}
    I_{\lambda, \, \rm{* \, (trans)}} = \delta_{\rm{ray}*} \, \mathcal{T}_{\lambda} \, I_{\lambda, \, \rm{*}}
\label{eq:intensity_ray}
\end{equation}
where the first factor
\begin{equation}
    \delta_{\rm{ray}*} = 
    \begin{cases}
        1 \ , &\text{if ray intersects star}\\
        0 \ , &\text{else}
    \end{cases}
\label{eq:delta_ray_appendix}  
\end{equation}
expresses that fact that only rays that trace back to the stellar disc have non-zero intensity. For the geometry in Figure~\ref{fig:schematic_diagram} and coordinate system in Figure~\ref{fig:coordinate_system}, one can analytically show in the geometric limit that
\begin{equation}
    \delta_{\rm{ray}*} (b, \phi) = 
    \begin{cases}
        1 \ , &\text{if} \ \ d^2 + b^2 + 2 \, b \, R_* \left(\sqrt{d^2/R_{*}^2 - b_{\rm{p}}^2 \sin\phi} +  b_{\rm{p}} \cos\phi \right) \leq R_{*}^2 \\
        0 \ , &\text{else}
    \end{cases}
\label{eq:delta_ray_geometric}  
\end{equation}
where $d$ is the projected distance between the centres of the star and planet and $b_{\rm{p}}$ is the planet's transit impact parameter. This equation states that only ray impact parameters and azimuthal angles with the stellar disc in the background (e.g. only illuminated atmospheric regions during ingress/egress) contribute to the transmission spectrum. For more general models including refraction, one cannot use Equation~\ref{eq:delta_ray_geometric} and must instead numerically compute $\delta_{\rm{ray}*} (b, \phi)$ via a ray tracing prescription. The second factor 
\begin{equation}
    \mathcal{T}_{\lambda} = e^{-\tau_{\lambda, \, \rm{path}}}
\label{eq:transmission_appendix}
\end{equation}
encodes the fraction of starlight in a ray transmitted through the planetary atmosphere. This functional form arises from solving the equation of radiative transfer for a ray with negligible forward scattering or scattering into the beam (i.e. all scattering is equivalent to extinction). The transmission is thus controlled by the path optical depth
\begin{equation}
    \tau_{\lambda, \, \rm{path}} = \displaystyle\int_{0}^{\infty} \kappa_{\lambda} (s) \, ds
\label{eq:path_optical_depth_appendix}
\end{equation}
where the path, $s$, need not be straight if refraction is considered.

In the photon treatment, the transmitted stellar intensity can be similarly expressed as
\begin{equation}
    I_{\lambda, \, \rm{* \, (trans)}} = \sum_{m=1}^{N_{\rm{phot}}} \delta_{\rm{ray}*, \, m} \, \mathcal{T}_{\lambda, \, m} \, I_{\lambda, \, \rm{*}, \, m} = \frac{1}{N_{\rm{phot}}} \left(\sum_{m=1}^{N_{\rm{phot}}} \delta_{\rm{ray}*, \, m} \, \mathcal{T}_{\lambda, \, m}\right) \, I_{\lambda, \, \rm{*}} = \overline{\delta_{\rm{ray}*} \, \mathcal{T}_{\lambda}} \, I_{\lambda, \, \rm{*}}
\label{eq:intensity_photon}
\end{equation}
where in the first equality we use the fact that each photon has the same energy at a given wavelength, such that the incident intensity is $I_{\lambda, \, \rm{*}} = N_{\rm{phot}} \, I_{\lambda, \, \rm{*}, \, m}$. We stress that Equation~\ref{eq:intensity_photon} does not state that the energy of an individual photon changes due to transmission through an atmosphere (as is the case for an attenuated ray). Rather, $\overline{\delta_{\rm{ray}*} \, \mathcal{T}_{\lambda}}$ can be interpreted as the probability that a packet of photons, injected into the atmosphere at a given location, are transmitted through the atmosphere \emph{and} trace back to the stellar disc. So Equation~\ref{eq:intensity_photon} accounts for the loss of photons due to both scattering (via $\delta_{\rm{ray}*, \, m}$) and absorption (via $\mathcal{T}_{\lambda, \, m}$). Since the transmission probability of a photon is locally controlled by absorption processes, the photon transmission is therefore
\begin{equation}
    \mathcal{T}_{\lambda, \, m} = e^{-\tau_{\lambda, \, \rm{path}, \, m}}
\label{eq:transmission_photon}
\end{equation}
where
\begin{equation}
    \tau_{\lambda, \, \rm{path}, \, m} = \displaystyle\int_{0}^{\infty} \tilde{\alpha}_{\lambda} (s_m) \, ds_m
\label{eq:path_optical_depth_photon}
\end{equation}
and $\tilde{\alpha}_{\lambda}$ is the absorption coefficient. For full multiple scattering calculations, the scattering coefficient does not enter Equation~\ref{eq:path_optical_depth_photon} directly (unlike for the ray treatment), as scattering only determines the path of each photon through the atmosphere, $s_m$, and its exit location (hence $\delta_{\rm{ray}*, \, m}$). The actual computation of the path of each photon is typically handled by a Monte Carlo tracing procedure \citep[see][for an excellent discussion]{Robinson2017a}. Since Equation~\ref{eq:intensity_photon} is the generalisation of Equation~\ref{eq:intensity_ray}, capable of including the physics of multiple scattering and refraction, we use the photon functional form for the transmitted stellar intensity in what follows.

We can now simplify our expression for transmission spectra. By substituting Equation~\ref{eq:intensity_photon} into Equation~\ref{eq:transit_depth_intensity_form_2}, then factoring out the constant stellar intensity, we have
\begin{equation}
    \Delta_{\lambda} = \frac{ \displaystyle\left( \frac{I_{\lambda, \, *} }{\overline{I}_{\lambda, \, \rm{*}}} \right) \int_{A_{\rm{p \, (overlap)}}} \, dA - \left( \frac{I_{\lambda, \, *} }{\overline{I}_{\lambda, \, \rm{*}}} \right) \int_{A_{\rm{p}}} \overline{\delta_{\rm{ray}*} \, \mathcal{T}_{\lambda}} \, dA }{\pi \, R_{*}^2}
    \left( \frac{1}{1 + \frac{F_{\lambda, \, \rm{p \, (night)}}}{F_{\lambda, \, *}} } \right)
\label{eq:transit_depth_intensity_form_3}
\end{equation}
Hence
\begin{equation}
    \Delta_{\lambda} = \frac{ A_{\rm{p \, (overlap)}} - \displaystyle\int_{A_{\rm{p}}} \overline{\delta_{\rm{ray}*} \, \mathcal{T}_{\lambda}} \, dA }{\pi \, R_{*}^2} \left( \frac{I_{\lambda, \, *} }{\overline{I}_{\lambda, \, \rm{*}}} \right)
    \left( \frac{1}{1 + \frac{F_{\lambda, \, \rm{p \, (night)}}}{F_{\lambda, \, *}} } \right)
\label{eq:transit_depth_intensity_form_4}
\end{equation}
The first term is simply the overlapping area between the planetary and stellar discs, given by
\begin{equation}
    A_{\rm{p \, (overlap)}} = R_{\rm{p, \, top}}^2 \tilde{\theta} + R_{*}^2 \tilde{\phi} - \frac{1}{2} R_{\rm{p, \, top}}^2 \sin(2\tilde{\theta}) - \frac{1}{2} R_{*}^2 \sin(2\tilde{\phi}) 
\label{eq:A_p_overlap}
\end{equation}
where
\begin{equation}
    \tilde{\theta} = \cos^{-1} \left(\frac{d^2 + R_{\rm{p, \, top}}^2 - R_{*}^2}{2 \, d \, R_{\rm{p, \, top}}}\right)
\label{eq:theta_tilde}
\end{equation}
\begin{equation}
    \tilde{\phi} = \cos^{-1} \left(\frac{d^2 + R_{*}^2 - R_{\rm{p, \, top}}^2}{2 \, d \, R_{*}}\right)
\label{eq:phi_tilde}
\end{equation}
Equation~\ref{eq:A_p_overlap} covers the general case of partial transits (ingress/egress or grazing transits). For the usually considered case of full transits, Equation~\ref{eq:A_p_overlap} reduces to $A_{\mathrm{p, \, (overlap)}} = \pi R_{\mathrm{p, \, top}}^2$.

Consider now the stellar intensity factor, $I_{\lambda, \, *} / \overline{I}_{\lambda, \, \rm{*}}$, in Equation~\ref{eq:transit_depth_intensity_form_4}. The intensity illuminating the planetary atmosphere will not, in general, be the same as the disc-averaged intensity. As shown in Figure~\ref{fig:schematic_diagram}, stellar heterogeneities (spots/faculae) may lie outside the transit chord. These active regions, with their own intrinsic intensities, will perturb the disc-averaged intensity from $I_{\lambda, \, *}$ and hence `contaminate' the transmission spectrum ($I_{\lambda, \, *} / \overline{I}_{\lambda, \, \rm{*}} \neq 1$). For a star with unocculted regions, each covering an area $A_{\rm{het}, \, i}$ with intensity $I_{\lambda, \, \rm{het}, \, i}$, Equation~\ref{eq:mean_intensity} yields
\begin{equation}
    \overline{I}_{\lambda, \, \rm{*}} = \frac{\displaystyle\sum_{i=1}^{N_{\rm{het}}} I_{\lambda, \, \rm{het}, \, i} \, A_{\rm{het}, \, i} + I_{\lambda, \, *} \left(A_* - \sum_{i=1}^{N_{\rm{het}}} A_{\rm{het}, \, i} \right)}{A_*} = \sum_{i=1}^{N_{\rm{het}}} f_{\rm{het}, \, i} \, I_{\lambda, \, \rm{het}, \, i} + I_{\lambda, \, *} \left(1 - \sum_{i=1}^{N_{\rm{het}}} f_{\rm{het}, \, i} \right)
\label{eq:mean_intensity_het}
\end{equation}
where in the last equality we defined the heterogeneity coverage fractions by $f_{\rm{het}, \, i} = A_{\rm{het}, \, i} / A_*$. Therefore, we have
\begin{equation}
    \left( \frac{I_{\lambda, \, *} }{\overline{I}_{\lambda, \, \rm{*}}} \right) = \frac{1}{1 - \displaystyle\sum_{i=1}^{N_{\rm{het}}} f_{\rm{het}, \, i} \left(1 - \frac{I_{\lambda, \, \rm{het}, \, i}}{I_{\lambda, \, *}} \right)}
\label{eq:contamination_het}
\end{equation}
and hence the stellar intensity factor encodes the well-known `transit light source effect' \citep[e.g.][]{Rackham2018}.

Finally, substituting Equation~\ref{eq:contamination_het} into Equation~\ref{eq:transit_depth_intensity_form_4}, we arrive at
\begin{equation}
    \Delta_{\lambda} = \left( \frac{ A_{\rm{p \, (overlap)}} - \displaystyle\int_{A_{\rm{p}}} \overline{\delta_{\rm{ray}*} \, \mathcal{T}_{\lambda}} \, dA }{\pi \, R_{*}^2} \right) \left( \frac{1}{1 - \displaystyle\sum_{i=1}^{N_{\rm{het}}} f_{\rm{het}, \, i} \left(1 - \frac{I_{\lambda, \, \rm{het}, \, i}}{I_{\lambda, \, *}} \right)} \right)
    \left( \frac{1}{1 + \displaystyle\frac{F_{\lambda, \, \rm{p \, (night)}}}{F_{\lambda, \, *}} } \right)
\label{eq:unified_equation_transmission_spectra}
\end{equation}
which completes the derivation of Equation~\ref{eq:transmission_spectrum_general}, our unified equation for transmission spectra.

\newpage

\section{Elements of the Path Distribution Tensor} \label{appendix_B}

We offer here a derivation of the elements of the path distribution tensor in the geometric limit. We first reproduce the 1D case introduced by \citet{Robinson2017a}, before presenting our derivation of the elements of the path distribution tensor for a 3D atmosphere.

\subsection{1D Atmosphere} \label{appendix_B_1D}

\begin{figure*}[htb!]
    \centering
    \includegraphics[width=\textwidth]{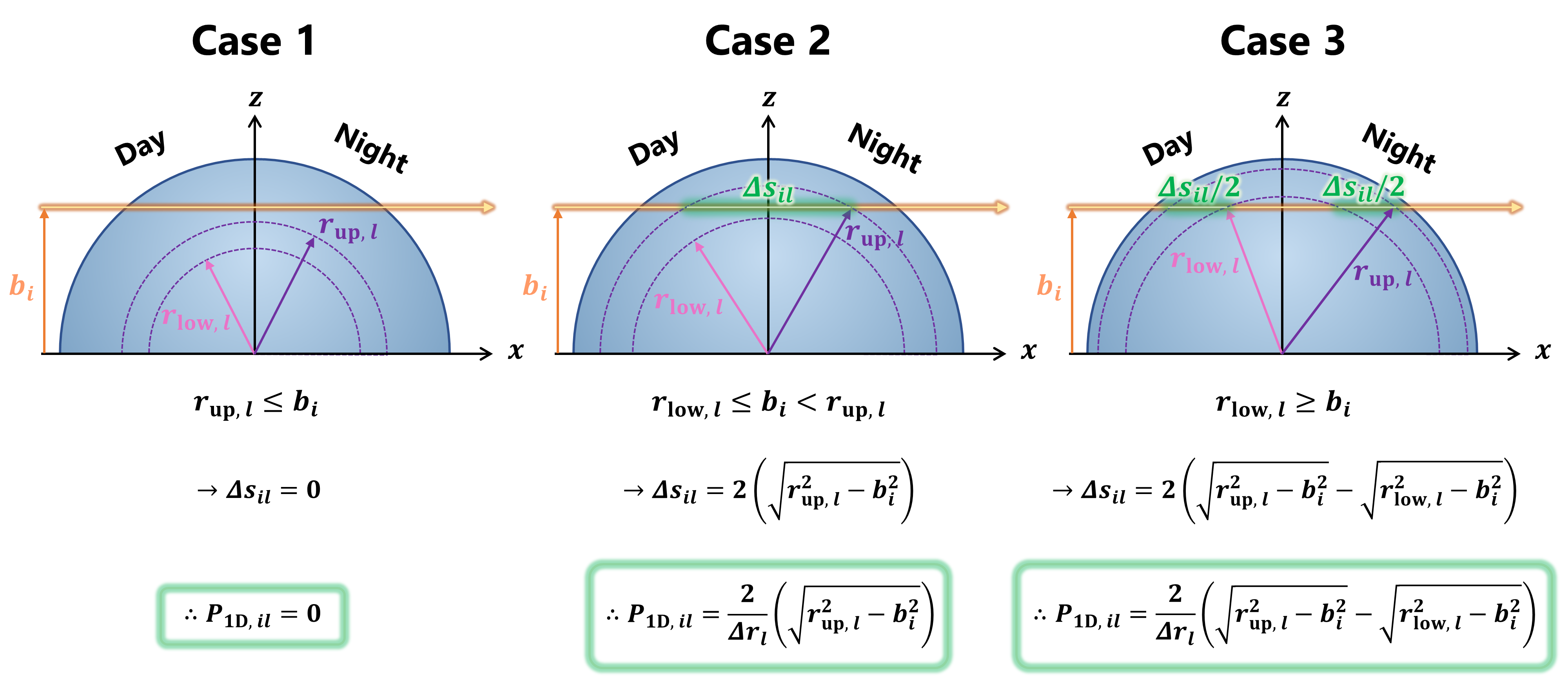}
    \caption{Derivation of the elements of the 1D path distribution matrix. For a given layer, $l$, and impact parameter, $b_i$, there are three ways the ray can intersect (or miss) the layer. The 1D path distribution encodes the slant distance travelled by a ray ($\Delta s_{il}$) per radial increment ($\Delta r_{l}$).}
\label{fig:path_distribution_derivation_1D}
\end{figure*}

For a 1D atmosphere, the path distribution tensor is a matrix, $\boldsymbol{\mathcal{P}}_{\rm{1D}}$, that encodes the slant distance travelled by a ray in a given layer per radial layer vertical extent. We show a visual derivation of the 1D path distribution elements (Equations~\ref{eq:1D_path_distribution_elements} and \ref{eq:1D_path_distribution_conditions}) in Figure~\ref{fig:path_distribution_derivation_1D}.

\subsection{3D Atmosphere} \label{appendix_B_3D}

\begin{figure*}[ht!]
    \centering
    \includegraphics[width=0.945\textwidth]{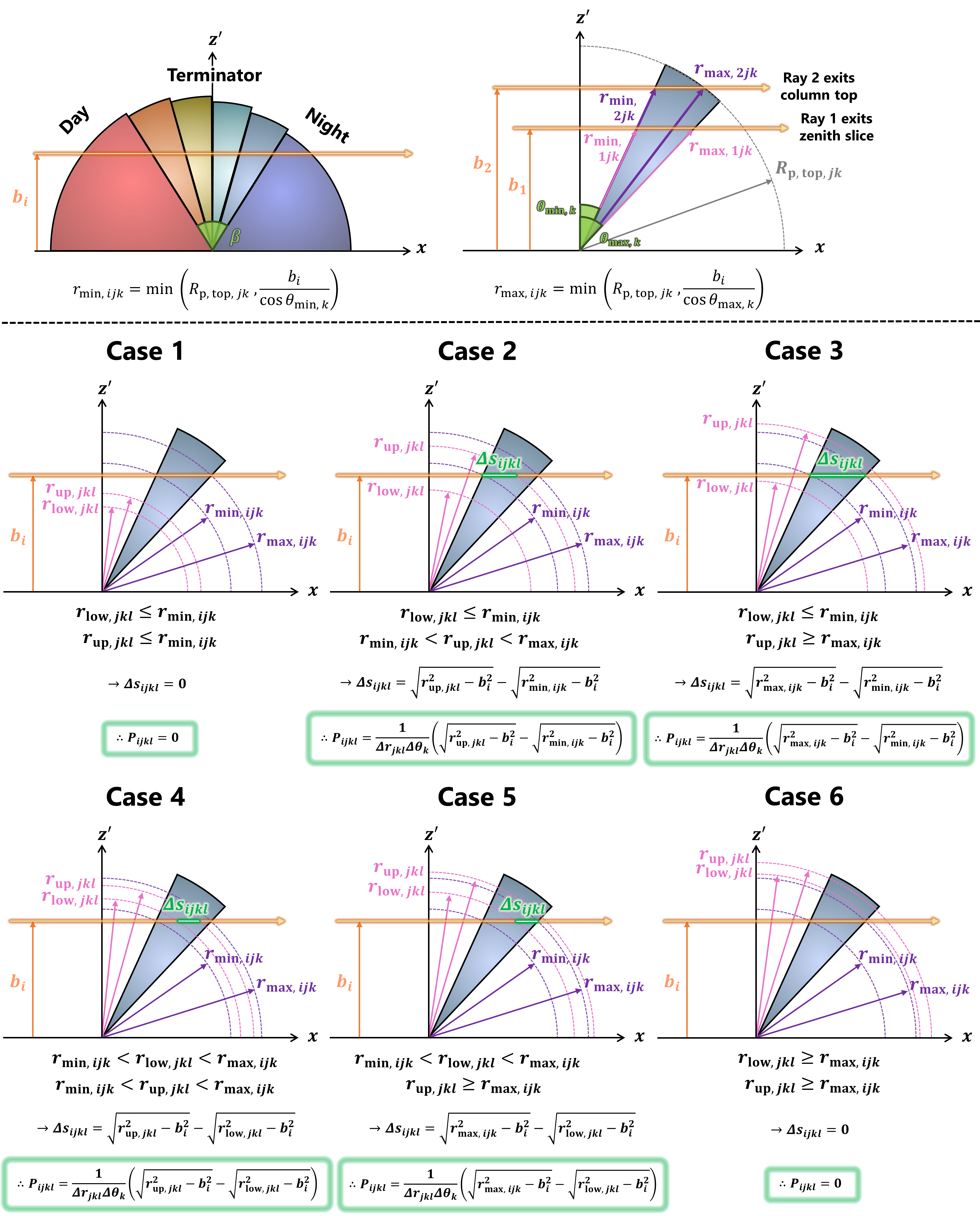}
    \caption{Derivation of the elements of the 3D path distribution tensor. For a given layer, $l$, impact parameter, $b_i$, axial sector, $j$, and zenith slice, $k$, there are six ways the ray can intersect (or miss) the layer. The 3D path distribution encodes the slant distance travelled by a ray ($\Delta s_{ijkl}$) per zenith increment ($\Delta \theta_k$) per radial increment ($\Delta r_{jkl}$).}
\label{fig:path_distribution_derivation_3D}
\end{figure*}

For a 3D atmosphere, the path distribution tensor gains two ranks to cover inhomogeneities for discrete axial sectors and zenith slices. For a given sector, $j$, one can rotate the $z$ axis by $\phi_j$ to define a new vertical axis $z^{'}$. The geometry then reduces to a 2D problem of computing the distances travelled by a ray through circular sectors defined by starting and ending zenith angles, $\theta_{\rm{min}, \, k}$ and $\theta_{\rm{max}, \, k}$. Figure~\ref{fig:path_distribution_derivation_3D} (top) demonstrates the geometry of the problem for a ray passing through the day-night transition region. This defines the maximum and minimum radii, $r_{\rm{max}, \, ijk}$ and $r_{\rm{min}, \, ijk}$, encountered by a ray with impact parameter $b_i$ in atmospheric column $jk$. We derive the elements of the 3D path distribution tensor (Equations~\ref{eq:3D_path_distribution_elements} and \ref{eq:3D_path_distribution_conditions}) in Figure~\ref{fig:path_distribution_derivation_3D} (bottom).

\newpage

\section{Opacity Sources} \label{appendix_C}

Here we summarise the molecular, atomic, ionic, and continuum opacities in the current version of the POSEIDON \citep{MacDonald2017a} opacity database (shared with TRIDENT). Table~\ref{table:molecular_line_lists} lists our molecular line list sources, Table~\ref{table:atomic_line_lists} lists our atomic and ionic line list sources, and Table~\ref{table:continuum_opacity} lists our continuum opacity sources.

\begin{table}[h!]
    \ra{0.78}
    \caption{Molecular line lists in the POSEIDON opacity database}
    \begin{tabular*}{\columnwidth}{l@{\extracolsep{\fill}} cccl@{}} \toprule 
    Molecule & Line list / data source & Reference & Broadening \\ \midrule
    H$_2$O & POKAZATEL & \citet{Polyansky2018} & H$_2$ + He \\
    CH$_4$ & 34to10 & \citet{Yurchenko2017} & H$_2$ + He \\
    NH$_3$ & CoYuTe & \citet{Coles2019} & H$_2$ + He \\
    HCN & Harris & \citet{Barber2014} & H$_2$ + He \\
    CO & Li15 & \citet{Li2015} & H$_2$ + He \\
    CO$_2$ & CDSD-4000 & \citet{Tashkun2011} & H$_2$ + He \\
    C$_2$H$_2$ & ASD-1000 & \citet{Lyulin2017} & H$_2$ + He \\
    PH$_3$ & SAlTY & \citet{Sousa-Silva2015} & H$_2$ + He \\
    SO$_2$ & ExoAmes & \citet{Underwood2016} & H$_2$ + He \\
    H$_2$S & AYT2 & \citet{Azzam2016} & Air \\
    SH & SNaSH & \citet{Yurchenko2018d} & Air \\
    OH & Brooke16 & \citet{Brooke2016} & Air \\
    NO & NOname & \citet{Wong2017} & Air \\
    N$_2$O & NOSD-1000 & \citet{Tashkun2016} & Air \\
    NO$_2$ & NDSD-1000 & \citet{Lukashevskaya2016} & Air \\
    TiO & ToTo & \citet{McKemmish2019} & SB07 \\
    VO & VOMYT & \citet{McKemmish2016} & SB07 \\
    AlO & ATP & \citet{Patrascu2015} & SB07 \\
    SiO & EBJT & \citet{Barton2013} & SB07 \\
    CaO & VBATHY & \citet{Yurchenko2016} & SB07 \\
    TiH & Burrows05 & \citet{Burrows2005} & SB07 \\
    CrH & Burrows02 & \citet{Burrows2002} & SB07 \\
    FeH & Wende & \citet{Wende2010} & SB07 \\
    ScH & LYT& \citet{Lodi2015} & SB07 \\
    AlH & AlHambra & \citet{Yurchenko2018b} & SB07 \\
    SiH & SiGHTLY & \citet{Yurchenko2018c} & SB07 \\
    BeH & Darby-Lewis & \citet{Darby-Lewis2018} & SB07 \\
    CaH & Yadin-CaH & \citet{Yadin2012} & SB07 \\
    `` '' & Li12 & \citet{Li2012} & SB07 \\
    MgH & Yadin-MgH & \citet{Yadin2012} & SB07 \\
    `` '' & Gharib-Nezhad & \citet{Gharib-Nezhad2013} & SB07 \\
    LiH & CLT & \citet{Coppola2011} & SB07 \\
    NaH & Rivlin & \citet{Rivlin2015} & SB07 \\
    CH & Masseron & \citet{Masseron2014} & SB07 \\
    NH & Brooke14 & \citet{Brooke2014} & SB07 \\
    PN & YYLT & \citet{Yorke2014} & SB07 \\
    PO & POPS & \citet{Prajapat2017} & SB07 \\
    PS & POPS & \citet{Prajapat2017} & SB07 \\
    H$_{3}^{+}$ & MiZATeP & \citet{Mizus2017} & ExoMol default \\
    \midrule
    N$_2$ (low-T) & HITRAN2016-N$_2$ & \citet{Gordon2017} & Air \\
    O$_2$ (low-T) & HITRAN2016-O$_2$ & \citet{Gordon2017} & Air \\
    O$_3$ (low-T) & HITRAN2016-O$_3$ & \citet{Gordon2017} & Air \\
    `` '' & Serdyuchenko14 & \citet{Serdyuchenko2014} & N/A \\
    H$_2$O (low-T) & HITRAN2016-H$_2$O & \citet{Gordon2017} & Air \\
    CH$_4$ (low-T) & HITRAN2016-CH$_4$ & \citet{Gordon2017} & Air \\
    CO (low-T) & HITRAN2016-CO & \citet{Gordon2017} & Air \\
    CO$_2$ (low-T) & HITRAN2016-CO$_2$ & \citet{Gordon2017} & Air \\
    N$_2$O (low-T) & HITRAN2016-N$_2$O & \citet{Gordon2017} & Air \\
    CH$_3$Cl (low-T) & HITRAN2016-CH$_3$Cl & \citet{Gordon2017} & Air \\
    \bottomrule
    \vspace{-0.1em}
    \end{tabular*}
    \textbf{Notes}: partition functions are obtained from the same references. Air broadening parameters are obtained from averaging HITRAN $\gamma_{\rm air}$ and $n_{\rm air}$ values over $J^{\rm low}$. SB07 is the \citet{Sharp2007} prescription for metal oxides (their Eq.15). Entries below the midline constitute a separate opacity database for low-temperatures (e.g. terrestrial planets). 
    \label{table:molecular_line_lists}
\end{table}

\begin{table}[ht!]
    \ra{0.84}
    \caption{Atomic and ionic line lists in the POSEIDON opacity database}
    \begin{tabular*}{\columnwidth}{l@{\extracolsep{\fill}} cccl@{}} \toprule 
    Atom & Line list & Partition function & Broadening & \\ \midrule
    Na & VALD3 & \citet{Barklem2016} & H$_2$ + He \\
    K & VALD3 & \citet{Barklem2016} & H$_2$ + He \\
    Li & VALD3 & \citet{Barklem2016} & H$_2$ + He \\
    Rb & VALD3 & \citet{Barklem2016} & H$_2$ + He \\
    Cs & VALD3 & \citet{Barklem2016} & H$_2$ + He \\
    Fe & VALD3 & \citet{Barklem2016} & H$_2$ + He \\
    Fe$^{+}$ & VALD3 & \citet{Barklem2016} & H$_2$ + He \\
    Ti & VALD3 & \citet{Barklem2016} & H$_2$ + He \\
    Ti$^{+}$ & VALD3 & \citet{Barklem2016} & H$_2$ + He \\
    Mg & VALD3 & \citet{Barklem2016} & H$_2$ + He \\
    Mg$^{+}$ & VALD3 & \citet{Barklem2016} & H$_2$ + He \\
    Ca & VALD3 & \citet{Barklem2016} & H$_2$ + He \\
    Ca$^{+}$ & VALD3 & \citet{Barklem2016} & H$_2$ + He \\
    Mn & VALD3 & \citet{Barklem2016} & H$_2$ + He \\
    \bottomrule
    \vspace{-0.2em}
    \end{tabular*}
    \textbf{Notes}: VALD3 \citep{Ryabchikova2015} provides compilations of atomic line lists from many reference sources. A sub-Voigt prescription is used for the Na and K resonance lines \citep{Burrows2003,Baudino2015}. 
    \label{table:atomic_line_lists}
\end{table}

\begin{table}[ht!]
    \ra{0.84}
    \caption{Continuum opacity sources in the POSEIDON opacity database}
    \begin{tabular*}{\columnwidth}{l@{\extracolsep{\fill}} ccl@{}} \toprule 
    \textbf{Rayleigh} & Reference for $\eta\,(\lambda)$ or $\bar{\alpha}\,(\lambda)$ & Reference for $F_{\rm{King}}\,(\lambda)$ \\ \midrule
    H$_2$ & \citet{Hohm1994} & \citet{Hohm1994} \\
    He & \citet{Cuthbertson1932} & 1.0  \\
    `` '' & \citet{Mansfield1969} & 1.0 \\
    H$_2$O & \citet{Hill1986} & \citet{Hinchliffe2006} \\
    CO$_2$ & \citet{Hohm1994} & \citet{Hohm1994} \\
    CH$_4$ & \citet{Sneep2005} & \citet{Sneep2005} \\
    NH$_3$ & \citet{Hohm1994} & \citet{Hohm1994} \\
    N$_2$ & \citet{Sneep2005} & \citet{Sneep2005} \\
    O$_2$ & \citet{Hohm1994} & \citet{Hohm1994} \\
    N$_2$O & \citet{Hohm1994} & \citet{Hohm1994} \\
    O$_3$ & \citet{Haynes2014} & \citet{Brasseur1986} \\
    CO & \citet{Haynes2014} & \citet{Bogaard1978} \\
    H$_2$S & \citet{Haynes2014} & \citet{Bogaard1978} \\
    SO$_2$ & \citet{Haynes2014} & \citet{Bogaard1978} \\
    C$_2$H$_2$ & \citet{Haynes2014} & \citet{Bogaard1978} \\ \midrule
    \textbf{CIA} (all from \citealt{Karman2019}) & \textbf{Other} (\citealt{John1988}) \vspace{0.2em} \\ \midrule
    H$_2$-H$_2$, H$_2$-He, H$_2$-H, H$_2$-CH$_4$ & H$^{-}$ (bound-free) \\ 
    O$_2$-O$_2$, O$_2$-N$_2$, O$_2$-CO$_2$ & \hspace{-0.5cm} H$^{-}$ (free-free) \\
    N$_2$-N$_2$, N$_2$-H$_2$O, N$_2$-H$_2$  \\
    CO$_2$-CO$_2$, CO$_2$-H$_2$, CO$_2$-CH$_4$ \\
    \bottomrule
    \vspace{-0.2em}
    \end{tabular*}
    \textbf{Notes}: He has $F_{\rm{King}} = 1$ due to spherical symmetry. All species without $F_{\rm{King}}\,(\lambda)$ data are taken to have $F_{\rm{King}} = 1.0 \; \forall \; \lambda $. The opacity outside of the tabulated wavelength range for a given species is set to zero.
    \label{table:continuum_opacity}
\end{table}

\newpage

\section{Model Computation Times with TRIDENT} \label{appendix_D}

Here we offer representative transmission spectra computation times with TRIDENT for 1D, 2D, and 3D models. We consider the 3D atmospheric model defined in Section~\ref{subsec:multidimensional_clouds}, which includes opacity from Na, K, H$_2$O, CO$_2$, Rayleigh scattering, and CIA. This model covers 25,257 wavelengths from 0.4--5\,$\micron$ ($R =$ 10,000) and has $N_{\rm{layer}} = 100$. In Table~\ref{table:TRIDENT_runtime}, we show the computation times (in seconds) for a single Intel$^{\circledR}$ Core™ i7-8565U CPU @ 1.80\,GHz for different user choices in the number of zenith slices, $N_{\rm{zenith}}$, and azimuthal sectors, $N_{\rm{sector}}$. The 3D model reduces to a 2D model when we set either $N_{\rm{zenith}}$ or $N_{\rm{sector}}$ to 1, and a 1D model when we set both to 1.

TRIDENT is sufficiently fast to allow 1D, 2D, and 3D atmospheric retrievals when coupled with a Bayesian sampling algorithm. We see that a 1D model, with sufficient spectral resolution and wavelength range for JWST near-IR datasets, takes just 70\,ms. We recommend a minimum of $N_{\rm{zenith}} = 6$ for 2D retrievals with day-night gradients (i.e. 4 zenith slices over the day-night opening angle alongside the dayside and nightside), which is $\approx 4 \times$ slower than the equivalent 1D model. We recommend a minimum of $N_{\rm{sector}} = 4$ for 2D retrievals with morning-evening gradients (i.e. 2 azimuthal sectors over the morning-evening opening angle alongside the morning and evening terminator sectors), which is $\approx 6 \times$ slower than the equivalent 1D model. We find these recommendations mitigate the impact of numerical errors to $<$ 10\,ppm for model spectra binned to $R=100$. Consequently, we propose $N_{\rm{zenith}} \geq 6$ and $N_{\rm{sector}} \geq 4$ for 3D retrievals of JWST data. The resulting model evaluation time of 1.69\,s on a single CPU core of a standard laptop readily scales to allow 3D retrievals on a modest multi-core machine or cluster.

\begin{table}[ht!]
    \aboverulesep=0ex
    \belowrulesep=0ex
    \ra{1.0}
    \caption{Transmission spectra computation times for 1D, 2D, and 3D models with TRIDENT}
    \begin{tabular*}{\columnwidth}{l@{\extracolsep{\fill}} c|ccccccl@{}} \toprule 
    & & \multicolumn{5}{c}{$N_{\rm{zenith}}$} \\
    & & 1 & 4 & 6 & 8 & 10 \\
    \midrule
    \multirow{5}{*}{$N_{\rm{sector}}$} & 1 & 0.07 & 0.20 & 0.29 & 0.36 & 0.49 \\
    & 4 & 0.42 & 1.22 & 1.69 & 2.31 & 2.67 \\
    & 6 & 0.69 & 1.88 & 2.62 & 3.45 & 4.09 \\
    & 8 & 0.91 & 2.61 & 3.90 & 5.08 & 5.79 \\
    & 10 & 1.18 & 3.24 & 4.94 & 6.25 & 8.38 \\
    \bottomrule
    \vspace{-0.2em}
    \end{tabular*}
    \textbf{Notes}: all computation times are in seconds on a single CPU core. Multi-core machines can attain a linear speedup via parallel model computation (e.g. within a retrieval code).  The 1D model has $N_{\rm{zenith}} = N_{\rm{sector}} = 1$. The 2D models with day-night gradients have $N_{\rm{zenith}} > 1$ and azimuthal symmetry ($N_{\rm{sector}} = 1$). The 2D models with morning-evening gradients have $N_{\rm{sector}} > 1$ and uniform day-night properties ($N_{\rm{zenith}} = 1$). The 3D models have $N_{\rm{zenith}} > 1$ and $N_{\rm{sector}} > 1$.
    \label{table:TRIDENT_runtime}
\end{table}

\bibliography{TRIDENT}{}
\bibliographystyle{aasjournal}

\end{document}